# Physical machine learning outperforms "human learning" in Quantum Chemistry

*Anton V. Sinitskiy, Vijay S. Pande*

*Department of Bioengineering, Stanford University, Stanford CA 94305*

**Two types of approaches to modeling molecular systems have demonstrated high practical efficiency. Density functional theory (DFT), the most widely used quantum chemical method, is a physical approach predicting energies and electron densities of molecules. Recently, numerous papers on machine learning (ML) of molecular properties have also been published. ML models greatly outperform DFT in terms of computational costs, and may even reach comparable accuracy, but they are missing physicality — a direct link to Quantum Physics — which limits their applicability. Here, we propose an approach that combines the strong sides of DFT and ML, namely, physicality and low computational cost. By generalizing the famous Hohenberg-Kohn theorems, we derive general equations for exact electron densities and energies that can naturally guide applications of ML in Quantum Chemistry. Based on these equations, we build a deep neural network that can compute electron densities and energies of a wide range of organic molecules not only much faster, but also closer to exact physical values than current versions of DFT. In particular, we reached a mean absolute error in energies of molecules with up to eight non-hydrogen atoms as low as 0.9 kcal/mol relative to CCSD(T) values, noticeably lower than those of DFT (down to ~3 kcal/mol on the same set of molecules) and ML (down to ~1.5 kcal/mol) methods. A simultaneous improvement in the accuracy of predictions of electron densities and energies suggests that the proposed approach describes the physics of molecules better than DFT functionals developed by "human learning" earlier. Thus, physics-based ML offers exciting opportunities for modeling, with high-theory-level quantum chemical accuracy, of much larger molecular systems than currently possible.**

From the Schrödinger equation (and relativistic corrections to it), one could in principle compute any physico-chemical property of any molecular system. A multitude of quantum chemical methods were developed to approximately solve the Schrödinger equation, with density functional theory (DFT) being nowadays the most widely used method.[1-4] DFT computes approximate values of energies $E$ and electron densities $\rho(\mathbf{r})$ of molecules (section S1), providing the basis for explanations and predictions of various properties of chemical compounds. DFT is based on the Hohenberg-Kohn (HK) theorems[1] proving that $E$ and $\rho(\mathbf{r})$ for the ground state of a molecule can be computed variationally, by minimizing a certain functional $E[\cdot]$ of $\rho(\mathbf{r})$. An explicit analytical expression for $E[\cdot]$ is unknown, and a multitude of versions of DFT, based on various approximations, have been proposed (in other words, *found by "human learning"*). Unfortunately, the accuracy of existing DFT methods in predicting $E$ (down to ~ 3 kcal/mol, see Section S2) does not reach the levels required for competing with experiments in most cases (e.g., the thermochemical accuracy of ~ 1 kcal/mol). Also, approximate DFT functionals are usually fitted and tested on $E$, but not $\rho(\mathbf{r})$ datasets, which may have ended up with many DFT functionals overfitted to $E$,[5,6] and therefore not transferable (that is, not capable to extrapolate) to some other classes of molecules or conformations. In addition to the concerns about the accuracy, practical applications of DFT are limited to molecular systems with only hundreds of atoms because of a quick polynomial growth in the cost of DFT computations with the size of a modeled system.

Recently, many authors used machine learning (ML) for much faster computations of $E$ and some other molecular properties (polarizabilities, HOMO-LUMO gaps, etc.).[7-19] However, neither currently used descriptors of molecules, nor ML models have a transparent connection to the basic physical concepts derived from the Schrödinger equation. For example, the vast majority of ML models cannot be extended to compounds with new elements without retraining, making them methodologically inferior even to least accurate DFT functionals, let alone the absolutely transferable Schrödinger equation. As a result, such approaches might be devoid of physics, and the criticism of ML[20] as "black boxes" and "curve-fitting" seems totally applicable to them.

In this work, we put forward a theoretically strict approach to ML in Quantum Chemistry. By generalizing the HK theorems,[1,21] we prove that the ground state energy $E$ and electron density $\rho(\mathbf{r})$ of an arbitrary molecular system, as well as all exact energies and electron densities of its excited states, are uniquely determined by its approximate ground state electron density $\rho_0(\mathbf{r})$ computed with a specific DFT functional or Hartree-Fock (HF) theory (Section S3). Combining this result with the idea of delta-ML, we arrive at the following exact equations:

$$\rho(\mathbf{r}) = \rho_0(\mathbf{r}) + \Delta\rho[\rho_0(\mathbf{r})],$$

$$E = E_0 + \Delta E[\rho_0(\mathbf{r})],$$

where $E_0$ is the approximate energy corresponding to $\rho_0(\mathbf{r})$ and computed with the same DFT or HF method, and $\Delta\rho[\cdot]$ and $\Delta E[\cdot]$ are a function-to-function and function-to-scalar (functional) mappings, respectively, each of which can be found with ML. These equations have two implications of fundamental importance for ML in Quantum Chemistry. First, it is an approximate electron density $\rho_0(\mathbf{r})$ computed with HF or a specific DFT method that can serve as a natural universal descriptor of a molecular system for the purpose of ML. Second, it is not necessary to use ML to learn the HK functional $E[\cdot]$ itself[22] and iteratively minimize it, as earlier suggested; instead, one can learn $\Delta\rho[\cdot]$ and $\Delta E[\cdot]$, and use them for direct, non-iterative, single-shot predictions of $E$ and $\rho(\mathbf{r})$.

Three-dimensional (3D) convolutional (Conv) deep neural networks (DNNs) are a natural ML model for this purpose (Fig. 1). Recently, DNNs have been successfully applied to various tasks in ML, in particular, computer vision and other manipulations with 2D and 3D data.[23] The input density $\rho_0(\mathbf{r})$ and the output correction to the density $\Delta\rho[\rho_0(\mathbf{r})]$ are 3D objects naturally admitting a 3D grid representation (so-called "cube representation"), a format that a 3D Conv DNN can directly work with. A 3D cube representation is more flexible than finite basis set expansions typically used in Quantum Chemistry, because it is not restricted to specific analytical expressions, and may provide higher accuracy in predicting $\Delta\rho[\rho_0(\mathbf{r})]$, and thereby $\rho(\mathbf{r})$, especially in the regions of low electron density important for non-covalent interactions. The "conv3d" transformation in the first layer of a DNN can compute, among other entities, gradients and hessians of the density with respect to $\mathbf{r}$, the variables that often appear in analytical approximations in Quantum Chemistry, and play key roles in previously developed advanced (GGA, meta-GGA and hybrid) DFT functionals.[24] At the last layer of the DNN, the use of "conv3d" (for $\Delta\rho$) or "average_pooling3d" (for $\Delta E$) operations is a natural way to estimate integrals present in analytical expressions for $\Delta\rho[\cdot]$ and $\Delta E[\cdot]$ [Fig. 1c,d; Eqs. (S73), (S74)]. Finally, the mentioned analytical expressions consist of similar building blocks, or "precursors" [schematically denoted as $\delta E/\delta\rho$ and $(\delta^2 E/\delta\rho^2)^{-1}$ in Fig. 1] that need to be computed first, and only after that do computations fork into two paths leading to $\Delta\rho$ and $\Delta E$. The corresponding forking architecture of DNN is not only physically motivated, but also, from the viewpoint of ML, prevents overfitting of the network to the specific task of predicting $E$, and uses the rich information containing in $\rho(\mathbf{r})$ as a physics-based regularizer of the DNN. Overall, a physically driven approach to Quantum Chemistry can mesh well with DNN structures. Intermediate layers in DNNs may learn physically well-defined concepts or "precursors" of the output, and first and last layers can perform operations (such as differentiation or integration) that are present in "humanly learned" approaches to the problem.

To validate this research philosophy, we have developed a "physical machine learning in Quantum Chemistry" (PML-QC) model, which is a 3D Conv DNN that can compute $E$ and $\rho(\mathbf{r})$ of a large set of organic molecules (Fig. 1; see also Section S4). The model was trained in two stages: first, on cheaper and not-so-accurate DFT data, resulting in the PML-QC$_{DFT}$ version of the model[25] (Fig. 2), and after that, on more accurate, but much more expensive (and therefore more scarce) CCSD [for $\rho(\mathbf{r})$] and CCSD(T) (for $E$) data, resulting in the PML-QC$_{CCSD(T)}$ version of the model (Fig. 3, Table 1). On test sets of molecules, not used for training or validation, PML-QC$_{DFT}$ demonstrated the accuracy comparable to that of best versions of DFT, and PML-QC$_{CCSD(T)}$ a performance significantly better than DFT (Table 1, Sections S4 and S5). In particular, the mean absolute error in $E$ predicted with PML-QC$_{CCSD(T)}$ was as low as ~0.9 kcal/mol (Table 1), below the psychological threshold of "thermochemical accuracy" of ~1 kcal/mol required for practical applications, and below the accuracy of 1.5 kcal/mol reported[18] for an ML model that was trained and tested on CCSD(T)-level $E$ (and not DFT-level $E$) for a large dataset of molecules (Section S5). We attribute this excellent performance of the model primarily to the use of an approximate electron density as the input to a neural network, which, even at the HF level, is a more physically meaningful and information rich representation of a molecule than previously used featurizations.

Thus, in this work we demonstrate that physics-based ML in Quantum Chemistry can outperform "humanly learned" versions of DFT, the most widely used quantum chemical method nowadays, in the accuracy of $E$ and $\rho(\mathbf{r})$ predictions and the speed of computations, all at the same time. Our theoretical analysis (a generalization of the HK theorems leading to the formulas given above) explains the astonishing fact that a single input channel with an approximate density $\rho_0(\mathbf{r})$ was sufficient for the network to predict both $E$ and $\rho(\mathbf{r})$, and no other input channels [e.g., with atomic numbers, geometries, electrostatic potentials, gradients and Hessians of $\rho_0(\mathbf{r})$[17]] were required. Not only does our framework predict $\rho(\mathbf{r})$ as a physical measurable, but also uses $\rho(\mathbf{r})$ for regularization of the neural network, pushing it towards learning physics rather than curve fitting. Similarly to post-HF methods, such as MP2 or CCSD(T), PML-QC can build on an approximate HF solution; unlike post-HF methods, it may also build on DFT solutions, and accounting for correlations does not create a computational bottleneck in PML-QC. We foresee wide-scale applications of the approach proposed here (after certain technical developments) to computer modeling of various large molecular systems, including biomolecular or artificial material systems of much practical interest, and highly accurate and fast modeling of excited states of molecules and chemical reactions involving bond breaking and formation.

**Conflict of interests.** V.S.P. is a consultant & SAB member of Schrodinger, LLC and Globavir, sits on the Board of Directors of Apeel Inc, Asimov Inc, BioAge Labs, Freenome Inc, Omada Health, Patient Ping, Rigetti Computing, and is a General Partner at Andreessen Horowitz.

**Acknowledgements**. The authors would like to thank other current and past members of the Pande lab, in particular, Dr. Peter Eastman, Prof. Lee-Ping Wang, and Prof. Joseph Gomes, for valuable feedback on our results and manuscript. The Pande Group acknowledges the generous support of Dr. Anders G. Frøseth and Mr. Christian Sundt for our work on machine learning. The Pande Group is broadly supported by grants from the NIH (R01 GM062868 and U19 AI109662) as well as gift funds and contributions from Folding@home donors.

**Table 1.** Mean absolute errors (MAEs) in energies predicted by PML-QC after the first stage of training (PML-QC$_{DFT}$), in comparison to CCSD(T) energies (section S2), were nearly the same as MAEs in the original QM9 energies computed with DFT (~5 kcal/mol). After transfer learning on CCSD(T) energies, the MAEs of the neural network (PML-QC$_{CCSD}$) dropped to ~0.9 kcal/mol.

| Subset / molecule | QM9 error, kcal/mol | PML-QC$_{DFT}$ error, kcal/mol | PML-QC$_{CCSD}$ error, kcal/mol |
|---|---|---|---|
| Training (3830 molecules, indices up to 6000) | 6.0 | 6.0 | **0.82** |
| Validation (470 molecules, indices up to 6000) | 5.4 | 5.5 | **0.80** |
| Test (462 molecules, indices up to 6000) | 4.4 | 4.5 | **0.87** |

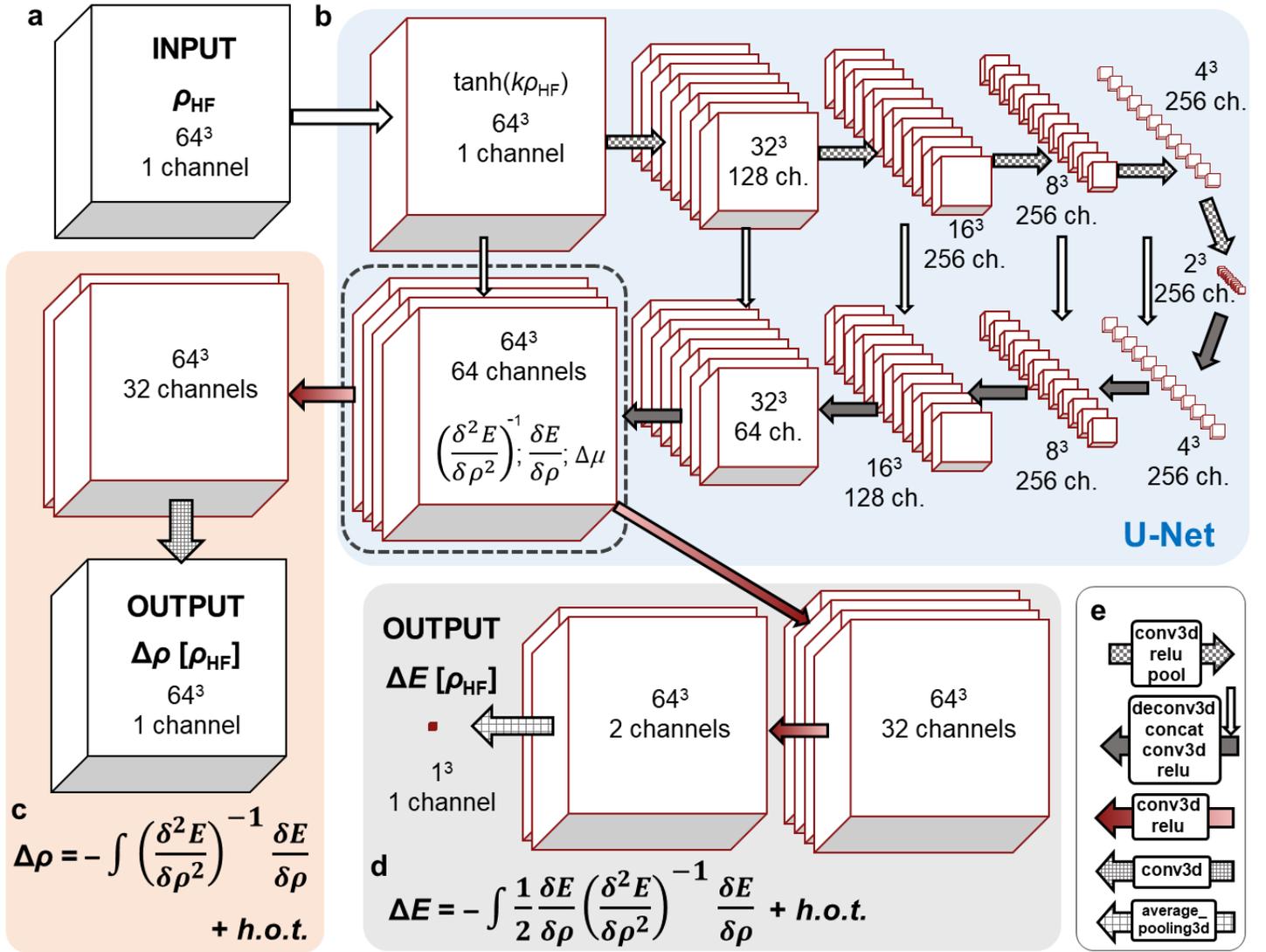

**Fig. 1. Proposed machine learning approach to Quantum Chemistry has deep physical motivation.** (a) The only input to the network, according to the generalization of the HK theorem proven here, is an approximate electron density [in this work, a density computed with the Hartree-Fock (HF) theory $\rho_{HF}(\mathbf{r})$, in a cube file representation on a 64×64×64 grid]. (b) U-Net[26,27] part, *blue*, serves to learn high-level features characterizing the modeled molecule, *gray dashed box* [presumably features related to the first-order functional derivative $\delta E/\delta\rho$, the resolvent $(\delta^2 E/\delta\rho^2)^{-1}$, and the chemical potential $\Delta\mu$, see Eqs. (S73) and (S74)]. Then, the network forks into two paths, one of which leads to the output electron density, *peach*, and the other one to the output energy, *gray*. This forking architecture is justified by the fact that the equations for $\Delta\rho$ and $\Delta E$ (see c,d) consist of similar building blocks. "ch.", shorthand for "channels". (c) The correction to the electron density $\Delta\rho(\mathbf{r})$ can be perturbatively written as the value of a function-to-function mapping $\Delta\rho[\cdot]$ evaluated at $\rho_{HF}(\mathbf{r})$. This mapping $\Delta\rho[\cdot]$ is uniquely defined by $E[\cdot]$; *h.o.t.*, higher order terms. (d) Similarly, the correction to the energy $\Delta E = E[\rho] - E[\rho_{HF}]$ can be computed as the value of a functional $\Delta E[\cdot]$ at $\rho_{HF}(\mathbf{r})$. This functional $\Delta E[\cdot]$ is also uniquely defined by $E[\cdot]$. (b,c) Instead of finding $E[\cdot]$ better than in the existing DFT methods, we suggest that a deep neural network (DNN) be trained to directly approximate $\Delta\rho[\cdot]$ and $\Delta E[\cdot]$. Panels (c,d) show sketches of formulas; for mathematically strict expressions, see Eqs. (S73) and (S74). (e) Legend for the arrows in the DNN architecture.

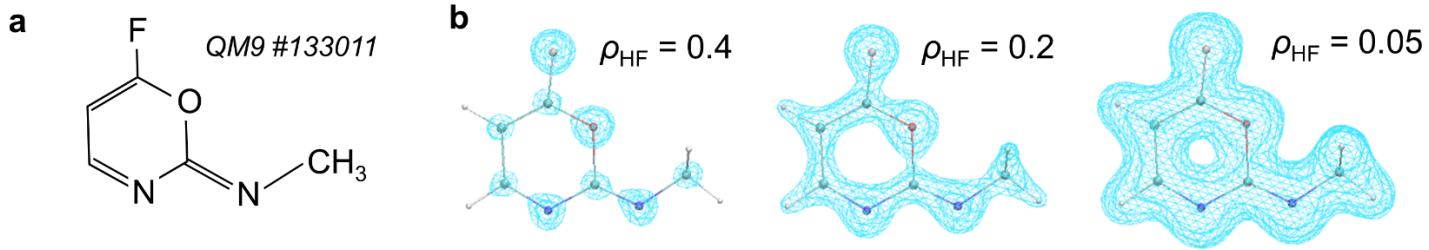

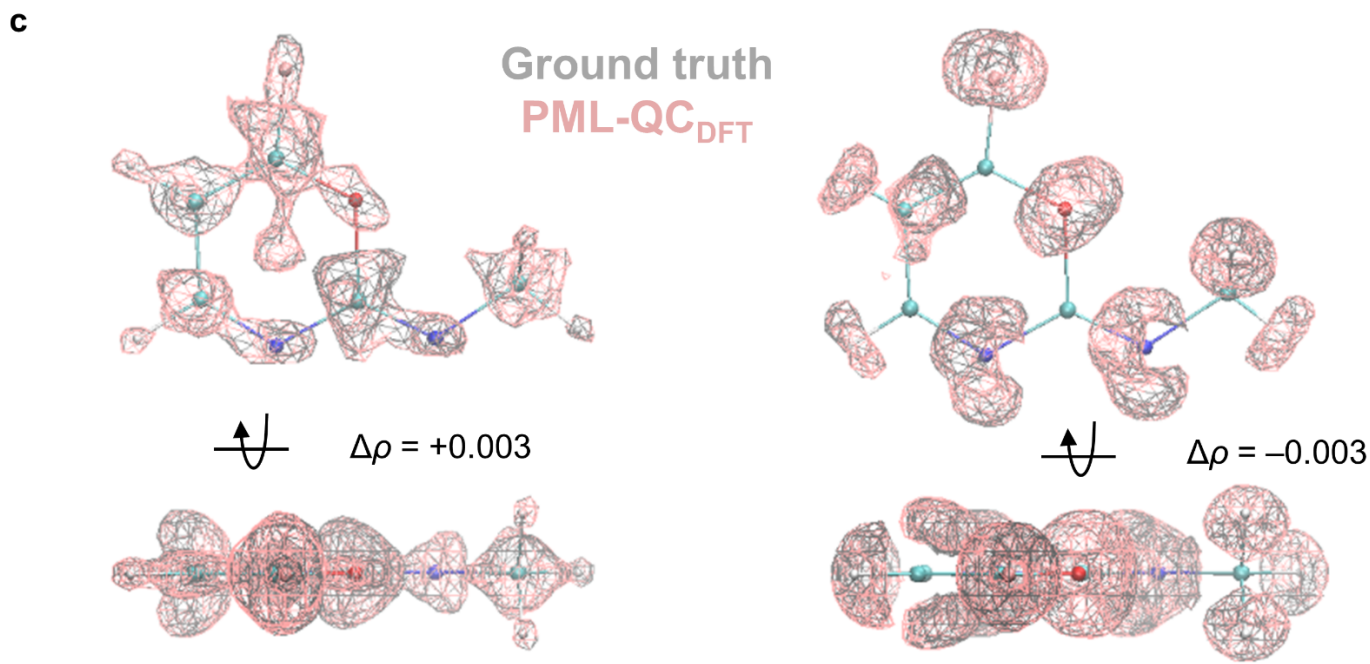

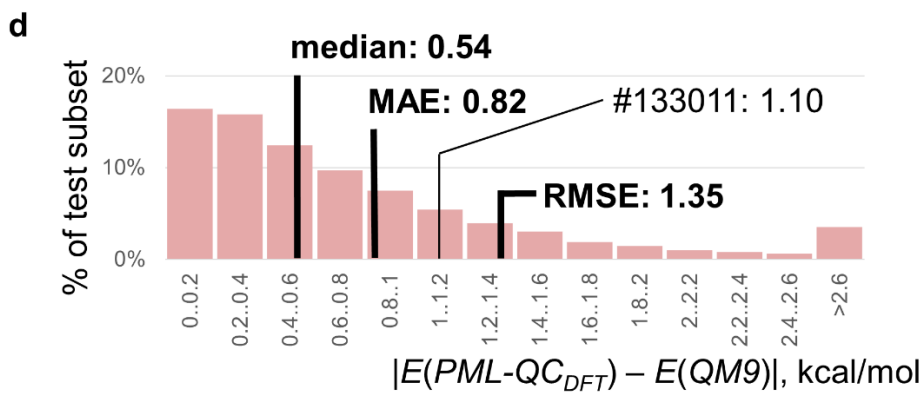

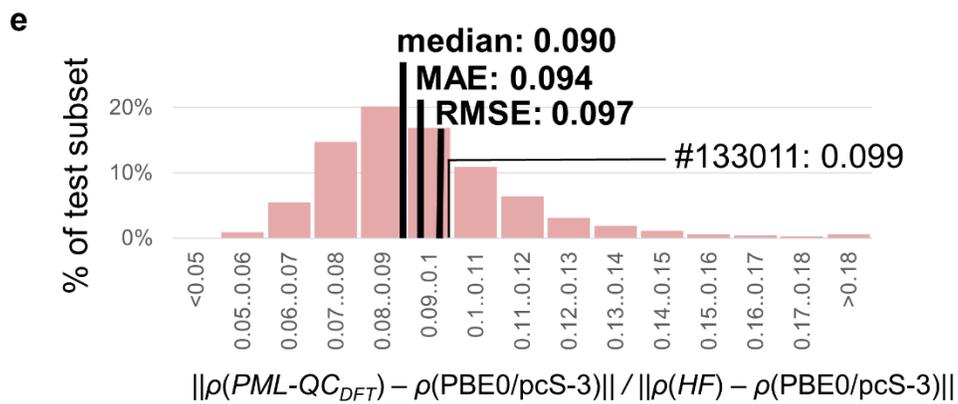

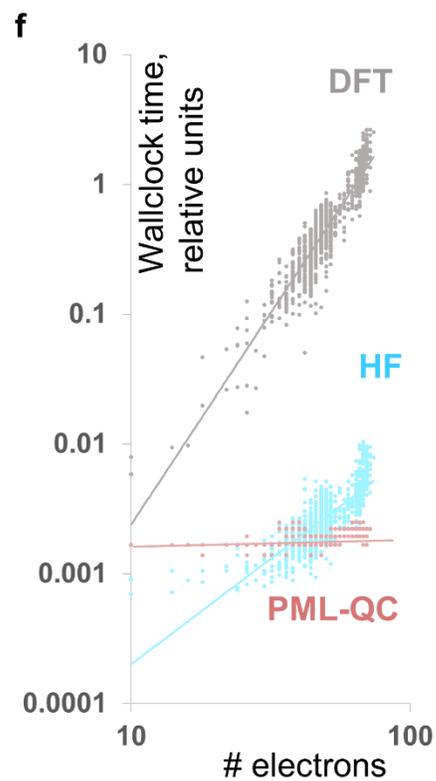

◄ **Fig. 2. PML-QC$_{DFT}$, the network we trained on DFT data, predicts electron densities and energies with the accuracy comparable to DFT, but does so several orders of magnitude faster**. (a) Molecules from the QM9 database[28] (nearly 134,000 organic molecules) were used for training, validation and testing. Performance of PML-QC$_{DFT}$ is illustrated here for the molecule with QM9 index 133011. This molecule belongs to the test subset, has a large QM9 index, is diverse in terms of elemental composition and functional groups, and can serve as a conservative example of the excellent performance of PML-QC$_{DFT}$ (the accuracies of the energy and density predictions for it are somewhat worse than on average for the test set). (b) Approximate electron density $\rho_{HF}(\mathbf{r})$ is quickly computed with HF in a small basis set (HF/cc-pVDZ). Isosurfaces for $\rho_{HF}(\mathbf{r})$ = 0.4, 0.2, and 0.05 shown in *light blue* (here and below, electron densities are given in Hartree units). See also Movie S1. (c) Correction to the electron density $\Delta\rho(\mathbf{r})$ predicted by PML-QC$_{DFT}$, *pink*, is very close to the ground truth value (computed with PBE0[29]/pcS-3[30]), *gray*. See also Movie S2. (d) Distribution diagram for the absolute differences of predicted and ground truth [from DFT, B3LYP/6-31G(2df,p)[28]] values of $\Delta E$, for molecules in the test subset (test set 1, see Sections S4.3-4.4). In a half of all molecules, the differences are below 0.54 kcal/mol, and the distribution density decays fast with energy differences. (e) The same, for the *L*1 measure of the deviations of predicted electron densities from PBE0/pcS-3 densities, divided by the *L*1 measure of the deviations of HF/cc-pVDZ densities from PBE0/pcS-3 densities. In most cases, the electron densities predicted with PML-QC$_{DFT}$ are ~10–12 times closer to the ground truth densities that the HF densities. (f) Scaling of the wallclock time to run DFT, PML-QC and HF computations with the number of electrons for various QM9 molecules (note log scales on both axes) shows that the cost of PML-QC computations is not a computational bottleneck for molecules of practically important sizes (Section S4.5).

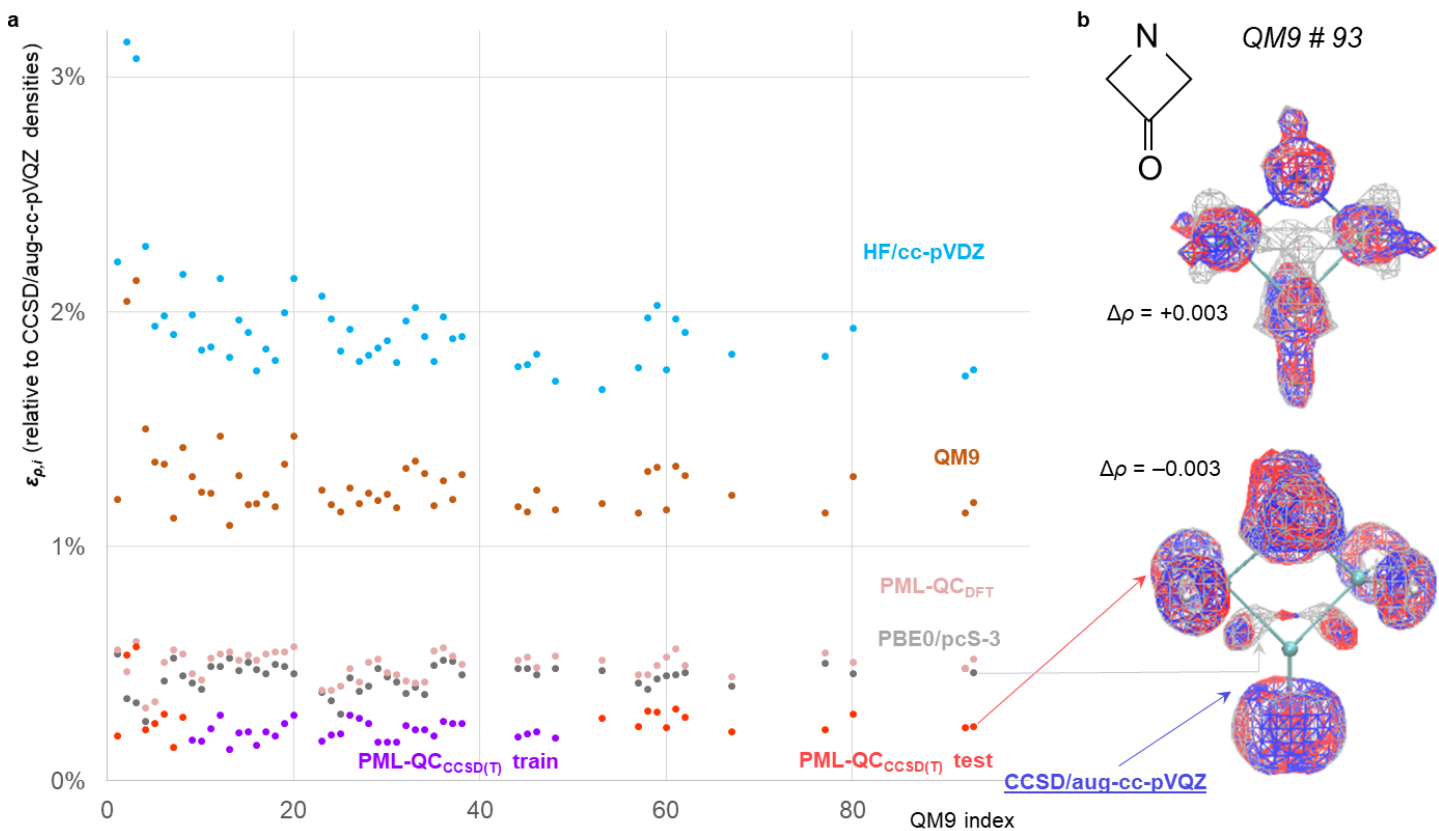

**Fig. 3**. **Transfer learning on CCSD electron densities significantly improves the performance of the neural network (PML-QC$_{CCSD(T)}$)**. (a) *L*1 measure of the difference of electron densities [for a strict definition,[16,31] see Eq. (S13)] predicted with HF/cc-pVDZ (*light blue*), B3LYP/6-31G(2df,p) (*orange*), PBE0/pcS-3 (*gray*), PML-QC$_{DFT}$ (*pink*) and PML-QC$_{CCSD(T)}$ (training set, *violet*, test set, *red*) from CCSD/aug-cc-pVQZ electron densities. (b) Comparison of the electron densities of molecule with QM9 index 93 computed with DFT (PBE0/pcS-3, here and below $\Delta\rho(\mathbf{r})$ shown for clarity, *gray*), CCSD/aug-cc-pVQZ (*blue*), and PML-QC$_{CCSD(T)}$ (*red*) demonstrates that the output of PML-QC$_{CCSD(T)}$ is much closer to CCSD than DFT results (see also Movie S3).

# Physical machine learning outperforms "human learning" in Quantum Chemistry

*Anton V. Sinitskiy, Vijay S. Pande*

*Department of Bioengineering, Stanford University, Stanford CA 94305*

# SUPPLEMENTARY INFORMATION

## S1. Definitions and basic relationships

In general, an isolated quantum mechanical system (for example, a molecule in vacuum) can be characterized by a wave function $\Psi$. A non-relativistic electronic wave function of a molecule with $N$ electrons $\Psi(\mathbf{r}_1, …, \mathbf{r}_N, \tau_1, …, \tau_N)$ depends on $3N$ spatial coordinates $\mathbf{r}_1, …, \mathbf{r}_N$ and $N$ spin coordinates $\tau_1, …, \tau_N$ and can be found, together with the corresponding electronic energy $E$, from the stationary Schrödinger equation (assuming the Born–Oppenheimer approximation to separate the motion of atomic nuclei):

$$\widehat{H}_e \Psi(\mathbf{r}_1, …, \mathbf{r}_N, \tau_1, …, \tau_N) = E \Psi(\mathbf{r}_1, …, \mathbf{r}_N, \tau_1, …, \tau_N), \tag{S1}$$

where $\widehat{H}_e$ is the electronic Hamiltonian for a given molecular system. For systems of practical interest, $N \gg 1$, making it difficult to work with wave functions because of their high dimensionality. On the contrary, the electron density, defined as

$$\rho(\mathbf{r}) = N \sum_{\tau_1} … \sum_{\tau_N} \int d\mathbf{r}_2 … \int d\mathbf{r}_N |\Psi(\mathbf{r}_1, …, \mathbf{r}_N, \tau_1, …, \tau_N)|^2, \tag{S2}$$

is a function of only a three-dimensional vector $\mathbf{r}$, regardless of the size of the molecule. Hence, the electron density is much easier to deal with than the wave function from various viewpoints, including intuitive appreciation, graphical visualization and basis set expansions. This function shows how electrons are delocalized in a molecule, namely, where the probability to find an electron per unit volume is higher and where it is lower.

According to the Hohenberg–Kohn theorems,[1] the ground state energy $E$ of a molecular system can be found by a minimization of a certain density functional $E[n(\mathbf{r})]$; the density $n$ at which the minimum is achieved yields the ground state electron density $\rho(\mathbf{r})$:

$$E = \min_{n(\mathbf{r})} E[n(\mathbf{r})], \quad \rho(\mathbf{r}) = \operatorname*{argmin}_{n(\mathbf{r})} E[n(\mathbf{r})]. \tag{S3}$$

The Hohenberg–Kohn theorems serve as the motivation for Density Functional Theory (DFT), one of the main methods in Quantum Chemistry.[1,3,4,24,32,33] An exact expression for the density functional $E[\cdot]$ is not known. Numerous approximate functional have been proposed, leading to multiple versions of DFT. Overall, DFT has been widely used by the scientific community because it offers a favorable tradeoff between the cost of computations and the accuracy of results, and deals with electron densities, which seem simpler for common (that is, based on macroscopic classical physics) intuition than multidimensional wave functions.

In practice, DFT has been commonly used to predict energies, but seldom to predict electron densities. It was even argued that a DFT model does not have to predict exact densities in order to predict accurate energies.[21,34,35] For example, an approximate electron density computed with an exchange-only DFT functional in principle suffices for computing its exact ground state energy[21] (below in Section S3 we prove a more general result on approximate electron densities). However, we do not share an opinion that approximate densities computed with DFT should be considered only "as a (very useful) label for the different potentials".[36] Computations of electron densities in DFT provide a powerful tool to check the physicality of a DFT functional, and researchers should employ this opportunity,[5,6,37,38] especially for DFT potentials fitted purely on energies. In this work, we also compare electron densities predicted with ML to those predicted with accurate *ab initio* methods as one of possible ways to estimate the physicality of ML models.

## S2. Comparison of the performance of various DFT functionals and *ab initio* methods

### S2.1. On the choice of measures of accuracy

How to measure the accuracy of a quantum chemical method (or an ML model predicting the same physical variables, such as energies and electron densities)? Several measures of error are widely used in practice. For energy, the commonly used measures are mean absolute error (MAE) and root mean square error (RMSE), which, in the terminology accepted in the field of ML, correspond to $L1$ and $L2$ measures of the error, respectively:

$$MAE = \frac{1}{N_M} \sum_{i=1}^{N_M} |E_i - E_{ref,i}|, \quad RMSE = \sqrt{\frac{1}{N_M} \sum_{i=1}^{N_M} (E_i - E_{ref,i})^2}, \quad (S4)$$

where $N_M$ is the number of molecules in the test set, $E_i$ is an approximate energy computed with the evaluated method for molecule $i$, $E_{ref,i}$ is the exact (reference) energy of the same molecule. Since larger molecules tend to have larger errors, MAE and RMSE values are sometimes computed for energy errors per "heavy" (non-hydrogen) atom. For example, MAE per heavy atom can be computed as:

$$MAE\ per\ heavy\ atom = \frac{1}{N_M} \sum_{i=1}^{N_M} \frac{|E_i - E_{ref,i}|}{n_{ha,i}}, \quad (S5)$$

where $n_{ha,i}$ is the number of heavy atoms in molecule $i$. Such averaging can also be performed with weights proportional to the number of heavy atoms in each molecule, by analogy with $\varepsilon_\rho$ measure of density error introduced below in Eq. (S13). The corresponding error measure, which we denote here $\varepsilon_E$, can be easily found from the MAE value:

$$\varepsilon_E = \sum_{i=1}^{N_M} \frac{n_{ha,i}}{\sum_{i=1}^{N_M} n_{ha,i}} \cdot \frac{|E_i - E_{ref,i}|}{n_{ha,i}} = \frac{MAE}{\bar{n}_{ha}}, \tag{S6}$$

where $\bar{n}_{ha} = \sum_{i=1}^{N_M} n_{ha,i}/N_M$ is the average number of heavy atoms per molecule in the dataset. Note, however, that physical accuracy of a model does not necessarily reduces to its accuracy in predicting energies (or their differences). For example, in classical statistical mechanics, the value of an observable $O$ can be written as an average over a relevant statistical ensemble, say a canonical ensemble:

$$O = \frac{\int d\mathbf{r}\, O(\mathbf{r}) e^{-U(\mathbf{r})/kT}}{\int d\mathbf{r}\, e^{-U(\mathbf{r})/kT}}, \tag{S7}$$

where $O(\mathbf{r})$ is the microscopic counterpart of $O$ (here we assume it depends only on the coordinates $\mathbf{r}$, and not the momenta $\mathbf{p}$) and $U(\mathbf{r})$ is the potential energy. Evidently, the error in a predicted value of $O$ cannot be exhaustively expressed in terms of MAE or RMSE in the energy $U(\mathbf{r})$, because this equation involves Boltzmann factors nonlinearly depending on $U(\mathbf{r})$. Thus, even though MAE or RMSE values are widely used, they do not provide unique measures of errors of models in predicting physical observables.

The same refers to measuring errors in predicted electron densities. It was pointed out that no single measure of these errors exists, and even argued that any quantum chemical method could be shown to be the best at predicting electron densities with a carefully (though artificially) chosen measure of error.[39] If the only purpose of quantum chemical modeling is to predict energies, then the so-called ideal metric for the error in an approximate electron density $\rho_0$

$$\Delta E^{id}[\rho_0] = E[\rho_0] - E[\rho], \tag{S8}$$

where $E[\cdot]$ is the exact density functional and $\rho$ is the exact electron density, can serve to this end, accounting for the sensitivity of the target variable $E$ to the electron density.[34,35,39,40] In practice, it is difficult to compute $\Delta E^{id}$, but a good approximation to it is provided by the density driven error $\Delta E_D$ defined as

$$\Delta E_D[\rho_0] = E_0[\rho_0] - E_0[\rho], \tag{S9}$$

where $E_0[\cdot]$ is an approximate density functional used to variationally compute $\rho_0$.[34,35,39,40] The analysis of $\Delta E_D$ values revealed important aspects of the relationship between the accuracy of DFT methods in predicting energies and electron densities, in particular, allowed to distinguish normal and abnormal DFT calculations.[34,35,39,40] However, the electron energy is not the only property of interest for molecular systems. For example, a practically important problem of predicting NMR shifts implies a different measure of error in terms of electron densities.[39] More generally, for a molecular system in a state described by wave function $\Psi$, the observable $O$ corresponding to a quantum mechanical operator $\hat{O}$ is

$$O = \int d\mathbf{p} d\mathbf{r}\, \Psi^*(\mathbf{r},\mathbf{p}) \hat{O} \Psi(\mathbf{r},\mathbf{p}), \tag{S10}$$

with $\Psi$ uniquely determined (up to a phase), according to the Hohenberg–Kohn theorem, by the electron density $\rho$. In this general case, the error in the value of $O$ cannot be expressed in terms of any universal measure of error in the value of $\rho$. A more general expression from quantum statistical mechanics does not provide a universal expression for an error in an arbitrary physical observable in terms of errors in energies and electron densities either. Despite this absence of a unique measure of error in electron densities, one may follow the same approach as in the case of measuring the errors in energies, and use simple metrics of the error to get an overall estimate of the accuracy.

Previously, $L1$ measure of the error in electron density, normalized to the total number of electrons, has been used.[16,31] For a single molecule, the $L1$ measure can be written as

$$L1_i = \|\rho_i(\mathbf{r}) - \rho_{ref,i}(\mathbf{r})\| = \int d\mathbf{r} \, |\rho_i(\mathbf{r}) - \rho_{ref,i}(\mathbf{r})|, \tag{S11}$$

where $\rho_i(\mathbf{r})$ and $\rho_{ref,i}(\mathbf{r})$ are the evaluated and reference ('exact') electron densities for molecule $i$, respectively, and the integration is performed over the whole volume. Normalization to the total number of electrons in a molecule is required because these $L1$ measures are expected to grow with the size of a molecule:

$$\varepsilon_{\rho,i} = \frac{\int d\mathbf{r} \, |\rho_i(\mathbf{r}) - \rho_{ref,i}(\mathbf{r})|}{n_{e,i}}, \tag{S12}$$

where $n_{e,i} = \int d\mathbf{r} \, \rho_{ref,i}(\mathbf{r}) = \int d\mathbf{r} \, \rho_i(\mathbf{r})$. Averaging over $N_M$ molecules in the cited papers was performed with weights proportional to the number of electrons in each molecule:

$$\varepsilon_\rho = \sum_{i=1}^{N_M} \frac{n_{e,i}}{\sum_{i=1}^{N_M} n_{e,i}} \cdot \varepsilon_{\rho,i} = \frac{\sum_{i=1}^{N_M} \int d\mathbf{r} \, |\rho_i(\mathbf{r}) - \rho_{ref,i}(\mathbf{r})|}{\sum_{i=1}^{N_M} \int d\mathbf{r} \, \rho_{ref,i}(\mathbf{r})} = \frac{\overline{L1}}{\bar{n}_e}, \tag{S13}$$

where $\overline{L1} = \sum_{i=1}^{N_M} L1_i / N_M$ is the mean value of $L1_i$ over all molecules in the dataset, and $\bar{n}_e = \sum_{i=1}^{N_M} n_{e,i} / N_M$ is the mean number of electrons in a molecule.

These measures are not linked to a specific property of a system, even as important as energy. Like the above-mentioned MAE values for energies, $\varepsilon_\rho$ and $\varepsilon_{\rho,i}$ are in essence $L1$ measures of the error commonly used in ML, and provide an overall estimate of the accuracy of the results in various regions in space ($L2$ measures for electron densities would be less informative for predictions of chemical properties, because electron densities are much greater on nuclear cores than on chemical bonds). These measures are easy to compute for the electron densities predicted by the ML models we present here, while the corresponding computations of $\Delta E^{id}$ and $\Delta E_D$ would require a separate implementation. Finally, the use of $\varepsilon_\rho$ and $\varepsilon_{\rho,i}$ measures of errors in electron densities makes our results directly comparable to previous results in the literature on electron density predictions.

The focus of this paper is on building an ML model capable of learning physics, rather than interpolating data. Hence, measuring the accuracy in energy and electron density predictions pursues here the following specific purposes:

1) To choose reasonable quantum chemical methods and basis sets to generate ground truth data for two stages of training of our ML model (the first stage: initial training in a data rich regime, with a less expensive and therefore less accurate quantum chemical method; the second stage: transfer learning on results of a significantly more accurate method in a data-scarce regime). In either case, we do not have to find the best possible method and basis set; a reasonable choice that is not evidently worse than other good choices suffices.
2) To train and validate our ML models with the use of existing ML methodology and software, which implies using common (to the field of ML) error measures.
3) To preliminarily estimate the performance of the proposed models in comparison to previously suggested DFT and ML models with the use of simple criteria. ***An exhaustive evaluation of the performance of the proposed ML model can be based only on its large-scale applications to multiple molecular systems of practical interest to predict various properties***, which will involve the use of various case-specific measures of errors, and goes much beyond the scope of this paper.

We think that the $L1$ measures of errors in energies and densities used in this work suffice for these purposes. These measures are conceptually and computationally simple, and lead to physically reasonable results for comparisons of various quantum chemical methods to each other (see below). We invite the readers to use other physically motivated error measures to further analyze the relative performance of quantum chemical methods (including DFT) and ML models on large molecular datasets, including the datasets published with this work.

*S2.2. Comparison to highly accurate ab initio results for a few molecules*

We have been able to compute $\rho_{ref,i}(\mathbf{r})$ of the first six molecules in the QM9 database ($CH_4$, $NH_3$, $H_2O$, $C_2H_2$, HCN, $CH_2O$; for a description of the QM9 database, see Appendix A1) with a high level *ab initio* quantum chemical method (CCSD, without a frozen core approximation) often used as a source of 'exact' wave functions, in a large basis set (cc-pCV5Z; for technical details, see Appendix A2). Rapid increase of the cost of computations with the molecule size for high level *ab initio* quantum chemical methods prevents us from carrying out such computations for more than a few first entries in the QM9 database.

For the same six molecules, we also computed $\rho_i(\mathbf{r})$ using various combinations of DFT functionals and basis sets. In total, 2643 combinations were screened (excluding combinations for which computations did not converge for more than two molecules); for details, see Appendix A2. For each DFT functional, basis set and molecule, an $\varepsilon_\rho$ measure of the accuracy of the predicted electron density $\rho(\mathbf{r})$ was calculated by Eqs. (S11) and (S13). The combinations of DFT functionals and basis sets were sorted by $\varepsilon_\rho$ over these six molecules. The top of the list is given in Table S1. From the analysis of the list we conclude that:

- combinations of DFT functionals and basis sets with low values of $\varepsilon_{\rho,i}$ for some of these six molecules tend to have low values of $\varepsilon_{\rho,i}$ for the other molecules, and vice versa;
- the top of the list is dominated by two families of functionals (ωB97X-D[41] and related,[42] and PBE0[29,43] and related), and one family of basis sets (pcS sets[30]);
- pcS-3-level basis sets yield values of $\varepsilon_{\rho,i}$ as low as V5Z-level basis sets do; neither a shift to pcS-4-level basis sets, nor augmentation of pcS-3-level basis sets significantly improve $\varepsilon_{\rho,i}$.

**Table S1**. $\varepsilon_{\rho,i}$ measures (all in %, explicitly shown only in the top row) of the difference between 3D electron densities predicted by various DFT methods with various basis sets, and 'exact' (CCSD/cc-pCV5Z) electron densities [Eq. (S12)]. Data are given for the first six molecules in the QM9 database. Top 20 combinations of functionals and basis sets, in the order of increasing $\varepsilon_\rho$, are shown. Also, some other functionals are given for comparison, each with the basis set providing the highest $\varepsilon_\rho$ value for this functional. Double horizontal lines show that some rows were omitted. DFT functionals are named by the corresponding keywords from Gaussian or Q-Chem, except for PBE0 (Gaussian keyword 'PBE1PBE').

| DFT functional | basis set | 1 | 2 | 3 | 4 | 5 | 6 | $\varepsilon_\rho$ |
|---|---|---|---|---|---|---|---|---|
| ωB97X-D3 | pcS-3 | 0.45% | 0.41% | 0.39% | 0.31% | 0.25% | 0.36% | 0.35% |
| ωB97X-D3 | aug-pcS-3 | 0.45 | 0.42 | 0.41 | 0.31 | 0.26 | 0.37 | 0.36 |
| OHSE2PBE | pcS-3 | 0.52 | 0.42 | 0.40 | 0.29 | 0.24 | 0.44 | 0.38 |
| PBE0 | pcS-3 | 0.58 | 0.43 | 0.39 | 0.28 | 0.24 | 0.44 | 0.38 |
| OHSE2PBE | aug-pcS-3 | 0.52 | 0.43 | 0.41 | 0.30 | 0.25 | 0.45 | 0.38 |
| ωB97X-D3 | cc-pV5Z | 0.49 | 0.40 | 0.35 | 0.38 | 0.32 | 0.40 | 0.39 |
| PBE0 | aug-pcS-3 | 0.58 | 0.44 | 0.40 | 0.28 | 0.24 | 0.45 | 0.39 |
| ωB97X-D | pcS-3 | 0.48 | 0.47 | 0.42 | 0.35 | 0.28 | 0.42 | 0.39 |
| PBEh1PBE | pcS-3 | 0.57 | 0.43 | 0.41 | 0.31 | 0.26 | 0.45 | 0.39 |
| OHSE1PBE | pcS-3 | 0.57 | 0.43 | 0.41 | 0.32 | 0.26 | 0.46 | 0.40 |
| HSEH1PBE | pcS-3 | 0.57 | 0.43 | 0.41 | 0.32 | 0.26 | 0.46 | 0.40 |
| PBEh1PBE | aug-pcS-3 | 0.57 | 0.44 | 0.42 | 0.31 | 0.26 | 0.45 | 0.40 |
| OHSE1PBE | aug-pcS-3 | 0.57 | 0.45 | 0.42 | 0.31 | 0.26 | 0.46 | 0.40 |
| HSEH1PBE | aug-pcS-3 | 0.57 | 0.45 | 0.42 | 0.31 | 0.26 | 0.46 | 0.40 |
| ωB97X-D | aug-pcS-3 | 0.49 | 0.48 | 0.44 | 0.35 | 0.28 | 0.43 | 0.40 |
| revPBE0 | pcS-3 | 0.50 | 0.50 | 0.45 | 0.29 | 0.29 | 0.45 | 0.40 |
| ωB97X-D3 | pcS-4 | 0.48 | 0.46 | 0.45 | 0.36 | 0.32 | 0.40 | 0.41 |
| ωB97X-D3 | aug-pcS-4 | 0.49 | 0.46 | 0.45 | 0.37 | 0.32 | 0.41 | 0.41 |
| revPBE0 | aug-pcS-3 | 0.51 | 0.51 | 0.46 | 0.30 | 0.30 | 0.45 | 0.41 |
| ωB97X-D3 | pc-4 | 0.49 | 0.46 | 0.46 | 0.38 | 0.33 | 0.41 | 0.41 |
| ωB97X-V | pcS-3 | 0.48 | 0.45 | 0.47 | 0.41 | 0.37 | 0.39 | 0.42 |
| APFD | cc-pV5Z | 0.58 | 0.40 | 0.37 | 0.38 | 0.33 | 0.50 | 0.43 |
| B3PW91 | pcS-3 | 0.55 | 0.50 | 0.49 | 0.37 | 0.31 | 0.53 | 0.45 |
| ωB97X | pcS-3 | 0.53 | 0.49 | 0.49 | 0.45 | 0.39 | 0.44 | 0.46 |
| mPW3PBE | cc-pV5Z | 0.64 | 0.47 | 0.44 | 0.47 | 0.40 | 0.58 | 0.50 |
| TPSSh | pcS-3 | 0.63 | 0.66 | 0.64 | 0.36 | 0.36 | 0.54 | 0.52 |
| M062X | Apr-cc-pVQZ | 0.82 | 0.62 | 0.47 | 0.56 | 0.51 | 0.58 | 0.58 |
| B3LYP | pcS-3 | 0.93 | 0.93 | 0.87 | 0.84 | 0.73 | 0.79 | 0.84 |

These results on the accuracy of various DFT methods in predicting $\rho(\mathbf{r})$ cannot be considered as a benchmark study because of a small number of included molecules, which is a consequence of the use of a high-level *ab initio* method and a large basis set to compute the reference ('exact') electron densities, and the use of only one measure of error in electron densities (section S2.1). However, our major conclusions agree with previous results and theoretical considerations in the literature. In particular, a recent study compared the performance of several DFT functionals in predicting $\rho(\mathbf{r})$ of 30 organic molecules with two to ten heavy atoms (only aug-cc-pVTZ basis set was used). The best method was found to be TPSSh, with PBE0 being only slightly worse.[44] Another study compared the performance of numerous DFT functionals over a set of 14 atoms and monoatomic ions, with a measure of the accuracy in predicting $\rho(\mathbf{r})$ different from that used in this work. PBE0 was found to be one of the most accurate DFT methods.[5] Also, for a set of diatomic molecules, PBE0 was found to be one of the best performing DFT functionals in predicting electron densities in the vicinities of chemical bonds.[6] As for the pcS family basis sets, they are rarely included into comparisons, but when they are, they usually perform well.[45-47] Finally, there are some theoretical considerations supporting that ωB97X, PBE0 and related functionals and pcS basis sets may be good at predicting $\rho(\mathbf{r})$ of diverse sets of molecules. PBE0 functional is based to a significant degree on exact theoretical results on the energy functional, not fitting to experimental data.[5,29,43] ωB97X and related functionals include a relatively small number of fitted parameters.[24] pcS basis sets, unlike most other basis set families, were designed to fit experimentally measured nuclear magnetic shielding constants, not energies,[30] which may be the reason for their good performance in approximating $\rho(\mathbf{r})$.

We also carried out similar computations with Hartree-Fock (HF), MP2 and CCSD methods, in combinations with various basis sets, for the same first six molecules in the QM9 dataset (Table S2). As expected, in large basis sets the performance of these *ab initio* methods, in terms of $\varepsilon_\rho$ values, is as follows: CCSD < MP2 < HF (from the most accurate to the least accurate method).

**Table S2**. $\varepsilon_{\rho,i}$ measures (all in %, explicitly shown only in the top row) of the difference between 3D electron densities predicted by two main *ab initio* methods (MP2 and CCSD) with various basis sets, and 'exact' (CCSD/cc-pCV5Z) electron densities [Eq. (S12)]. Data are given for the first six molecules in the QM9 database. Several top basis sets are given for each theory level, in the order of increasing $\varepsilon_\rho$; the values for HF in a large basis set are also shown for comparison. Double horizontal lines between CCSD/cc-pVQZ and CCSD/pcJ-2 shows that some basis sets with intermediate $\varepsilon_\rho$ were omitted. "nan": computations not converged.

| method | basis set | 1 | 2 | 3 | 4 | 5 | 6 | $\varepsilon_\rho$ |
|---|---|---|---|---|---|---|---|---|
| MP2 | cc-pVQZ | 0.44% | 0.55% | 0.53% | 0.49% | 0.57% | 0.49 | *0.51%* |
| MP2 | cc-pCVQZ | 0.45 | 0.55 | 0.54 | 0.51 | 0.59 | 0.49 | *0.52* |
| MP2 | def2-QZVP | 0.44 | 0.61 | 0.64 | 0.49 | 0.57 | 0.51 | *0.54* |
| MP2 | pcS-3 | 0.43 | 0.60 | 0.61 | 0.53 | 0.59 | 0.50 | *0.54* |
| MP2 | aug-cc-pVQZ | 0.47 | 0.64 | 0.68 | 0.50 | 0.57 | 0.52 | *0.56* |
| MP2 | pc-3 | 0.44 | 0.62 | 0.63 | 0.56 | 0.61 | 0.52 | *0.56* |
| MP2 | pcJ-3 | 0.45 | 0.61 | 0.63 | 0.56 | 0.61 | 0.52 | *0.56* |
| MP2 | aug-cc-pCVQZ | 0.48 | 0.65 | 0.69 | 0.52 | 0.58 | 0.52 | *0.57* |
| MP2 | pcseg-3 | 0.46 | 0.63 | 0.63 | 0.55 | 0.60 | 0.54 | *0.57* |
| CCSD | pcJ-3 | 0.12 | 0.16 | 0.17 | 0.11 | 0.10 | 0.11 | *0.13* |
| CCSD | pcS-3 | 0.11 | 0.17 | 0.18 | 0.12 | 0.12 | 0.12 | *0.13* |
| CCSD | pc-3 | 0.13 | 0.18 | 0.19 | nan | 0.14 | 0.15 | *0.15* |
| CCSD | def2-QZVP | 0.15 | 0.20 | 0.21 | 0.12 | 0.12 | 0.14 | *0.15* |
| CCSD | def2-QZVPD | nan | 0.20 | 0.21 | 0.12 | 0.12 | 0.14 | *0.15* |
| CCSD | aug-pcS-3 | 0.22 | 0.18 | 0.19 | nan | 0.12 | 0.12 | *0.16* |
| CCSD | aug-pc-3 | nan | 0.18 | 0.21 | 0.14 | 0.14 | 0.15 | *0.16* |
| CCSD | aug-cc-pCVQZ | nan | 0.22 | 0.25 | nan | nan | 0.16 | *0.20* |
| CCSD | aug-cc-pVQZ | nan | 0.24 | 0.27 | 0.16 | 0.16 | 0.18 | *0.20* |
| CCSD | cc-pCVQZ | 0.17 | 0.40 | 0.37 | 0.15 | 0.14 | 0.18 | *0.22* |
| CCSD | cc-pVQZ | 0.19 | 0.42 | 0.39 | 0.17 | 0.17 | 0.20 | *0.24* |
| CCSD | pcJ-2 | 0.34 | 0.44 | 0.49 | 0.27 | 0.30 | 0.35 | *0.36* |
| CCSD | aug-cc-pCVTZ | 0.43 | 0.51 | 0.58 | 0.37 | 0.41 | 0.45 | *0.45* |
| CCSD | aug-cc-pVTZ | 0.46 | 0.53 | 0.61 | 0.39 | 0.43 | 0.47 | *0.47* |
| HF | cc-pCV5Z | 1.09 | 1.14 | 1.22 | 1.74 | 1.70 | 1.51 | *1.44* |

It is particularly interesting to compare the performance of different DFT functionals with *ab initio* methods. DFT methods from the top of the list in Table S1, such as ωB97X-D3 and PBE0, have slightly lower $\varepsilon_\rho$ values (~0.35-0.4%) than MP2 with large basis sets (~0.5%) for the six molecules under consideration, though the ratio of $\varepsilon_{\rho,i}$ for different molecules differ (e.g., for methane MP2 works better than the mentioned DFT functionals). However, many popular DFT functionals, including B3LYP and TPSSh ($\varepsilon_\rho$ ~ 0.5-0.8%), are outperformed (in terms of the $\varepsilon_\rho$ measure) by MP2 in large basis sets. The performance of CCSD with pcS-3- or VQZ-level basis sets for these six molecules ($\varepsilon_\rho$ down to 0.13%) is consistently better than that of DFT, but the use of smaller (VTZ- or pcS-2-level) basis sets kills this advantage of CCSD ($\varepsilon_\rho$ ~ 0.4%).

Note that the comparisons presented in this Section are limited by the accuracy of the reference electron densities computed with CCSD/cc-pCV5Z and the used *L*1 measure of the error $\varepsilon_\rho$. In particular, the reported accuracy of CCSD with smaller basis sets actually reflects the effect of the basis set size, but not the quantum chemical method, on the accuracy of the predicted electron densities. Unfortunately, we have not been able to find in the literature electron densities of these molecules computed with methods of higher level than CCSD, and the widespread quantum chemical software does not allow for computations of electron densities of these molecules with higher level methods either [full configuration interaction (FCI) is implemented in the existing software, but becomes prohibitively expensive with VQZ-level or larger basis sets even for small molecules], let alone experimental data on electron densities.

The advantages of the analysis presented in this section are as follows:
1) The comparisons are made relative to rather accurate *ab initio* electron densities.
2) The molecules for which the comparisons are made contain several types of chemical bonds that are very important as building blocks of larger organic molecules (C–H, N–H, O–H, C=O, C=C, C≡N).
3) The main conclusions on the relative performance of various *ab initio* and DFT methods are physically sound and match earlier results in the literature.

Even though only one measure of error in electron density was used here, the results are physically plausible and provide a good starting point for the choice of ground truth methods. Another limitation of this analysis, namely the small size of the dataset of molecules (which is a consequence of using a high-level quantum chemical method and a large basis set), is easier to overcome. In Section S2.3, we proceed to computations and comparisons for much larger sets of molecules, but in smaller datasets.

*S2.3. Comparison to ab initio results for larger datasets of molecules*

We have been able to compute energies of 4762 molecules from the QM9 dataset (indices 1 to 6000, with omissions due to computation not finished by the time limit) with RI-CCSD(T) method in the aug-cc-pVTZ basis set (for details on computations, see Appendix A2). The choice of the method was based on the following considerations. CCSD(T), often called 'a golden standard of Quantum Chemistry', demonstrates the performance superior to any DFT functional (given a sufficiently large basis set), and it often used as a source of 'exact' energies in benchmark studies on various DFT functionals and other quantum chemical methods. Resolution-of-identity [RI in "RI-CCSD(T)"] is a physically motivated approximation within the CCSD(T) framework, which radically improves scaling of the method (thus making it affordable for larger molecules) at a cost of minor errors introduced in the computed energies (at least an order of magnitude smaller than errors of the best DFT functionals, see below). The aug-cc-pVTZ basis set is one of a few for which an auxiliary basis set required for RI computations was implemented in standard quantum chemical software and used in the literature. This basis set provides the accuracy in predicted energies sufficient to estimate the errors of DFT methods, and allows for computing energies of a much larger set of molecules than the aug-cc-pVQZ basis set. The dataset of CCSD(T)/aug-cc-pVTZ energies created in this work is, to the best of our knowledge, one of the largest datasets of accurate reference energies, both in terms of the number of molecules and their size (some of the molecules have 8 heavy atoms).

The pairs of DFT functionals and basis sets providing the lowest errors in computed energies of these 4762 molecules are given in Table S3. The analysis of these results shows the following:

- TPSS, TPSSh and M06-2X functionals consistently demonstrates low errors with various basis sets (including relatively small ones, including pc-1-level and triple-zeta-level sets).
- ωB97M-V and related functionals also have relatively low errors with multiple basis sets.
- One of the lowest errors is demonstrated by ωB97M-V/pcJ-1. The ωB97M-V functional was previously reported as one of the best functionals.[24] However, it is surprising that the best result is achieved with a small and relatively exotic basis set (pcJ-1) optimized for computations of NMR parameters, not energies. Using larger basis sets from the same family increases the errors in energies (MAE, kcal/mol: 3.36 for pcJ-1, 4.71 for pcJ-2, 5.01 for pcJ-3). Also, widely used large basis sets have errors larger than that of pcJ-1 (MAE, kcal/mol: 4.39 for aug-cc-pVTZ, 4.71 for aug-cc-pVQZ). It is not clear to us what could be the reasons for ωB97M-V/pcJ-1, but not ωB97M-V with larger basis sets, to stay relatively more accurate for such a large and diverse set of molecules.
- The widespread practice of comparing DFT functionals (and, more generally, other quantum chemical methods) for a single basis set may keep important aspects uncovered. In particular, increases in energy errors for larger basis sets may signal that certain functionals might be not as physical as expected from a single smaller basis set comparisons.

We have also computed electron densities of 52 molecules from the QM9 dataset (indices 1 to 100, with omissions due to computation not finished by the time limit) with RI-CCSD method and aug-cc-pVQZ basis set. As discussed in Section S2.2, only CCSD (not MP2), and only with quadruple-zeta-level (not triple-zeta-level) basis sets provide consistently higher accuracy that DFT methods for electron densities of the first six molecules in the QM9 dataset. The third order correction in CCSD(T) is defined only for energies, not electron densities. The RI speed up is applicable to CCSD, and we used it to increase the number of molecules for which we could run computations. Basis set extrapolations are not commonly used for electron densities, and we did not perform such extrapolations here so that the densities we use do not become questionable. Altogether, these considerations limit our choice of the method to RI-CCSD/aug-cc-pVQZ, and these computations were affordable for 52 molecules. We have not been able to find any previous literature on comparisons of electron densities from high theory level methods for molecules of a size comparable to the largest of these 52 molecules. However, the stability of the relative difference between electron densities from various methods with the molecule size (Fig. 3a) suggests that the relative accuracy of these methods, including RI-CCSD, should stay more or less the same for larger molecules. Combinations of DFT functionals and basis sets with lowest errors in the electron densities (in terms of $\varepsilon_\rho$ measures, relative to RI-CCSD/aug-cc-pVQZ electron densities of 52 molecules) are listed in Table S4.

**Table S3**. DFT functionals and basis sets with the lowest errors in predicted energies of 4762 molecules from the QM9 dataset (indices 1 to 6000, with omissions). RI-CCSD(T)/aug-cc-pVTZ energies were used as reference ('exact') energies. MedAE, median absolute error; MAE, mean absolute error; RMSE, root mean squared error; all errors in *kcal/mol*. The list was sorted by decreasing MAE; MedAE values, which are more robust to outlying datapoints, closely correlate with MedAE values.

| DFT functional | Basis set | MedAE | MAE | RMSE |
|---|---|---|---|---|
| TPSSh | aug-pcS-1 | 2.66 | 3.21 | 4.08 |
| ωB97M-V | pcJ-1 | 2.81 | 3.36 | 4.23 |
| TPSSh | G3LARGE | 2.84 | 3.38 | 4.25 |
| M06-2X | def2-TZVPP | 3.08 | 3.52 | 4.42 |
| M06-2X | def2-TZVPPD | 2.96 | 3.57 | 4.46 |
| M06-2X | G3LARGE | 3.16 | 3.64 | 4.58 |
| M06-2X | G3MP2LARGE | 3.33 | 3.78 | 4.75 |
| TPSS | racc-pVTZ | 2.94 | 3.79 | 5.04 |
| B97M-rV | cc-pVQZ | 3.40 | 3.87 | 4.76 |
| ωB97X | pcseg-2 | 3.37 | 3.88 | 4.85 |
| M06-2X | cc-pCVQZ | 3.28 | 3.89 | 4.86 |
| M06-2X | pcseg-3 | 3.27 | 3.91 | 4.88 |

The analysis of the results on the accuracy of the electron densities computed with various DFT functionals and basis sets shows the following:
- As in the case of the first six molecules (Table S1), the top of the list is dominated by two families of DFT functionals (ωB97X-related and PBE-related), which, as discussed above, are based on exact physical results and/or were designed to minimize possible overfitting.
- pcS-3-level basis sets are often optimal or close to optimal for various DFT functionals (a comparison to V5Z-level or pcS-4-level basis sets, unlike the case of the first six molecules as in Table S1, is not always possible, because computations in larger basis sets are prohibitively expensive for some of these 52 molecules; however, for those molecules where a comparison is possible, V5Z- and pcS-4-level basis sets do not provide a clear improvement over pcS-3-level basis sets).
- DFT functionals and basis sets with low errors in energies as measured by MAE (Table S3) turn out to have relatively large errors in electron densities as measured by $\varepsilon_\rho$. For example, the combination of ωB97M-V/pcJ-1 that was one of the best at predicting energies (Table S3) has $\varepsilon_\rho$ as large as 1.4%, nearly 4 times worse that the best method/basis set combination in Table S4. Also, the M06-2X functional that provided relatively low errors with energies with several basis sets (Table S3) worked worse than many other DFT functionals at predicting the electron densities (the lowest $\varepsilon_\rho$ was achieved for M06-2X with the cc-pVQZ basis set: 0.59%, ~1.4 times worse than for ωB97X-V/pcS-3 or ωB97X-D3/pcS-3).

**Table S4**. DFT functionals and basis sets with the lowest errors in predicted electron densities of 52 molecules from the QM9 dataset (indices 1 to 93, with omissions). RI-CCSD(T)/aug-cc-pVQZ electron densities were used as reference ('exact') densities. The list was sorted by decreasing $\varepsilon_\rho$ computed over 52 molecules; median $\varepsilon_{\rho,i}$ measures, which should be more robust to outliers, are also given. Median values closely correlate with average values.

| DFT functional | Basis set | $\varepsilon_\rho$ | Median $\varepsilon_{\rho,i}$ |
|---|---|---|---|
| ωB97X-V | pcS-3 | 0.39% | 0.38% |
| ωB97X-D3 | pcS-3 | 0.42 | 0.44 |
| ωB97X-D3 | aug-pcS-3 | 0.42 | 0.44 |
| ωB97X-D3 | cc-pV5Z | 0.45 | 0.47 |
| ωB97X-D3 | pcS-4 | 0.45 | 0.47 |
| ωB97X-D3 | pc-4 | 0.46 | 0.48 |
| revPBE0 | pcS-3 | 0.47 | 0.49 |
| ωB97X | Jul-cc-pVQZ | 0.47 | 0.48 |
| revPBE0 | aug-pcS-3 | 0.47 | 0.49 |
| ωB97X | pcS-3 | 0.48 | 0.49 |
| PBE0 | pcS-3 | 0.48 | 0.49 |

Finally, it is curious to compare the performance of various DFT functional / basis set combinations at predicting energies and electron densities with the use of $L1$ measures of errors (Fig. S1). A straightforward conclusion from the Hohenberg–Kohn theorems would be that lower errors in electron densities should lead to lower errors in energies, and vice versa, and therefore the dots on the plot should be scattered near a line (or a curve) with a positive slope (Fig. S1, *green dotted line*). Instead, we observe a relatively independent scattering in terms of these two types of errors, with a negative slope of the convex hull of the dots in the region of low errors. Thus, for the best existing DFT functionals, lower errors in energies as measured by MAE are associated with higher errors in electron densities as measured by $\varepsilon_\rho$. This observation is in line with those in Refs. [5] and [6] for atomic and diatomic systems, respectively. It was suggested that this disagreement with the Hohenberg–Kohn theorems might be explained by the use of an inappropriate measure of error in electron densities, and using instead of them the ideal metric $\Delta E^{id}$, or the density driven error $\Delta E_D$ approximating it, might explain the disagreement.[35,39] Note, however, that all the molecules in the test set are expected to be "normal calculations", in the terminology of refs. [34,39,40], and the errors in predicting energies, according to the cited references, should mainly come from the functional error (the difference between the values of the exact and approximate DFT functionals, both evaluated on the same exact electron density), which in their turn should not depend on the choice of the basis set. On the contrary, our results demonstrate a significant dependence of MAE in energies on the basis set (Table S3, Fig. S1). This may point out at a significant role of errors in predicted densities in the molecules in this dataset. A detailed analysis of the functional and density driven errors goes beyond the scope of this work. Here, we stop with a reminder that the purposes of Section S2 are to choose a DFT functional and basis set to be used for the first stage of training, such that they are not evidently bad, and to establish a reference level of what "human learning" can do, so that we can compare the performance of DNN with it. As we can see from the presented data, this accuracy of "human learning" are on the order of ~3 kcal/mol for energies and ~0.4% for the $\varepsilon_\rho$ measure for electron densities, as computed over the used datasets (4762 and 52 molecules, respectively).

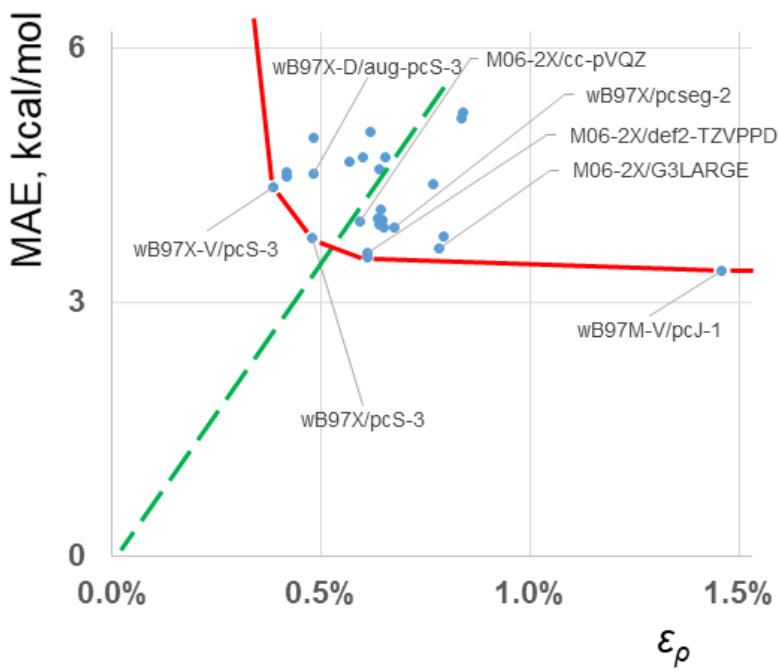

**Fig. S1**. DFT functionals and basis sets with the lowest MAE in predicted energies have relatively high errors in electron densities as measured by $\varepsilon_\rho$, and vice versa (*blue dots*). From the Hohenberg–Kohn theorems, one would expect lower errors in energies to be associated with lower errors in electron densities (*green*), but in practice we observe a relatively independent distribution of errors, with the convex hull (*red*) having a negative slope.

*S2.4. Choice of the method / basis set pairs to generate data for ML*

A rapid increase of the cost of computations with the molecule size necessitates two-stage training of a neural network: first, on a larger dataset generated with a computationally cheaper (and hence less accurate) method / basis set combination, and second, on a dataset of more accurate (and hence more scarce) results.

For the first stage, our choice of the method and basis set was based on two considerations: (1) computations should be fast enough to treat ~133,000 molecules in a reasonable amount of time (up to several months), given the computational resources we had, and (2) the accuracy of the computed electron densities and energies should be higher than or comparable to what was achievable with other methods and basis sets of comparable computational cost.

For 176 combinations of DFT functionals and basis sets that provided the lowest values of $\varepsilon_\rho$ for the first six molecules (Section S2.1), we ran a standard quantum chemical software for three larger molecules (QM9 entries 8 000, 16 000 and 32 000). (Here we discuss the data on the accuracy for the first six molecules, because we chose the DFT method and basis set for large-scale computations for the whole QM9 dataset long before we carried out the analysis presented in Section S2.3.) Then, we computed an aggregated indicator showing how much wallclock time is required to run DFT computations for a molecule of this size on a single GPU (further called 'effective time'; for exact definition and computation details, see Appendix A2). The plot for the computational cost (estimated by effective time) vs. accuracy in predicting $\rho(\mathbf{r})$ (estimated by $\varepsilon_\rho$ over the first six QM9 molecules) shows that some combinations of a DFT functional and a basis set have a favorable tradeoff between the cost and accuracy in predicting $\rho(\mathbf{r})$ (Fig. S2). PBE0/pcS-3, which is #4 on the list in Table S1, looks particularly attractive. Other good choices might be PBE1hPBE/pcS-3, PBE0/Def2-QZVP, and possibly ωB97X-D/cc-pV5Z (though the last combination is more expensive). As for the first three combinations on the list in Table S1, computations for ωB97X-D3[42] with pcS-3-level basis sets turned out to be prohibitively expensive to run them for the whole QM9 dataset, while OHSE2PBE (also known as HSE03) is less used in the literature than PBE0, and its gain in accuracy over PBE0/pcS-3 is marginal.

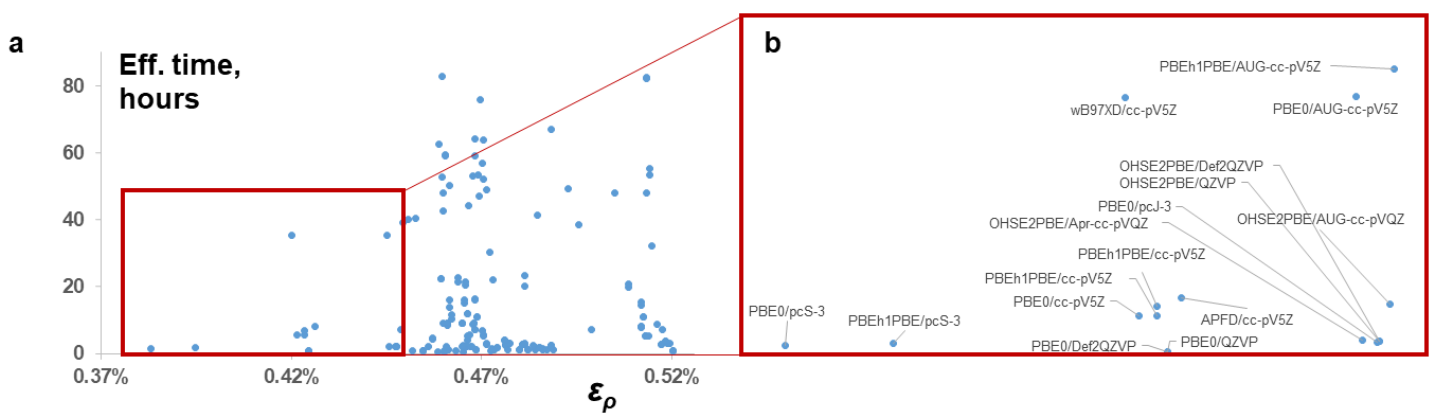

**Fig. S2**. Comparison of accuracy in predicting electron densities (measured by $\varepsilon_\rho$ over the first six molecules from the QM9 database) vs. computation cost (estimated by 'effective time', based on wallclock time for running computations for QM9 entries 8 000, 16 000 and 32 000, see Appendix A2) for various combinations of DFT functionals and basis sets. Attractive options are PBE0/pcS-3 (which is #4 on the list in Table S1), PBE1hPBE/pcS-3, PBE0/Def2-QZVP, and possibly ωB97X-D/cc-pV5Z. (a) All functional/basis set combinations with $\varepsilon_\rho < 0.52\%$ shown; (b) combinations with $\varepsilon_\rho < 0.45\%$, with labels.

Based on the provided data, we chose the combination of PBE0 functional and pcS-3 basis set to compute DFT electron densities for molecules in the QM9 dataset to be used at the first stage of training. This combination has a reasonable computational cost (it took us several months to compute electron densities of all molecules in the QM9 dataset with it). It also demonstrated good performance in predictions of electron densities of 52 molecules for which CCSD/aug-cc-pVQZ densities were available (Fig. 3a), though some other combinations were later found to be a bit more accurate (Fig. S6).

Unfortunately, energies predicted with PBE0/pcS-3 turned out to be significantly less accurate than those reported in the QM9 database [computed with B3LYP/6-31G(2df,p)]. In comparison to available 4762 RI-CCSD(T)/aug-cc-pVTZ energies, MAE of PBE0/pcS-3 was 10.9 kcal/mol, while that of B3LYP/6-31G(2df,p) energies was 5.8 kcal/mol. In order not to repeat computations for all ~133,000 molecules from the QM9 dataset to get energies with another DFT functional and basis set, which would be time- and resource-consuming, we chose to use B3LYP/6-31G(2df,p) energies reported in the original QM9 dataset for the first stage of ML.

Due to a high cost of CCSD computations in large basis sets, especially for molecules like the overwhelming majority of molecules in the QM9 database, it is currently impractical to use such methods to generate datasets of electronic densities comparable in size to QM9. As for smaller basis sets (e.g., VTZ-level sets),[44,49] from the viewpoint of $L$1 metric of the total electron densities, even the use of CCSD without a frozen core approximation does not offer an increase in performance in comparison to much faster and better scaling DFT methods, such as PBE0. The same is true for MP2 methods, regardless of the used basis set: in the best case scenario, they can offer only a marginal improvement in $\varepsilon_\rho$ values, not worth the increase in the computational cost (MP2 calculations are more expensive than those with PBE0).

We would like to emphasize that we do not claim here that a certain DFT functional or a basis sets is "the best". Different research goals (and hence, metrics of performance, not only the $\varepsilon_\rho$ and MAE measures used in this work) and different availability of computational resources may make different choices of quantum chemical methods and basis sets optimal under different circumstances.

As for the second stage of ML, we used RI-CCSD/aug-cc-pVQZ electron densities and RI-CCSD(T)/aug-cc-pVTZ energies for the reasons discussed above in Section S2.3. To briefly recap, CCSD and CCSD(T) (but not MP2) methods guarantee significantly higher accuracy than DFT for electron densities and energies, respectively, and this higher accuracy can be achieved with VQZ-level and VTZ-level (but not VDZ-level) basis sets, respectively. aug-cc-pVTZ and aug-cc-pVQZ are two of a few basis sets for which auxiliary basis sets required for RI speed up are available in standard quantum chemical packages and used in the literature.

**S3. Derivation of the main equations in this work**

*S3.1. General proof*

In this section, we present a generalization of the first Hohenberg-Kohn theorem to electron densities computed with DFT or Hartree-Fock (HF) theory. Existing versions of DFT are based on a minimization of an approximate density functional $E^{(0)}[n]$, which can be written as

$$E^{(0)}[n] = F_{approx}[n] + \int d\mathbf{r}\, n(\mathbf{r})V_{ext}(\mathbf{r}), \tag{S14}$$

where $F_{approx}$ is an approximation to the universal part of the density functional in a given DFT method, and $V_{ext}(\mathbf{r})$ is the external potential (Coulomb potential created by the nuclei in a given molecular system). If the corresponding approximate electron density for the ground state of this system exists, then it can be obtained by minimizing this approximate functional $E^{(0)}[n]$:

$$\rho_0 = \underset{n(\mathbf{r})}{\operatorname{argmin}} \left\{ F_{approx}[n] + \int d\mathbf{r}\, n(\mathbf{r})V_{ext}(\mathbf{r}) \right\}. \tag{S15}$$

This minimization is performed under the restriction imposed on the total number of electrons in the system:

$$\int d\mathbf{r}\, n(\mathbf{r}) = N. \tag{S16}$$

At the minimum,

$$\left(\frac{\delta F_{approx}[n]}{\delta n}\right)_{\rho_0} + V_{ext}(\mathbf{r}) - \mu = 0, \tag{S17}$$

where $\mu$ is a Lagrange multiplier enforcing Eq. (S16). From this equation, an explicit expression for the external potential can be easily obtained:

$$V_{ext}(\mathbf{r}) = \mu - \left(\frac{\delta F_{approx}[n]}{\delta n}\right)_{\rho_0}. \tag{S18}$$

If two ground state electron densities $\rho_{0,A}(\mathbf{r})$ and $\rho_{0,B}(\mathbf{r})$ computed with a certain approximate DFT functional for two external potentials $V_{ext,A}(\mathbf{r})$ and $V_{ext,B}(\mathbf{r})$ coincide,

$$\rho_{0,A}(\mathbf{r}) = \rho_{0,B}(\mathbf{r}), \tag{S19}$$

then these two potentials are the same (up to an additive constant):

$$V_{ext,B}(\mathbf{r}) - V_{ext,A}(\mathbf{r}) = \left[\mu_B - \left(\frac{\delta F_{approx}[n]}{\delta n}\right)_{\rho_{0,B}}\right] - \left[\mu_A - \left(\frac{\delta F_{approx}[n]}{\delta n}\right)_{\rho_{0,A}}\right] = \mu_B - \mu_A = const. \tag{S20}$$

Note that this proof is valid not only for orbital-free DFT, but also for Kohn-Sham DFT. In the latter case, $F_{approx}$ can be written as

$$F_{approx}[n] = T_s[n] + \frac{1}{2}\int d\mathbf{r}d\mathbf{r}' \frac{n(\mathbf{r})n(\mathbf{r}')}{|\mathbf{r}-\mathbf{r}'|} + E_{XC}[n], \tag{S21}$$

where $T_s[n]$ is the kinetic energy of a system of noninteracting electrons with density $n$, and $E_{XC}[n]$ is the exchange-correlation energy. As noted in the original paper by Kohn and Sham, it follows from the Hohenberg-Kohn theorem that $T_s$ is a unique functional of the density $n$.[2] The only approximate term in Eq. (S21) is $E_{XC}$; the other two terms do not contain any approximations and are the same for different DFT functionals. Despite the presence of Kohn-Sham orbitals in the practically used expression for $T_s$, conceptually, the external potential can be expressed in terms of the total electron density only, without referring to the orbitals, by Eq. (S18), and hence the resulting conclusion in Eq. (S20) is valid not only for orbital-free DFT, but also Kohn-Sham DFT.

The HF method can be reformulated in terms of a density functional minimization,[50] therefore the provided proof also holds for an approximate electron density computed with the HF theory $\rho_{HF}$.

Thus, **not only the exact electron density of the ground state, but also an approximate DFT or HF density of the ground state** (if this approximate density exists/converges) **uniquely determines the corresponding external potential (up to an additive constant)**:

$$V_{ext}(\mathbf{r}) = V_{ext}[\rho_0(\mathbf{r})] + const, \tag{S22}$$

where $\rho_0$ is a DFT or HF electron density. Furthermore, the exact Hamiltonian, which is a sum of universal terms for kinetic energy and electron-electron interactions, and a system-specific potential $V_{ext}(\mathbf{r})$, can also be found from $\rho_0$ (up to an additive constant):

$$\widehat{H}_e = \widehat{H}_e[\rho_0(\mathbf{r})] + const. \tag{S23}$$

together with all its eigenvalues and eigenfunctions, that is, exact energies $E^i$ and wave functions $\rho^i$ not only of the ground state, but also of all the excited states of the system:

$$\rho^i(\mathbf{r}) = \rho^i[\rho_0(\mathbf{r})], \quad E^i = E^i[\rho_0(\mathbf{r})] + const, \tag{S24}$$

where $i$ numerates the ground and excited states. In particular, the exact ground state electron density $\rho$ and energy $E$ can be written as function-to-function and function-to-scalar (functional) mappings of $\rho_0$, respectively:

$$\rho(\mathbf{r}) = \rho[\rho_0(\mathbf{r})], \quad E = E[\rho_0(\mathbf{r})]. \tag{S25}$$

Taking into account that a DFT or HF electron density $\rho_0$ and energy $E_0$ provide a reasonable approximation to the exact values,

$$\rho(\mathbf{r}) \approx \rho_0(\mathbf{r}), \quad E \approx E_0, \tag{S26}$$

we can rewrite Eq. (S25) in a "delta-correction" form:

$$\rho(\mathbf{r}) = \rho_0(\mathbf{r}) + \Delta\rho[\rho_0(\mathbf{r})], \quad E = E_0 + \Delta E[\rho_0(\mathbf{r})], \tag{S27}$$

where $\Delta\rho[\cdot]$ and $\Delta E[\cdot]$ are a function-to-function and function-to-scalar (functional) mappings, respectively. These are the equations provided in the main text of this paper.

Note that this proof is based on the fact that a DFT functional can be split into two contributions, one of which is universal, and the other is system-specific. Moreover, the system-specific part can be exactly and explicitly written in terms of the external potential $V_{ext}(\mathbf{r})$ [the second term on the right-hand side of Eq. (S14)], while all approximations are made only for the universal part $F_{approx}$ (the first term there). These conditions are very broad, ensuring a wide applicability of the results.

A particular case of the second equation in Eq. (S27) was proven back in 1995 by Levy and A. Görling.[21] Their proof, however, refers only to approximate electron densities $\rho_0$ computed with *exchange-only* functionals and to ground state energies $E$ (not energies of excited states), and does not include a proof of the corresponding equation for the electron density.

Recently, Ref. [36] claimed that an approximate DFT electron density determines the exact one. However, the reasoning in support of this statement, provided in SI section 2 of the cited paper, did not seem transparent to us. Specifically, Eq. (1) in the SI, written there as:

$$n(\mathbf{r}) = \frac{\delta E[v]}{\delta v(\mathbf{r})}, \tag{S28}$$

where $E[v]$ is the ground-state energy associated with one-body potential $v(\mathbf{r})$, should be clarified by saying that the right hand side of the equation is evaluated at the electron density written on the left hand side of the equation. Following the notations of the cited paper, this can be written as

$$n'(\mathbf{r}) = \left(\frac{\delta E[v]}{\delta v(\mathbf{r})}\right)_{n(\mathbf{r})=n'(\mathbf{r})}. \tag{S29}$$

Next, Eq. (2) in the SI in the cited work is written as

$$\Delta n(\mathbf{r}) = \frac{\delta \Delta E^{PBE}[v]}{\delta v(\mathbf{r})}, \tag{S30}$$

where $\Delta n$ is the difference between the exact (or CCSD(T), as written in the paper) and PBE electron densities, and $\Delta E^{PBE}$, not strictly defined in the paper, is presumably the difference between the exact and PBE density functionals. However, in fact the left hand side of this equation should equal

$$\Delta n(\mathbf{r}) = \left(\frac{\delta E^{exact}[v]}{\delta v(\mathbf{r})}\right)_{n(\mathbf{r})=n_{exact}(\mathbf{r})} - \left(\frac{\delta E^{PBE}[v]}{\delta v(\mathbf{r})}\right)_{n(\mathbf{r})=n_{PBE}(\mathbf{r})}, \tag{S31}$$

which, as far as we can see, cannot be simply rewritten as a single functional derivative, because these two functional derivatives are evaluated at two different electron densities. On the other hand, the right hand side of Eq. (2) from the cited work, as far as we can see, equals zero, because an approximate density functional involves an approximation for the universal part of the functional, but not for the term involving the external potential, which is known exactly:

$$E^{PBE}[n] = F^{PBE}[n] + \int d\mathbf{r}\, n(\mathbf{r})v(\mathbf{r}), \tag{S32}$$

and therefore the difference between an approximate and the exact functionals does not depend on $v(\mathbf{r})$:

$$\begin{aligned}\frac{\delta \Delta E^{PBE}[v]}{\delta v(\mathbf{r})} &= \frac{\delta}{\delta v(\mathbf{r})}\left[\left(F^{exact}[n] + \int d\mathbf{r}\, n(\mathbf{r})v(\mathbf{r})\right) - \left(F^{PBE}[n] + \int d\mathbf{r}\, n(\mathbf{r})v(\mathbf{r})\right)\right]\\ &= \frac{\delta(F^{exact}[n] - F^{PBE}[n])}{\delta v(\mathbf{r})} = 0.\end{aligned} \tag{S33}$$

*S3.2. Another proof for Hartree-Fock electron density*

Though the general proof provided in Section S3.1 also holds for HF densities, it is instructive to consider an alternative proof that explicitly includes key concepts of the HF theory. We perform the proof in three steps, the first of which is very similar to the known proof of the Hohenberg-Kohn theorem, while the other two steps are essentially different because of an approximate character of the HF solution.

**Lemma 1. If a Hartree-Fock solution for the ground state of a molecular system exists, then the Hartree-Fock electron density of this molecular system uniquely determines its Hartree-Fock wave function (up to an arbitrary complex phase).**

Proof: Consider two systems of electrons $A$ and $B$ trapped in external potentials $V_{ext,A}(\mathbf{r})$ and $V_{ext,B}(\mathbf{r})$, respectively, having the same ground state HF electron density $\rho_{HF}$ (and hence, the same number of electrons $N$). Denote the ground state HF wave functions for these molecular systems $\Psi_{HF,A}$ and $\Psi_{HF,B}$, respectively. Assume that these wave functions are different,

$$\Psi_{HF,A} \neq \Psi_{HF,B}, \tag{S34}$$

(or, more accurately,

$$\nexists \varphi: \Psi_{HF,A} = e^{i\varphi} \Psi_{HF,B}, \tag{S35}$$

because a wave function is defined only up to a complex phase). Then, due to the variational definition of an HF wave function,

$$E_{HF,A} < \langle \Psi_{HF,B} | \hat{H}_A | \Psi_{HF,B} \rangle = E_{HF,B} + \int d\mathbf{r}\, \rho_{HF}(\mathbf{r}) \left( V_{ext,A}(\mathbf{r}) - V_{ext,B}(\mathbf{r}) \right), \tag{S36}$$

$$E_{HF,B} < \langle \Psi_{HF,A} | \hat{H}_B | \Psi_{HF,A} \rangle = E_{HF,A} + \int d\mathbf{r}\, \rho_{HF}(\mathbf{r}) \left( V_{ext,B}(\mathbf{r}) - V_{ext,A}(\mathbf{r}) \right), \tag{S37}$$

where $E_{HF,A}$ and $E_{HF,B}$ are the ground state HF energies, and $\hat{H}_A$ and $\hat{H}_B$ are the Hamiltonians of the systems $A$ and $B$, respectively. Adding these two equations, we arrive at a contradiction:

$$E_{HF,A} + E_{HF,B} < E_{HF,A} + E_{HF,B}, \tag{S38}$$

which demonstrates that the initial assumption on the inequality of the wave functions $\Psi_{HF,A}$ and $\Psi_{HF,B}$, Eq. (S34), was not valid. Thus, the ground state HF electron density determines the HF wave function of the same system (up to an arbitrary complex phase):

$$\Psi_{HF}(\mathbf{r}) = \Psi_{HF}[\rho_{HF}(\mathbf{r})]. \tag{S39}$$

The strong version of the Hohenberg-Kohn theorem[51] can also be generalized to HF densities. For two systems in external potentials $V_{ext,A}(\mathbf{r})$ and $V_{ext,B}(\mathbf{r})$ and ground state HF electron densities $\rho_{HF,A}$ and $\rho_{HF,B}$, respectively, Eqs. (S36) and (S37) can be rewritten as

$$E_{HF,A} < \langle \Psi_{HF,B} | \hat{H}_A | \Psi_{HF,B} \rangle = E_{HF,B} + \int d\mathbf{r}\, \rho_{HF,B}(\mathbf{r}) \left( V_{ext,A}(\mathbf{r}) - V_{ext,B}(\mathbf{r}) \right), \tag{S40}$$

$$E_{HF,B} < \langle \Psi_{HF,A} | \hat{H}_B | \Psi_{HF,A} \rangle = E_{HF,A} + \int d\mathbf{r}\, \rho_{HF,A}(\mathbf{r}) \left( V_{ext,B}(\mathbf{r}) - V_{ext,A}(\mathbf{r}) \right). \tag{S41}$$

Adding these two equations and canceling out the energies, we arrive at the sought-for inequality:

$$\int d\mathbf{r} \left( \rho_{HF,B}(\mathbf{r}) - \rho_{HF,A}(\mathbf{r}) \right) \left( V_{ext,B}(\mathbf{r}) - V_{ext,A}(\mathbf{r}) \right) < 0. \tag{S42}$$

The derivations in this Subsection are direct generalizations of the known proofs of the first Hohenberg-Kohn theorem. Even though the HF energies $E_{HF,A}$ and $E_{HF,B}$ are not the global minima of $\langle \Psi | \hat{H}_A | \Psi \rangle$ and $\langle \Psi | \hat{H}_B | \Psi \rangle$ anymore, they are still the minima under the constraint that $\Psi$ could be only a Slater determinant built from one-electron spin-orbitals. This restriction is sufficient for the inequalities Eqs. (S36) and (S37) to hold, because the probe wave functions [$\Psi_{HF,B}$ in Eqs. (S36) and $\Psi_{HF,A}$ in Eqs. (S37)] are allowed to be only Slater determinants as well.

However, the remaining part of the proof of the Hohenberg-Kohn theorem does not directly generalize to HF solutions. Recall that knowing the exact wave function $\Psi$, we can reconstruct the corresponding potential $V(\mathbf{r})$ using the Schrodinger equation:

$$\left(\hat{T}\Psi(\mathbf{r}) + V(\mathbf{r})\right)\Psi(\mathbf{r}) = E\Psi(\mathbf{r}) \quad \Rightarrow \quad V(\mathbf{r}) = E - \frac{\hat{T}\Psi(\mathbf{r})}{\Psi(\mathbf{r})} \tag{S43}$$

and therefore, conclude that for two systems with the same ground state wave functions $\Psi_A(\mathbf{r}) = \Psi_B(\mathbf{r})$, the external potentials are the same (up to an additive constant):

$$V_{ext,B}(\mathbf{r}) - V_{ext,A}(\mathbf{r}) = V_B(\mathbf{r}) - V_A(\mathbf{r}) = \left(E_B - \frac{\hat{T}\Psi_B(\mathbf{r})}{\Psi_B(\mathbf{r})}\right) - \left(E_A - \frac{\hat{T}\Psi_A(\mathbf{r})}{\Psi_A(\mathbf{r})}\right) = E_B - E_A = const. \tag{S44}$$

For HF wave functions this proof does not work, because HF wave functions do not exactly satisfy the Schrodinger equation. In the remaining part of this Section S3.2, we present two lemmas proving that HF ground state densities nevertheless uniquely determine external potentials. The proof is based on equations that HF wave functions do satisfy exactly, namely Fock equations for the corresponding spin-orbitals.

**Lemma 2. A Hartree-Fock wave function determines the corresponding set of Hartree-Fock spin-orbitals (up to a unitary transformation).**

Assume that $\Psi_{HF,A}$ and $\Psi_{HF,B}$ are ground state HF wave functions of the same molecular system, built from two different sets of spin-orbitals $\{\Psi_{i,A}(i)\}_{i=1,...,N}$ and $\{\Psi_{i,B}(i)\}_{i=1,...,N}$, respectively:

$$\Psi_{HF,A} = \sqrt{N!}\,\hat{A}\left(\Psi_{1,A}(1)\Psi_{1,A}(2)\ldots\Psi_{1,A}(N)\right), \tag{S45}$$

$$\Psi_{HF,B} = \sqrt{N!}\,\hat{A}\left(\Psi_{1,B}(1)\Psi_{1,B}(2)\ldots\Psi_{1,B}(N)\right), \tag{S46}$$

$$\Psi_{HF,A} = e^{i\varphi}\,\Psi_{HF,B}, \tag{S47}$$

where $(i) = (\mathbf{r}_i, \sigma_i)$ is a short-hand notation for the full list of spatial $\mathbf{r}_i$ and spin $\sigma_i$ variables of particle $i$, and $\hat{A}$ is the antisymmetrizer. For given $i = 1, \ldots, N$, we choose a permutation $P$ for which $\pi(1) = i$, multiply the left and right hand sides of Eq. (S47) by $\Psi_{\pi(2),A}(2)\,\Psi_{\pi(3),A}(3)\,\ldots\,\Psi_{\pi(N),A}(N)$, and integrate over the spatial and spin coordinates of particles $2, \ldots, N$. The result is

$$(-1)^P \Psi_{i,A}(1) = \sum_{P'} (-1)^{P'} e^{i\varphi} \langle \Psi_{\pi(2),A} | \Psi_{\pi'(2),B}\rangle \ldots \langle \Psi_{\pi(N),A} | \Psi_{\pi'(N),B}\rangle \Psi_{\pi'(1),B}(1), \tag{S48}$$

where $(-1)^P$ and $(-1)^{P'}$ are the parities of permutations $P$ and $P'$, respectively. Therefore, each spin-orbital $\Psi_{i,A}$ is a linear combination of spin-orbitals $\{\Psi_{j,B}\}_{j=1,...,N}$:

$$\forall i = 1, \ldots, N: \quad \Psi_{i,A}(\mathbf{r},\sigma) = \sum_{j=1}^{N} \Psi_{j,B}(\mathbf{r},\sigma) u_{ji}. \tag{S49}$$

In order for $\{\Psi_{i,A}\}_{i=1,...,N}$ and $\{\Psi_{i,B}\}$ each to be orthonormal, matrix $U$ has to be unitary:

$$\delta_{ij} = \langle \Psi_{i,A} | \Psi_{j,A} \rangle = \sum_{kl} u_{ki}^T u_{lj} \langle \Psi_{k,B} | \Psi_{l,B} \rangle = \sum_{kl} u_{ki}^T u_{lj} \delta_{kl} = \sum_{k} u_{ki}^T u_{kj}. \tag{S50}$$

The unitarity of matrix $U$ is sufficient for the wave functions to be physically identical:

$$\Psi_{HF,A} = \frac{1}{\sqrt{N!}} \begin{vmatrix} \Psi_{1,A}(1) & \Psi_{2,A}(1) & \cdots & \Psi_{N,A}(1) \\ \Psi_{1,A}(2) & \Psi_{2,A}(2) & \cdots & \Psi_{N,A}(2) \\ \vdots & \vdots & \ddots & \vdots \\ \Psi_{1,A}(N) & \Psi_{2,A}(N) & \cdots & \Psi_{N,A}(N) \end{vmatrix} \tag{S51}$$

$$= \frac{1}{\sqrt{N!}} \begin{vmatrix} \sum_j \Psi_{j,B}(1) u_{j1} & \sum_j \Psi_{j,B}(1) u_{j2} & \cdots & \sum_j \Psi_{j,B}(1) u_{jN} \\ \sum_j \Psi_{j,B}(2) u_{j1} & \sum_j \Psi_{j,B}(2) u_{j2} & \cdots & \sum_j \Psi_{j,B}(2) u_{jN} \\ \vdots & \vdots & \ddots & \vdots \\ \sum_j \Psi_{j,B}(N) u_{j1} & \sum_j \Psi_{j,B}(N) u_{j2} & \cdots & \sum_j \Psi_{j,B}(N) u_{jN} \end{vmatrix}$$

$$= \frac{1}{\sqrt{N!}} \left| \begin{pmatrix} \Psi_{1,B}(1) & \Psi_{2,B}(1) & \cdots & \Psi_{N,B}(1) \\ \Psi_{1,B}(2) & \Psi_{2,B}(2) & \cdots & \Psi_{N,B}(2) \\ \vdots & \vdots & \ddots & \vdots \\ \Psi_{1,B}(N) & \Psi_{2,B}(N) & \cdots & \Psi_{N,B}(N) \end{pmatrix} \begin{pmatrix} u_{11} & u_{12} & \cdots & u_{1N} \\ u_{21} & u_{22} & \cdots & u_{2N} \\ \vdots & \vdots & \ddots & \vdots \\ u_{N1} & u_{N2} & \cdots & u_{NN} \end{pmatrix} \right| =$$

$$= \frac{1}{\sqrt{N!}} \begin{vmatrix} \Psi_{1,B}(1) & \Psi_{2,B}(1) & \cdots & \Psi_{N,B}(1) \\ \Psi_{1,B}(2) & \Psi_{2,B}(2) & \cdots & \Psi_{N,B}(2) \\ \vdots & \vdots & \ddots & \vdots \\ \Psi_{1,B}(N) & \Psi_{2,B}(N) & \cdots & \Psi_{N,B}(N) \end{vmatrix} \begin{vmatrix} u_{11} & u_{12} & \cdots & u_{1N} \\ u_{21} & u_{22} & \cdots & u_{2N} \\ \vdots & \vdots & \ddots & \vdots \\ u_{N1} & u_{N2} & \cdots & u_{NN} \end{vmatrix} = \Psi_{HF,B} \det U$$

$$= e^{i\varphi} \Psi_{HF,B}.$$

Therefore, HF spin-orbitals can be reconstructed, up to an arbitrary unitary transformation, Eq. (S49), from the HF wave function.

**Lemma 3. The full set of Hartree-Fock spin-orbitals uniquely determines the external potential (up to an additive constant).**

The Fock equations for spin-orbitals { $\Psi_j(\mathbf{r},\sigma)$ }$_{j=1,\ldots,N}$ corresponding to the external potential $V_{ext}(\mathbf{r})$ can be written in the general case as

$$\left( \hat{F}_u[\{\Psi_l\}] + V_{ext}(\mathbf{r}) \right) \Psi_i(\mathbf{r}, \sigma) = \sum_{j=1}^{N} \Psi_j(\mathbf{r}, \sigma) \varepsilon_{ji}, \tag{S52}$$

where $\hat{F}_u = \hat{F} - V_{ext}(\mathbf{r})$ is the universal part of the Fockian $\hat{F}$ that parametrically depends on the set of all spin-orbitals { $\Psi_l$ }$_{l=1,\ldots,N}$, and $\varepsilon_{ji}$ are coefficients. The usual way to proceed with solving Eqs. (S52) would be to perform a unitary transformation of spin-orbitals:

$$\Psi_j(\mathbf{r}, \sigma) = \sum_{k} \Psi_k'(\mathbf{r}, \sigma) u_{kj} \tag{S53}$$

such that the unitary matrix $U$ formed by elements $u_{kj}$ diagonalizes the matrix E formed by elements $\varepsilon_{ji}$:

$$UEU^T = \Lambda_\varepsilon, \tag{S54}$$

where $\Lambda$ is a diagonal matrix (we denote its diagonal elements $\varepsilon_i$). For this set of spin-orbitals, the Fock equations would assume the following diagonal form:

$$\left(\hat{F}_u[\{\Psi'_l\}] + V_{ext}(\mathbf{r})\right)\Psi'_i(\mathbf{r},\sigma) = \Psi'_i(\mathbf{r},\sigma)\varepsilon_i \tag{S55}$$

(as is well-known, the Fockian $\hat{F}$, and likewise its universal part $\hat{F}_u$, are invariant with respect to a unitary transformation of spin-orbitals, $\hat{F}_u[\{\Psi_l\}] = \hat{F}_u[\{\Psi'_l\}]$). Therefore, $V_{ext}(\mathbf{r})$ could be explicitly written in terms of such spin-orbitals, by analogy with Eq. (S43):

$$V_{ext}(\mathbf{r}) = \varepsilon_i - \frac{\hat{F}_u[\{\Psi'_l\}]\Psi'_i(\mathbf{r},\sigma)}{\Psi'_i(\mathbf{r},\sigma)}. \tag{S56}$$

Any spin-orbital – any value of $i = 1, \ldots, N$ – must yield the same external potential $V_{ext}(\mathbf{r})$. However, Eq. (S56) holds only for a specific set of spin-orbitals, while from the HF wave function, according to Lemma 2, we know a set of spin-orbitals only up to an arbitrary unitary transformation. The value of $U$ could be found by diagonalization of matrix E, and to compute it, we need to know $V_{ext}(\mathbf{r})$. But in the problem under consideration, unlike the usual problem of solving HF equations, $V_{ext}(\mathbf{r})$ is unknown.

To prove the uniqueness of $V_{ext}(\mathbf{r})$ (up to an additive constant), we have to follow another path. Assume that for a given set of spin-orbitals $\{\Psi_l\}$ there exist two different external potentials, $V_{ext,A}(\mathbf{r})$ and $V_{ext,B}(\mathbf{r})$, each satisfying Eq. (S52); the sets of coefficients on the right hand side, $\{\varepsilon^A_{ji}\}$ and $\{\varepsilon^B_{ji}\}$, respectively, may be different, while the universal part of the Fockian has to be the same (because it is invariant with respect to a unitary transformation of spin-orbitals):

$$\forall i = 1, \ldots, N: \quad \left(\hat{F}'[\{\Psi_l\}] + V_{ext,A}(\mathbf{r})\right)\Psi_i(\mathbf{r},\sigma) = \sum_{j=1}^{N} \Psi_j(\mathbf{r},\sigma)\varepsilon^A_{ji}, \tag{S57}$$

$$\forall i = 1, \ldots, N: \quad \left(\hat{F}'[\{\Psi_l\}] + V_{ext,B}(\mathbf{r})\right)\Psi_i(\mathbf{r},\sigma) = \sum_{j=1}^{N} \Psi_j(\mathbf{r},\sigma)\varepsilon^B_{ji}. \tag{S58}$$

Hence, the difference between the two potentials equals

$$\forall i = 1, \ldots, N: \quad V_{ext,B}(\mathbf{r}) - V_{ext,A}(\mathbf{r}) = \Delta\varepsilon_{ii} + \sum_{j=1,j\neq i}^{N} \Delta\varepsilon_{ji}\frac{\Psi_j(\mathbf{r},\sigma)}{\Psi_i(\mathbf{r},\sigma)}, \tag{S59}$$

where we introduced a short-hand notation: $\Delta\varepsilon_{ji} = \varepsilon^B_{ji} - \varepsilon^A_{ji}$. These $N$ equations, Eq. (S59), impose $N-1$ different restrictions on the spin-orbitals $\{\Psi_l\}$. At the same time, the right hand sides of Eq. (S59) can be written in terms of $N-1$ independent variables, for example, variables

$$r_j(\mathbf{r},\sigma) = \frac{\Psi_j(\mathbf{r},\sigma)}{\Psi_1(\mathbf{r},\sigma)}, \quad j = 2, \ldots, N. \tag{S60}$$

In terms of these variables,

$$\frac{\Psi_j(\mathbf{r},\sigma)}{\Psi_i(\mathbf{r},\sigma)} = \frac{r_j(\mathbf{r},\sigma)}{r_i(\mathbf{r},\sigma)}, \quad i = 2, \ldots, N, \ j = 2, \ldots, N. \tag{S61}$$

Solving Eq. (S59), we can express the values of all $r_j(\mathbf{r},\sigma)$ ($j = 2, \ldots, N$) in terms of coefficients $\{\Delta\varepsilon_{kl}\}$. Hence, the values of $r_j$ do not depend on $(\mathbf{r},\sigma)$:

$$\frac{\Psi_j(\mathbf{r},\sigma)}{\Psi_1(\mathbf{r},\sigma)} = r_j(\{\Delta\varepsilon_{kl}\}), \quad j = 2, \ldots, N. \tag{S62}$$

By multiplying each of Eqs. (S62) by $|\Psi_1(\mathbf{r},\sigma)|^2$ and integrating over $(\mathbf{r},\sigma)$, we obtain, due to the orthonormality of the spin-orbitals,

$$r_j(\{\Delta\varepsilon_{kl}\}) = r_j(\{\Delta\varepsilon_{kl}\})\int d\mathbf{r}d\sigma \,|\Psi_1(\mathbf{r},\sigma)|^2 = \int d\mathbf{r}d\sigma \,\Psi_1^*(\mathbf{r},\sigma)\Psi_j(\mathbf{r},\sigma) = 0, \quad j = 2, \ldots, N, \tag{S63}$$

and hence,

$$\Psi_j(\mathbf{r},\sigma) = 0, \quad j = 2, \ldots, N. \tag{S64}$$

which contradicts to the orthonormality of the spin-orbitals. This contradiction does not appear only if all off-diagonal $\Delta\varepsilon_{kl}$ in Eq. (S59) equal zero, and hence

$$\forall i = 1, \ldots, N: \quad V_{ext,B}(\mathbf{r}) - V_{ext,A}(\mathbf{r}) = \Delta\varepsilon_{ii} = const, \tag{S65}$$

which proves the sought-for statement: The external potential is determined by the set of HF spin-orbitals up to an additive constant.

For example, in the case of $N = 2$ Eqs. (S59) assume the following form:

$$V_{ext,B}(\mathbf{r}) - V_{ext,A}(\mathbf{r}) = \Delta\varepsilon_{11} + \Delta\varepsilon_{21}\frac{\Psi_2(\mathbf{r},\sigma)}{\Psi_1(\mathbf{r},\sigma)} = \Delta\varepsilon_{22} + \Delta\varepsilon_{12}\frac{\Psi_1(\mathbf{r},\sigma)}{\Psi_2(\mathbf{r},\sigma)}. \tag{S66}$$

Taking into account the definition of $r_2$ given by Eq. (S60), namely $r_2(\mathbf{r},\sigma) = \Psi_2(\mathbf{r},\sigma)/\Psi_1(\mathbf{r},\sigma)$, we can rewrite this equation as:

$$\Delta\varepsilon_{11} + \Delta\varepsilon_{21}r_2(\mathbf{r},\sigma) = \Delta\varepsilon_{22} + \Delta\varepsilon_{12}\frac{1}{r_2(\mathbf{r},\sigma)}. \tag{S67}$$

If at least one of the off-diagonal elements $\{\Delta\varepsilon_{12}, \Delta\varepsilon_{21}\}$ were non-zero, then we could express $r_2$ in terms of coefficients $\{\Delta\varepsilon_{kl}\}$, ending up with expressions that do not depend on $(\mathbf{r},\sigma)$:

$$r_2(\mathbf{r},\sigma) = -\frac{\Delta\varepsilon_{11} - \Delta\varepsilon_{22} \pm \sqrt{(\Delta\varepsilon_{11} - \Delta\varepsilon_{22})^2 + 4\Delta\varepsilon_{12}\Delta\varepsilon_{21}}}{2\Delta\varepsilon_{21}} = const, \tag{S68}$$

leading to a contradiction to the orthonormality of spin-orbitals, as explained above.

In the case of $N = 3$ Eqs. (S59) assume the following form:

$$V_B(\mathbf{r}) - V_A(\mathbf{r}) = \Delta\varepsilon_{11} + \Delta\varepsilon_{21}\frac{\Psi_2(\mathbf{r},\sigma)}{\Psi_1(\mathbf{r},\sigma)} + \Delta\varepsilon_{31}\frac{\Psi_3(\mathbf{r},\sigma)}{\Psi_1(\mathbf{r},\sigma)} = \Delta\varepsilon_{22} + \Delta\varepsilon_{12}\frac{\Psi_1(\mathbf{r},\sigma)}{\Psi_2(\mathbf{r},\sigma)} + \Delta\varepsilon_{32}\frac{\Psi_3(\mathbf{r},\sigma)}{\Psi_2(\mathbf{r},\sigma)} \quad (S69)$$

$$= \Delta\varepsilon_{33} + \Delta\varepsilon_{13}\frac{\Psi_1(\mathbf{r},\sigma)}{\Psi_3(\mathbf{r},\sigma)} + \Delta\varepsilon_{23}\frac{\Psi_2(\mathbf{r},\sigma)}{\Psi_3(\mathbf{r},\sigma)},$$

which can be rewritten in terms of $r_2$ and $r_3$ as

$$\begin{cases} \Delta\varepsilon_{11} + \Delta\varepsilon_{21} r_2 + \Delta\varepsilon_{31} r_3 = \Delta\varepsilon_{22} + \Delta\varepsilon_{12}\frac{1}{r_2} + \Delta\varepsilon_{32}\frac{r_3}{r_2} \\ \Delta\varepsilon_{11} + \Delta\varepsilon_{21} r_2 + \Delta\varepsilon_{31} r_3 = \Delta\varepsilon_{33} + \Delta\varepsilon_{13}\frac{1}{r_3} + \Delta\varepsilon_{23}\frac{r_2}{r_3} \end{cases} \quad (S70)$$

These two independent restrictions imposed on two variables $r_2$ and $r_3$ would fix their values in terms of coefficients $\{\Delta\varepsilon_{kl}\}$, leading to the contradiction given by Eq. (S64), unless all off-diagonal elements $\Delta\varepsilon_{kl}$ are zeros. Hence, the external potential is also determined in this case up to an additive constant, Eq. (S65).

To sum up, the full set of $N$ HF spin-orbitals, like the exact wave function, uniquely determines the external potential (up to an additive constant), even if these spin-orbitals are known only up to a unitary transformation.

Taken together, Lemmas 1, 2 and 3 demonstrate that **the approximate Hartree-Fock electron density of the ground state of a molecular system $\rho_{HF}$ (if exists) uniquely determines the corresponding external potential (up to an additive constant), hence the exact Hamiltonian, hence all the exact energies and electron densities (not only of the ground state, but also of all excited states) of a given molecular system.**

The proofs provided in Sections S3.1 and S3.2 are not constructive, and do not shed light on the nature of $\Delta\rho[\cdot]$ and $\Delta E[\cdot]$. This gap can be filled out with a perturbative analysis as given below, in Section S3.3.

*S3.3. Perturbative analysis of corrections*

For the exact density functional $E[n]$ of the electron density $n(\mathbf{r})$, the Taylor series expansion[35,52] (up to the second order) in the vicinity of some density $\rho_0(\mathbf{r})$ close to the ground state density can be written as:

$$E[\rho_0 + \Delta\rho] \cong E[\rho_0] + \int d\mathbf{r}\left(\frac{\delta E}{\delta n(\mathbf{r})}\right)_{\rho_0} \Delta\rho(\mathbf{r}) + \frac{1}{2}\int d\mathbf{r} d\mathbf{r}'\left(\frac{\delta^2 E}{\delta n(\mathbf{r})\delta n(\mathbf{r}')}\right)_{\rho_0} \Delta\rho(\mathbf{r})\Delta\rho(\mathbf{r}'), \quad (S71)$$

where $\Delta\rho(\mathbf{r}) = n(\mathbf{r}) - \rho_0(\mathbf{r})$, and $\delta E/\delta n(\mathbf{r})$ and $\delta E^2/\delta n(\mathbf{r})\delta n(\mathbf{r}')$ are the first- and second-order functional derivatives of the density functional $E[\cdot]$, respectively. The minimum of $E[\rho_0 + \Delta\rho]$ under the constraint that

$$\int d\mathbf{r}\,\Delta\rho(\mathbf{r}) = 0 \quad (S72)$$

(which maintains a constant number of electrons in the molecule) is reached at the following values of $\Delta\rho$ and $\Delta E$:

$$\Delta\rho(\mathbf{r}) = -\int d\mathbf{r}' \left(\frac{\delta^2 E}{\delta n(\mathbf{r})\delta n(\mathbf{r}')}\right)^{-1}_{\rho_0} \left[\left(\frac{\delta E}{\delta n(\mathbf{r}')}\right) - \Delta\mu\right]_{\rho_0}, \tag{S73}$$

$$\Delta E = -\frac{1}{2}\int d\mathbf{r} d\mathbf{r}' \left[\left(\frac{\delta E}{\delta n(\mathbf{r}')}\right) - \Delta\mu\right]_{\rho_0} \left(\frac{\delta^2 E}{\delta n(\mathbf{r})\delta n(\mathbf{r}')}\right)^{-1}_{\rho_0} \left[\left(\frac{\delta E}{\delta n(\mathbf{r}')}\right) - \Delta\mu\right]_{\rho_0}, \tag{S74}$$

where $\Delta\mu$ is a Lagrange multiplier ensuring the electron number conservation, $(\delta^2 E/\delta n(\mathbf{r})\delta n(\mathbf{r}'))^{-1}$ is the resolvent for the second-order functional derivative of the density functional, also known as the static density–density response function,[34,40,52] and $\Delta E$ is defined as $E[\rho_0 + \Delta\rho] - E[\rho_0]$. The value of $\Delta\mu$ is implicitly set by Eqs. (S72) and (S73), Thus, $\Delta\rho$ can be considered as a correction to the approximate electron density $\rho_0$, similar to the analysis in Sections S3.1 and S3.2, and $\Delta E$ as a correction to an approximate energy $E[\rho_0]$.

These expressions for $\Delta\rho$ and $\Delta E$ can be further generalized to include higher order functional derivatives. For example, a perturbative inclusion of the third order functional derivative $\delta^3 E/\delta n(\mathbf{r})\delta n(\mathbf{r}')\delta n(\mathbf{r}'')$ leads to the following expression for $\Delta\rho$:

$$\Delta\rho(\mathbf{r}) = -\int d\mathbf{r}' \left(\frac{\delta^2 E}{\delta n(\mathbf{r})\delta n(\mathbf{r}')}\right)^{-1}_{\rho_0} \left[\left(\frac{\delta E}{\delta n(\mathbf{r}')}\right) - \Delta\mu \right. \tag{S75}$$
$$\left. + \frac{1}{2}\int d\mathbf{r}' d\mathbf{r}'' \left(\frac{\delta^3 E}{\delta n(\mathbf{r})\delta n(\mathbf{r}')\delta n(\mathbf{r}'')}\right) \Delta\rho_{lo}(\mathbf{r}')\Delta\rho_{lo}(\mathbf{r}'')\right]_{\rho_0},$$

where

$$\Delta\rho_{lo}(\mathbf{r}) = -\int d\mathbf{r}' \left(\frac{\delta^2 E}{\delta n(\mathbf{r})\delta n(\mathbf{r}')}\right)^{-1}_{\rho_0} \left[\left(\frac{\delta E}{\delta n(\mathbf{r}')}\right) - \Delta\mu\right]_{\rho_0}. \tag{S76}$$

The idea of this generalization is illustrated in more detail in Appendix A3 for a simple case of a function of one variable. Overall, if functional derivatives of $E[n]$ of all orders exist, then an exact series expansion for $\Delta\rho$ and $\Delta E$ could be written in terms of the values of these functional derivatives at $\rho_0$, in a way similar to Eq. (S75). These series for $\Delta\rho$ and $\Delta E$ can be interpreted as values of two mappings evaluated at $\rho_0$, providing explicit expressions for $\Delta\rho[\cdot]$ and $\Delta E[\cdot]$, which were non-constructively proven to exist in Sections S3.1 and S3.2.

So far, no restrictions on $\rho_0$ in Eq. (S71) were imposed, besides that it should be somewhat close to the true ground state electron density. However, in the interests of practical efficiency, a more specific choice of $\rho_0$ for perturbative expansions can be made. Indeed, the function-to-function mapping $\Delta\rho[\cdot]$ as introduced in this Subsection S3.3 [but not in previous Subsections S3.1 and S3.2, where $\rho_0(\mathbf{r})$ was uniquely determined by variational principle from the approximate density functional, Eq. (S15)] has to be very sensitive to the input. Regardless of whether $\rho_0(\mathbf{r})$ or $\rho_0(\mathbf{r}) + \delta\rho(\mathbf{r})$, where $\delta\rho(\mathbf{r})$ is an arbitrary function, is used as the reference density, the predicted values of $\rho(\mathbf{r})$ should be the same:

$$\rho(\mathbf{r}) = \rho_0(\mathbf{r}) + \Delta\rho[\rho_0(\mathbf{r})] = [\rho_0(\mathbf{r}) + \delta\rho(\mathbf{r})] + \Delta\rho[\rho_0(\mathbf{r}) + \delta\rho(\mathbf{r})], \tag{S77}$$

and therefore,

$$\Delta\rho[\rho_0(\mathbf{r}) + \delta\rho(\mathbf{r})] = \Delta\rho[\rho_0(\mathbf{r})] - \delta\rho(\mathbf{r}) \tag{S78}$$

It would be very difficult to ensure this property of $\Delta\rho[\cdot]$ without using an overcomplicated DNN and a very large training set. Instead, a simple prescription on the choice of $\rho_0$ can ensure that Eq. (S78) is automatically satisfied.

We start from the fact that relatively good analytical approximations to the exact density functional $E[n]$ are provided by HF or existing versions of DFT (at least, some of them; see Section S2). For a specific approximate density functional $E^{(0)}[n]$ previously found by "human learning", the exact functional $E[n]$ can be written as

$$E[n] = E^{(0)}[n] + E^{(1)}[n], \tag{S79}$$

where $E^{(1)}[n]$ is the correction to the density functional defined by this equation, Eq. (S79). Then, the correction to the electron density $\Delta\rho$ can be split into two contributions, matching the splitting of $E[\cdot]$ in Eq. (S79), namely:

$$\Delta\rho = \Delta\rho^{(0)} + \Delta\rho^{(1)}, \tag{S80}$$

where the first term is the difference between the *exact* solution for the *approximate* density functional $E^{(0)}[\cdot]$ and the reference density $\rho_0$:

$$\Delta\rho^{(0)} = \underset{n(\mathbf{r})}{\mathrm{argmin}}\, E^{(0)}[n(\mathbf{r})] - \rho_0(\mathbf{r}), \tag{S81}$$

and the second term is the difference between the exact solutions for $E[\cdot]$ and $E^{(0)}[\cdot]$:

$$\Delta\rho^{(1)} = \underset{n(\mathbf{r})}{\mathrm{argmin}}\, E[n(\mathbf{r})] - \underset{n(\mathbf{r})}{\mathrm{argmin}}\, E^{(0)}[n(\mathbf{r})]. \tag{S82}$$

In these notations, the exact electron density $\rho(\mathbf{r})$ can be written in the following form:

$$\rho(\mathbf{r}) = \rho_0(\mathbf{r}) + \Delta\rho^{(0)}[\rho_0(\mathbf{r})] + \Delta\rho^{(1)}\left[\rho_0(\mathbf{r}) + \Delta\rho^{(0)}[\rho_0(\mathbf{r})]\right]. \tag{S83}$$

With this equation, the condition given by Eq. (S78) is automatically satisfied (Fig. S3). Since the correction $\Delta\rho^{(0)}$ depends only on the known functional $E^{(0)}[\cdot]$, it can be found analytically or numerically with the traditional methods and existing software. Only the second correction, $\Delta\rho^{(1)}$, needs to be learned with ML, and the corresponding function-to-function mapping is pretty robust in the sense that it does not need to be as sensitive to the input as $\Delta\rho$ in Eq. (S78).

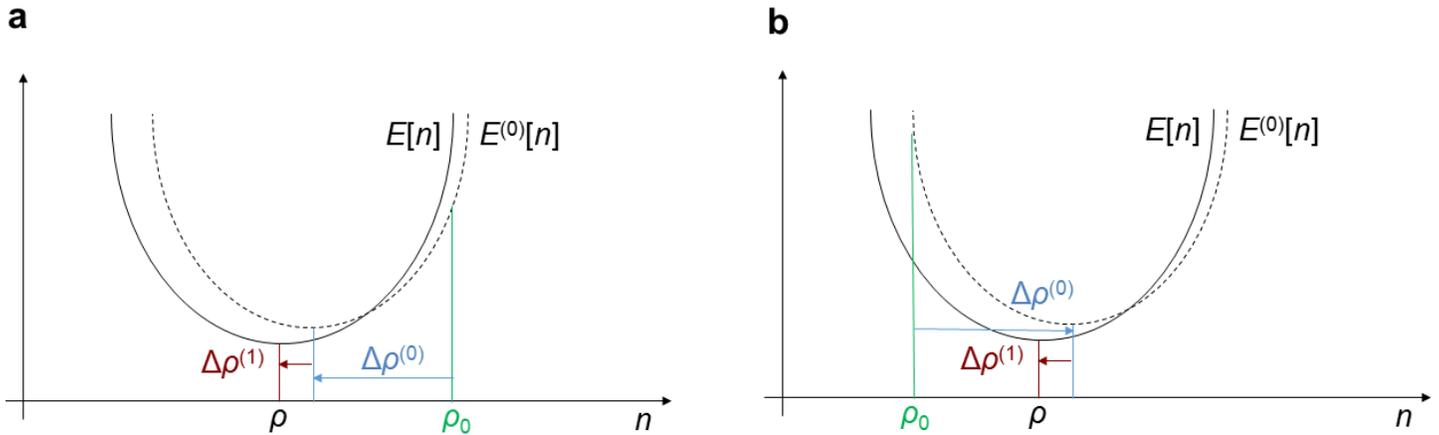

**Fig. S3. A schematic illustration of Eq. (S83) in two dimensions.** [In fact, $\rho(\mathbf{r})$ and $\rho_0(\mathbf{r})$ are functions, and $E[n]$ and $E^{(0)}[n]$ are functionals, and cannot be strictly represented in 2D.] (a) To go from an approximate electron density $\rho_0$ (*green*) to the exact density $\rho$ that minimizes $E[n]$, we first go to the minimum of $E^{(0)}[n]$ (*blue*), and after that to the minimum of $E[n]$ (*cardinal*). (b) If we start from another approximate electron density $\rho_0$ (*green*), the first step is different (*blue*), but the second step is the same (*cardinal*). This splitting of $\Delta\rho$ into $\Delta\rho^{(0)}$ and $\Delta\rho^{(1)}$ ensures the robustness of the mapping to be approximated with ML ($\Delta\rho^{(1)}$ only, *cardinal*).

If $\rho_0$ is chosen to be the electron density that minimizes $E^{(0)}[n]$, in other words, is the solution for a given molecule provided by a specific version of DFT (or HF, which can also be formulated in the form of a density functional minimization[50]), then $\Delta\rho^{(0)}(\mathbf{r})$ vanishes, and Eq. (S83) simplifies to

$$\rho(\mathbf{r}) = \rho_0(\mathbf{r}) + \Delta\rho^{(1)}[\rho_0(\mathbf{r})], \tag{S84}$$

where $\Delta\rho^{(1)}(\mathbf{r})$ is supposed to be learned with ML. This equation, together with the corresponding equation for the exact energy $E$, matches the results of non-perturbative derivations in Sections S3.1 and S3.2, Eq. (S27), with the notational simplification that $\Delta\rho^{(1)} = \Delta\rho$, valid in the case of the choice of $\rho_0$ as the solution provided by a specific DFT or HF method (as assumed in Sections S3.1 and S3.2). Without the restriction that $\rho_0$ is a solution from HF or a specific DFT method, the perturbative approach of Section S3.3 would still be valid, but we would not have been able to use Eq. (S84); instead, a more general and more computationally expensive Eq. (S83) would have to be used.

Unlike the strict proofs given in Sections S3.1 and S3.2, the derivation presented in this Section S3.3 is perturbative, and hence is based on the assumptions that the functional derivatives of the density functional $E[\cdot]$ of arbitrarily high orders exist, and that the exact electron density $\rho(\mathbf{r})$ is within the radius of convergence of the functional Taylor series built about the approximate density $\rho_0(\mathbf{r})$. To the best of our knowledge, the existence of all higher order functional derivatives of the density functional $E[\cdot]$ has not been strictly proven in DFT yet, though, on the other hand, no symptoms of their non-existence have ever been reported either (an expression for the static density–density response function in terms of Kohn-Sham orbitals diverges for small HOMO-LUMO gaps in the KS orbital energies[34,40]). If only a limited number of higher order functional derivatives of $E[\cdot]$ exists, then the equations like Eqs. (S73)-(S74) can be used as approximations. In practice, this should not limit the approach proposed in this work from the practical viewpoint, because mappings $\Delta\rho[\cdot]$ and $\Delta E[\cdot]$ are in any case supposed to be found by ML, therefore, only approximately.

Thus, the perturbative proof given in this Section S3.3, and the non-perturbative proofs presented in Sections S3.1 and S3.2, complement each other in that the perturbative derivation provides explicit, interpretable expressions for the target mappings $\Delta\rho[\cdot]$ and $\Delta E[\cdot]$, such as Eqs. (S73)-(S75), while the non-perturbative proofs guarantee the existence of these mappings $\Delta\rho[\cdot]$ and $\Delta E[\cdot]$ even in non-perturbative regimes (for example, if very rough DFT functionals were used). Though these general theorems might be of less practical value, because in strongly non-perturbative regimes the simplicity of Eqs. (S73)-(S74) is lost, it is important to have them to be confident in the robustness of this approach. Unlike previously published multiple papers aiming at learning the unknown exact DFT functional $E[\cdot]$ with ML, our approach – and this perturbative analysis in particular – naturally leads to the idea to compute the *corrections* to the electron density and energy as the values of mappings $\Delta\rho[\cdot]$ and $\Delta E[\cdot]$ at $\rho_0$, avoiding the computationally expensive procedure of iterative minimization of $E[n]$. The forking architecture of the proposed DNN is justified by the fact that the expressions for $\Delta\rho$ and $\Delta E$ in Eqs. (S73)-(S75) are built of similar blocks, and therefore, can be found from a shared set of intermediate results.

*S3.4. Concluding remarks on the use of approximate electron densities as molecular descriptors in ML*

The derivations in Sections S3.1-3.3 prove that an approximate electron density $\rho_0(\mathbf{r})$ computed with HF or a certain approximate DFT method can serve as a universal descriptor of a molecular system (or, more generally, of a system of electrons in an external potential), uniquely determining exact energies and electron densities of the ground state and all excited states of the system, and therefore, all quantum physical observables $O$ for that system:

$$O = O[\rho_0(\mathbf{r})]. \tag{S85}$$

However, $\rho_0(\mathbf{r})$ is not unique in having this property. The same statements can be made, for example, about the external potential $V_{ext}(\mathbf{r})$ that, together with the total number of electrons $N$, uniquely determines the Hamiltonian, and hence all quantum physical observables:

$$O = O[V_{ext}(\mathbf{r}), N]. \tag{S86}$$

Respectively, the electrostatic potential could be used as a descriptor of a molecule in ML models.[8] Furthermore, for Coulomb potentials, $V_{ext}(\mathbf{r})$ is uniquely determined by the list of coordinates and charges of all the nuclei in the molecular system:

$$V_{ext}(\mathbf{r}) = V_{ext}(\mathbf{r}; \{Z_a, \mathbf{r}_a\}_{a=1,\ldots,N_a}), \tag{S87}$$

where $Z_a$ and $\mathbf{r}_a$ are the charge and coordinates of nucleus $a$, and $N_a$ is the total number of the nuclei in the system. Hence, all the observables can in principle be computed just from a list of scalar and vector variables, without a need to use expensive 3D representations:

$$O = O(\{Z_a, \mathbf{r}_a\}, N). \tag{S88}$$

The same conclusion also holds for any set of internal variables $\{\xi_b\}_b$ that determine the geometry of the molecule and its charge. Various descriptors previously used for ML in Quantum Chemistry (though often formalizable in the spirit of dimensionality reduction, rather than a one-to-one mapping to/from the coordinates), such as Behler-Parrinello symmetry functions[53] or their generalizations,[10,18] invariant two-body and three-body interaction descriptors,[54], encoded bonds descriptors,[55] lists of distances to and charges of the closest atoms,[56] and many other featurization schemes, follow this approach:

$$O = O(\{\xi\}_b). \tag{S89}$$

Why in this case should we care to prove the nontrivial theorems in Sections S3.1-3.3 to arrive at Eq. (S85), if a conceptually much simpler statement, Eq. (S89), is available? Why should we use 3D electron densities as the input to ML models, if much cheaper 1D descriptors $\{\xi_b\}_b$ could also work? Before answering these questions, let us push this line of reasoning to its logical end. As is known from set theory, one can build a one-to-one correspondence between a set $\{\xi_b\}_b$, which includes only a finite number of real and integer numbers, and a single real number $\Xi$. Then, any quantum physical observable would be just a function of that single number:

$$O = O(\Xi). \tag{S90}$$

Evidently, this approach is impractical: It would take an enormous amount of training data and time for an ML model to learn such indirect mappings. In the cases of physically meaningful internal variables $\{\xi_b\}_b$ or the electrostatic potential $V_{ext}(\mathbf{r})$, this problem is not so severe, and ML with such descriptors turned out to be practically feasible. However, the corresponding mappings are still indirect from the physical viewpoint, and attempts to learn them with ML on training datasets of practically affordable sizes may result in overfitting and a lack of transferability to types of molecular systems absent from training sets. Eqs. (S86), (S88), and (S89), unlike Eqs. (S73) and (S74), do not provide guidance for the choice of the architecture of ML models. On the other hand, explicit expressions for $\rho$ and $E$ in terms of $\rho_0$, such as those following from Eqs. (S73)-(S75), are more straightforward than those for $\rho$ and $E$ in terms of $V_{ext}(\mathbf{r})$, let alone the other descriptors. Hence, we may expect that a simpler ML model could learn (not just interpolate over a limited domain) such relationships. The resulting ML models should be less prone to overfitting and more transferable to new classes of molecular systems not included in training datasets. Moreover, our approach naturally leads to delta-learning, that is, using ML to predict *corrections* to results of approximate physical models. In such cases, a typical accuracy of regression ML models, maybe ~1-10%, would be sufficient for ML models to compete in the accuracy of predictions with the best physical models, such as CCSD(T). For a further discussion of the advantages of using $\rho_0(\mathbf{r})$ as a universal descriptor of molecular systems in the light of our numerical results, see Section S5.

**S4. DNN architecture and training**

*S4.1. Input and output format*

The only **input** to the reported PML-QC models is an approximate electron density of a molecule of interest computed with HF method $\rho_{HF}$. Preparation of an input file is carried out in three steps:

(1) For a given molecule, HF calculations are run in standard quantum chemical software.
(2) Using the fchk file generated in step (1), a cube file for the total electron density is generated. This cube file contains numerical values of the electron density on $256 \times 256 \times 256$ grid points with a grid spacing of 0.1 bohr (~0.05 Å). This grid spacing is necessary to approximate integrals over the volume, as in Eq. (S11), by summation over grid points. The number of grid points in each direction was chosen following a tradition in the field of deep learning to use power-of-two sizes, which simplifies architectures of neural networks (a $128 \times 128 \times 128$ grid with a grid spacing of 0.1 bohr is not large enough to contain some of the largest QM9 molecules).

(3) The 256 × 256 × 256 cube file generated in step (2) is coarse grained to a 64 × 64 × 64 cube file by summation of the electron density values in non-overlapping 4 × 4 × 4 cubes. We perform this transformation to speed up DNN training, and make training possible on a single GPU. We checked that this coarse-graining quantitatively preserves the spatial behavior of the electron density. Coarse-graining of a 256 × 256 × 256 cube into a 64 × 64 × 64 cube ensures that the integral of the electron density computed from a sum of the values on all grid points has the right value, and mitigates artifacts of discrete representation of the electron density near nuclei where the gradient of the density is large. A 64 × 64 × 64 cube file with a grid spacing of 0.4 bohr directly generated from the fchk file does not satisfy either of these two conditions, hence the need for separate steps (2) and (3).

The immediate **output** from the DNN is $\Delta\rho$, which is the difference between $\rho$ and $\rho_{HF}$:

$$\Delta\rho(\mathbf{r}) = \rho(\mathbf{r}) - \rho_{HF}(\mathbf{r}), \tag{S91}$$

and $\Delta E'$, defined as:

$$\Delta E' = \left( E - \sum_{a \in H,C,N,O,F} E_a n_a \right) - \left( E_{HF} - \sum_{a \in H,C,N,O,F} E_{HF,a} n_a \right) - \left[ c_0 + \sum_{a \in H,C,N,O,F} c_a n_a \right], \tag{S92}$$

where $E_a$ is the energy of an isolated atom $a$ computed with the same quantum chemical method as the energy of the molecule $E$, $n_a$ is the number of atoms of element $a$ in the molecule, $E_{HF}$ and $E_{HF,a}$ are energies of the molecule and of an isolated atom $a$, respectively, computed with the HF method in the cc-pVDZ basis set, and $c_0$ and $c_a$ are empirical coefficients found from the least square fit of the equation

$$\left( E - \sum_{a \in H,C,N,O,F} E_a n_a \right) - \left( E_{HF} - \sum_{a \in H,C,N,O,F} E_{HF,a} n_a \right) \approx c_0 + \sum_{a \in H,C,N,O,F} c_a n_a, \tag{S93}$$

over the molecules in the training set (for the PML-QC$_{DFT}$ model: $c_0 = 0.002720$, $c_H = -0.020721$, $c_C = -0.076106$, $c_N = -0.108908$, $c_O = -0.087452$, $c_F = -0.058814$; for the PML-QC$_{CCSD(T)}$ model: $c_0 = 0.008006$, $c_H = -0.020997$, $c_C = -0.079304$, $c_N = -0.101827$, $c_O = -0.085634$, $c_F = -0.052935$; all values in hartree). The use of the linear correction term in Eq. (S92) allows us to increase the accuracy of the model, because the DNNs focuses in this case on finding relatively small, but complicated contributions to the energy of chemical binding, and not larger, but conceptually simple corrections per atom (similar corrections were used, for example, in Refs. [18] and [57]). In other words, the DNNs in the PML-QC$_{DFT}$ and PML-QC$_{CCSD(T)}$ models are used to predict the discrepancies, Eq. (S92), between the right and left hand sides of Eq. (S93). Note that this approach does not contradict to the general theory presented in Section 3. Comparing this definition of $\Delta E'$ to $\Delta E$ from the main text, one can see that

$$\Delta E' = \Delta E - \left[ c_0 + \sum_{a \in H,C,N,O,F} (E_a - E_{HF,a} + c_a) n_a \right], \tag{S94}$$

which is also a functional of $\rho_{HF}(\mathbf{r})$, because the numbers of atoms of each type in a given molecule can be found from an approximate electron density,

$$n_a = n_a[\rho_{HF}(\mathbf{r})], \tag{S95}$$

and all other entities inside the square brackets in Eq. (S94) are constants. Thus, $\Delta E'$ can be considered as a functional of $\rho_{HF}$, like $\Delta E$.

From $\Delta\rho$, the desired values of $\rho$ can be easily computed, because $\rho_{HF}$ is already known. The values of $\Delta\rho$ used for training the PML-QC$_{DFT}$ and PML-QC$_{CCSD(T)}$ models were computed with PBE0/pcS-3 and RI-CCSD/aug-cc-pVQZ, respectively, similar to $\rho_{HF}$ as described above, and represented after coarse-graining in the form of 64 × 64 × 64 cube files. The output of the DNN follows the same format of $\Delta\rho$ representation. Similarly, the value of $\Delta E'$ predicted by the DNN is sufficient to compute the desired value of $E$, because $E_{HF}$ is known from the HF/cc-VDZ computation that we perform anyway to get the input to the DNN, and the linear correction term is easy to compute from the molecular formula.

*S4.2. Architecture*

The architecture of the DNN is shown in Fig. 1b in the main text. The only input channel is the HF/cc-VDZ density given on a 64 × 64 × 64 grid. First, this input is processed elementwise with a tanh function [namely, tanh($1.28\rho_{HF}$)], such that $\rho_{HF}$ in the regions where $\rho_{HF} \gtrsim 0.8$ (typical of atomic cores) saturates to 1, while in other regions (including covalent bonds and regions responsible for non-covalent interactions) the input is only linearly rescaled. This transformation ensures that the information on chemical bonding is not dwarfed by the atomic core densities, and artifacts of a discrete representation of the density near the atomic cores are removed. Next, the information is processed by ten hidden layers with a U-Net architecture. This type of architecture was originally proposed to process medical images,[26] and proved efficient in other physics-related tasks.[27] The first five hidden layers encode step-by-step the input into a very coarse spatial representation, the subsequent five hidden layers decode it back to the original resolution, and there is also a direct flow of information from encoding hidden layers to decoding hidden layers having the same spatial resolution (hence the term 'U-Net' for the architecture). Along this path, spatial resolution goes from 64 × 64 × 64 to 2 × 2 × 2 and then back to 64 × 64 × 64, and intermediate representations include up to 256 channels. At the end of the U-Net block, a tensor with 64 channels, each of which has a 64 × 64 × 64 spatial resolution, is computed and concatenated with the input (after tanh transformation), yielding a 65-channel tensor. After this concatenation, computations fork into two paths – one for the electron density and the other for the energy calculations. The correction $\Delta\rho$ to the density is computed by a convolution of the concatenated tensor to 32 channels with the same spatial resolution, rectified linear unit (ReLu) activation, and convolution of the result to a single channel yielding the predicted $\Delta\rho$ values on a 64 × 64 × 64 grid. The other path of computations includes two subsequent 3D convolution operations, each of which is followed by ReLu activation, such that the first convolution decreases the number of channels to 32, and the second convolution to two. Finally, the output value of $\Delta E$ is computed as a linear function of the elements in these two channels and on all 64 × 64 × 64 grid points. The summation over 3D grid points is motivated by integration over space in the perturbative Eqs. (S73)-(S74); also, it was shown that the correlation energy in various methods, including CCSD(T), can be represented as an integral of an energy density over the 3D space.[17] The forked architecture of the network is designed to push the model to learn first some general high-level features of molecules, such as the first-order functional derivative of the density functional, or the resolvent for the second-order derivative (Fig. 1c,d), and only after than to employ these general features to solve the specific tasks of computing the energy or electron density of a given molecule.

*S4.3. Loss functions and training curves*

The loss function $L$ to be minimized to train the PML-QC$_{DFT}$ model was chosen as a linear combination of the $L1$ measures of the performance of the DNN in predicting the electron densities and energies, with an additional $L2$ regularization term:

$$L = \sum_{i \in \text{training set}} \|\Delta\rho_{DNN}[\rho_{HF,i}(\mathbf{r})] - \Delta\rho_i(\mathbf{r})\| + w_E \sum_{i \in \text{training set}} |\Delta E'_{DNN}[\rho_{HF,i}(\mathbf{r})] - \Delta E'_i| + w_{Ereg} \sum_{j \in K} k_j^2, \quad (S96)$$

where $i$ numerates molecules in the training set, $\Delta\rho_i(\mathbf{r})$ and $\Delta E'_i$ are the ground truth (i.e., computed from DFT for training the PML-QC$_{DFT}$ model) values of $\Delta\rho$ and $\Delta E'$ [Eqs. (S91) and (S94)] for the $i$-th molecule, $\rho_{HF,i}(\mathbf{r})$ is the input (HF/cc-VDZ) electron density for the $i$-th molecule, $\Delta\rho_{DNN}$ and $\Delta E'_{DNN}$ are the values of $\Delta\rho(\mathbf{r})$ and $\Delta E'$ predicted by the DNN for the $i$-th molecule, $w_E$ is a coefficient defining a relative weight of the electron density and energy discrepancies in the overall estimate of the performance of the DNN, $w_{Ereg}$ is a prefactor for the regularization term, $K$ is the set of kernel coefficients in convolution operations within the path to the energy prediction from the bifurcation, $j$ enumerates these coefficients, and $k_j$ are the values of these coefficients. No regularization for the kernel coefficients from the path to the electron density predictions was included in $L$, because we did not observe any significant overfitting for $\Delta\rho$ predictions. The $L1$ norm of the error in $\Delta\rho$ in Eq. (S96) is interpreted as in Eq. (S11) and estimated numerically by summation of $|\Delta\rho_{DNN}[\rho_{HF,i}(\mathbf{r})] - \Delta\rho_i(\mathbf{r})|$ over all points on the $64 \times 64 \times 64$ grid.

To train the PML-QC$_{DFT}$ model, we split the QM9 database into training, validation and testing subsets based on the following rules. All molecules with 9 heavy atoms (QM9 indices 21989 to 133885) and indices matching the mask "???[2-9]??", where "[2-9]" stands for any digit from 2 to 9, were assigned to the training set. All molecules with 9 heavy atoms and indices "???1??" were assigned to the validation set, and all molecules with 9 heavy atoms and indices "???0??" were assigned to test set 1. Finally, all molecules with 1 to 8 heavy atoms (indices 1 to 21988) were considered as a separate test set 2. This approach allows us to test the transferability of the PML-QC$_{DFT}$ model between atoms with different numbers of heavy atoms (9 vs. 1 to 8), and to compare predictions of the PML-QC$_{DFT}$ to high quantum theory level results in large basis sets (available only for molecules with a few atoms) for molecules not included in the training set. In this way, 89432 molecules (66.8% of the whole QM9 dataset) were assigned to the training set, 11191 molecules (8.4%) to the validation set, 11194 molecules (8.4%) to test set 1, and 21963 molecules (16.4%) to test set 2. Besides that, 55 molecules originally present in the QM9 database (0.04%) were excluded because PBE0/pcS-3 computations for them did not converge, and 50 more molecules (0.04%) were excluded because DFT results for them did not seem reliable (specifically, two different functional / basis set combinations – PBE0/pcS-3 and B3LYP/6-31G(2df,p) – led to $\Delta E'$ values that differed by more than 50 kcal/mol).

The PML-QC$_{DFT}$ model was trained in four stages. First, $w_E$ and $w_{Ereg}$ were set to 0, the learning rate was set to $2 \cdot 10^{-4}$, and 12 epochs of training were performed. During each epoch, all molecules in the training set, randomly sorted, were processed in minibatches of 16 molecules, such that each molecule was used once and only once during each epoch. Second, $w_E$ was changed to $10^4$, and 12 more epochs of training were performed, with $w_{Ereg}$ equal to $d_{Ereg}i_{Epoch}$ at each epoch, where $i_{Epoch}$ increased from 1 to 12, and the value of $d_{Ereg}$ optimal for the regularization was found to be 10 (Fig. S4). Third, 12 more epochs were carried out with the learning rate of $2 \cdot 10^{-5}$, and $w_{Ereg}$ equal to $d_{Ereg}i_{Epoch}$, where $i_{Epoch}$ ran from 13 to 24. Finally, one more epoch was carried out, with the learning rate of $2 \cdot 10^{-6}$, and $w_{Ereg} = d_{Ereg} i_{Epoch}$, where $i_{Epoch} = 25$. The value of $w_E = 10^4$ was chosen such that the contribution of the energy term to the total loss function $L$, Eq. (S96), was ~10% of $L$ by the end of training (as measured on the validation set). An attempt to significantly increase $w_E$ did not lead to noticeable changes in the performance of the trained model in terms of $\Delta E$ or $\Delta \rho$ prediction; significantly smaller $w_E$ increase the gap between the performance of the model on the training and validation sets. Attempts to use other training schedules, including ones with a gradual decrease in the learning rate and a gradual increase in $w_E$ and $w_{Ereg}$ values throughout training epochs, did not yield better models (data not shown).

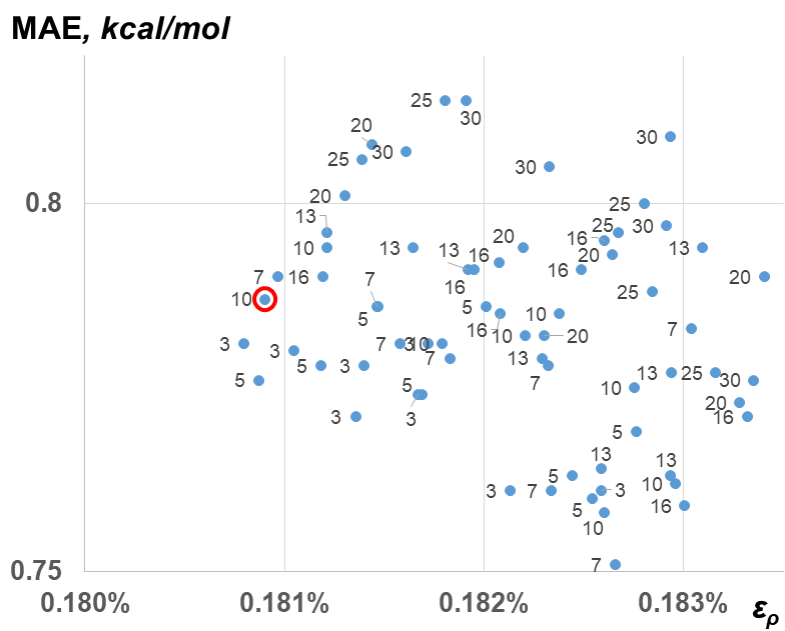

**Fig. S4**. Optimal value of $d_{Ereg}$ for the network regularization for the PML-QC$_{DFT}$ model was chosen by comparing the performance (on the validation set) of models trained with different $d_{Ereg}$ values. Final $\varepsilon_\rho$ vs. MAE values over the validation set for each independent training session shown, with the corresponding values of $d_{Ereg}$ used as labels. Multiple independent training sessions were run for each $d_{Ereg}$. Models trained with $d_{Ereg} > 10$ demonstrated deteriorated performance on the validation test set, so we chose $d_{Ereg} = 10$ as the optimal value. One of the models learned with $d_{Ereg} = 10$ was chosen as the final PML-QC$_{DFT}$ model (*red circle*). This model has the lowest error in predicting densities (among models with $d_{Ereg} = 10$), and a relatively low error in energies.

As for training the PML-QC$_{CCSD(T)}$ model, we had CCSD(T) data on energies only for molecules with QM9 indices up to 6000 (with omissions, because some of the computations did not finish with the computational resources available to us), and CCSD data on electron densities only for molecules with QM9 indices up to 93 (with omissions, for the same reason). All of these molecules were in test set 2 when we trained the PML-QC$_{DFT}$ model. For training the PML-QC$_{CCSD(T)}$ model, we divided the available set of 4762 molecules with known CCSD(T) energies into new training, validation and test sets: 3830 molecules with indices "???[2-9]??" were assigned to the training set, 470 molecules with indices "???1??" to the validation set, and 462 molecules with indices "???0??" to the test set. CCSD electron densities were available only for 52 molecules with QM9 indices up to 93 (with omissions). This dataset was split in the following way: molecules with 3 or 4 heavy atoms (indices 9 to 48, with omissions) were assigned to the training set, and molecules with 1, 2 or 5 heavy atoms (indices 1 to 8 and 49 to 93) to the test set.

When choosing the test sets, we kept in mind that, as previously noted, training ML models on datasets like QM9, where most molecules are large and few are small, larger errors are observed for smaller molecules in QM9.[9] For this reason, we chose to put the smallest molecules to test sets, not only because more accurate reference energies and densities are known for smallest molecules, but also because this provides a harder test for the transferability of the models. For example, the closest molecules to water (which was always in test sets) in the validation set for energies in the PML-QC$_{CCSD(T)}$ model were n-butanol ($C_4H_9OH$, QM9 index 134) and 1,3-propanediol [$HO(CH_2)_3OH$, QM9 index 138], and in the training set isobutanol [$(CH_3)_2CHCH_2OH$, QM9 index 272) and glycerol [$HOCH_2CH(OH)CH_2OH$, QM9 index 138], all molecules very different from water. As for the PML-QC$_{DFT}$ model, all molecules in the training set had 9 heavy atoms, which corresponds to an even more distant extrapolation to water.

The PML-QC$_{CCSD(T)}$ model reported here was trained in three stages, each of which included 5 epochs. Learning rates during the first, second and the third stages were kept constant and equal to $2 \cdot 10^{-4}$, $1 \cdot 10^{-4}$, and $2 \cdot 10^{-5}$, respectively. Each epoch included 100 sub-epochs of minimization of $\sum_{i \in \text{training set}} \|\Delta\rho_{DNN}[\rho_{HF,i}(\mathbf{r})] - \Delta\rho_i(\mathbf{r})\|$ and 1 sub-epoch of minimization of $\sum_{i \in \text{training set}} |\Delta E'_{DNN}[\rho_{HF,i}(\mathbf{r})] - \Delta E'_i|$ (simultaneous minimization of $L$ defined by Eq. (S96) was impossible, because training sets for the energies and electron densities were different). The proportion of 100 sub-epochs vs. 1 sub-epoch was found empirically, as that ensuring comparable relative rates of decay of the loss functions for the energies and densities. No regularization for the energy term was performed, because there were no noticeable signs of overtraining in terms of energy predictions for the PML-QC$_{CCSD(T)}$ model (unlike the PML-QC$_{DFT}$ model). The parameters to be changed during training were the kernel coefficients in convolution operators in the paths leading to the energy and density predictions, as well as in the last layer before forking; earlier layers before forking of the computational paths were kept unchanged.

We performed data augmentation on the dataset for the PML-QC$_{CCSD(T)}$ model by rotating and/or reflecting the molecules, because the corresponding high-theory-level data were more scarce that those used to train the PML-QC$_{DFT}$ model. For each molecule, three different random orientations in the 3D box were generated in the dataset for densities, and five different random orientations in the dataset for energies. For each orientation, in its turn, reflections in the xy, xz and/or xz planes were performed (in total, $2^3 = 8$ possible combinations, all leading to nontrivial results, because the axes and planes of symmetry in these molecules were randomly oriented relative to the x, y and z axes), and permutations x, y and/or z coordinate axis were performed (in total, 6 possible permutations), in total producing $3 \cdot 6 \cdot 8 = 144$ different pairs ($\rho_{HF}$, $\Delta\rho$) per molecule in the density dataset and $5 \cdot 6 \cdot 8 = 240$ different pairs ($\rho_{HF}$, $\Delta E'$) per molecule in the energy dataset. Training was performed with minibatches, each of which including 48 ($\rho_{HF}$, $\Delta\rho$) or ($\rho_{HF}$, $\Delta E'$) pairs produced by reflections and axes permutations of the same orientation of the same molecule to enforce the symmetry of the trained model.

To analyze the progress in training, we recorded the MAE and $\varepsilon_\rho$ values, defined by Eqs. (S4) and (S13), as measures of errors in the energy and electron density, respectively. During training, MAE and $\varepsilon_\rho$ computed on the training set, as expected, gradually decreased for both models (Fig. S5). MAE and $\varepsilon_\rho$ computed on the validation set also tended to decrease. Note that the values of MAE and $\varepsilon_\rho$ for the validation set were computed at the end of each training epoch (with the enumeration of epochs starting from 1) using the same state of the DNN (after training on all minibatches in the corresponding epoch) and all molecules from the validation set, while the values of MAE and $\varepsilon_\rho$ for the training set were computed differently: the DNN was updated after each minibatch, the contributions to MAE and $\varepsilon_\rho$ for the molecules in the current minibatch were computed with the current DNN and averaged over all minibatches. Respectively, the values of MAE and $\varepsilon_\rho$ for the training set are shown in Fig. S5 in the middle between the indices of the previous and next epochs.

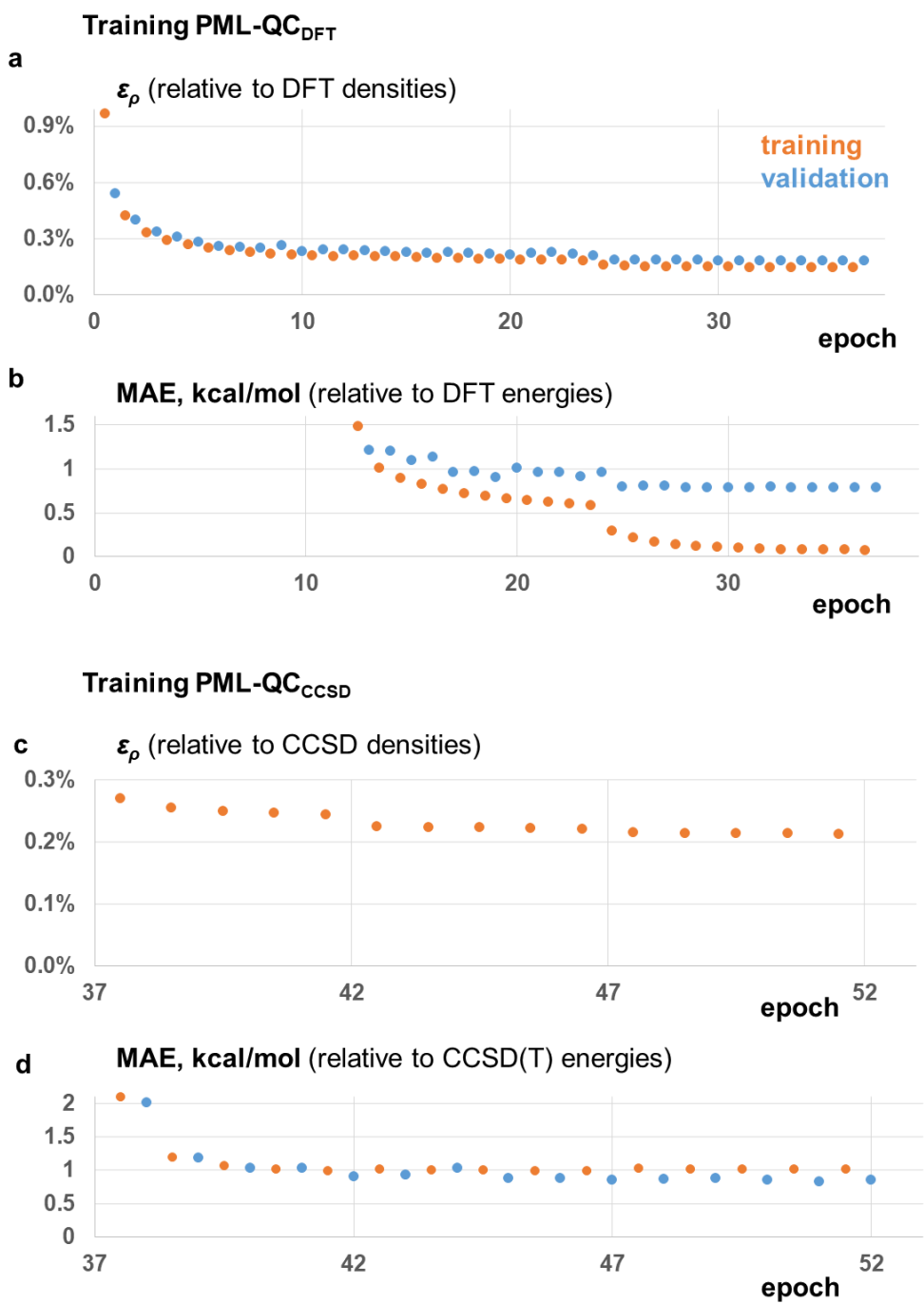

**Fig. S5**. Learning curves for (a,b) PML-QC$_{DFT}$ and (c,d) PML-QC$_{CCSD(T)}$ models in terms of predictions of electron densities (a,c) and energies (b,d). Validation for the electron densities predicted by PML-QC$_{CCSD(T)}$ in (c) was not performed, because we did not observe overfitting on electron density data in (a). For comparison of the performance of the model on training and test data for each molecule in the dataset, see Fig. 3a in the main text and Fig. S6 below.

*S4.4. Performance of the resulting models on training, validation and test sets*

By the end of training of the **PML-QC$_{DFT}$ model**, $\varepsilon_\rho$ on the validation set decreased to 0.18% and stabilized at this level for 3-5 last training epochs (Fig. S5a). The difference of $\varepsilon_\rho$ over the training (0.14%) and validation (0.18%) sets is minor, implying that there is no significant overfitting of the PML-QC$_{DFT}$ model on the electron density data. Over test set 1 (~10% of QM9 molecules with nine heavy atoms), $\varepsilon_\rho$ was found to be 0.14%, confirming the conclusion that there was no overfitting of the model on electron densities. For test set 2 (all molecules with up to eight heavy atoms), $\varepsilon_\rho$ was nearly the same (0.15%), suggesting that the PML-QC$_{DFT}$ model is well transferrable to molecules of sizes different from those in the training set.

MAE for the PML-QC$_{DFT}$ model computed over the validation set decreased to 0.79 kcal/mol by the end of training, staying stable at this level over the last 4 training epochs (Fig. S5b). The corresponding value for the training set was much lower, 0.07 kcal/mol, even with the regularization as described above, Eq. (S96). With $d_{Ereg}$ = 10, the value of the regularization term per molecule, in energy terms, reached 0.21 kcal/mol by the end of training; larger values of $d_{Ereg}$ only increased the values of MAE over the validation set, and did not reduce the gap between the results on the training and validation sets (data not shown). The performance of the model on test set 1 (~10% of molecules with nine heavy atoms) is close to that for the validation set: MAE = 0.82 kcal/mol, while on test set 2 (all molecules with up to eight heavy atoms) MAE was even lower, 0.58 kcal/mol. Median absolute errors were 0.54 and 0.38 kcal/mol for test sets 1 and 2, respectively. Most of the molecules in test set 2 have exactly eight heavy atoms, the mean error per heavy atom is even lower for test set 2 than for test set 1 or the validation set. Therefore, the PML-QC$_{DFT}$ model extrapolates well to molecules with a different number of atoms than in the training set, both in terms of energy and electron density. The values of MAE and $\varepsilon_\rho$ demonstrated by the PML-QC$_{DFT}$ model are comparable to those of other ML models over the QM9 dataset reported in the literature[9,12,31,54,55,58-65] (see also discussion below in Section S5.1).

The values of MAE and $\varepsilon_\rho$ given above refer to a comparison of predicted electron densities and energies to DFT values [PBE0/pcS-3 densities and B3LYP/6-31G(2df,p) energies, respectively]. Comparisons to more physical CCSD electron densities (Fig. 3a, *pink*) and CCSD(T) energies (Table 1) demonstrates that the accuracy of the PML-QC$_{DFT}$ model is nearly the same (actually, a little bit worse) than the accuracy of the DFT methods used to generate the training sets.

The performance of the **PML-QC$_{CCSD(T)}$ model** on the datasets with CCSD electron densities and CCSD(T) energies was as follows. The value of $\varepsilon_\rho$ decreased by the end of training to 0.21% and stayed stable over the last ~2 epochs (Fig. S5c). Due to a limited number of molecules for which CCSD electron densities were computed, a comparison of the errors of the PML-QC$_{CCSD(T)}$ model, HF, various DFT and other methods can be visually performed molecule-by-molecule (Fig. 3a, Fig. S6). The errors for molecules in the training set are comparable to those in the test set, which we explain by suppression of overfitting by data augmentation (an early version of the model was trained on a dataset without augmentation and demonstrated a noticeable difference of $\varepsilon_\rho$ over the training and test datasets, see an earlier version of this manuscript[66]). In 50 out of 52 cases, they stay below the corresponding errors of various DFT methods (Fig. 3a, Fig. S6). The value of MAE by the end of training fell down to 0.82 kcal/mol for the training set and 0.80 kcal/mol for the validation set, staying stable over the last ~3 epochs (Fig. S5d). (During late stages of training, MAE over the training set fluctuates instead of monotonously decreasing, because training is performed not only on energies, but also on densities.) Over the test set, MAE was as low as 0.87 kcal/mol (median absolute error: 0.70 kcal/mol, RMSE: 1.23 kcal/mol). For a further discussion of the performance of our models in comparison to those in the literature, see Section S5.1 below.

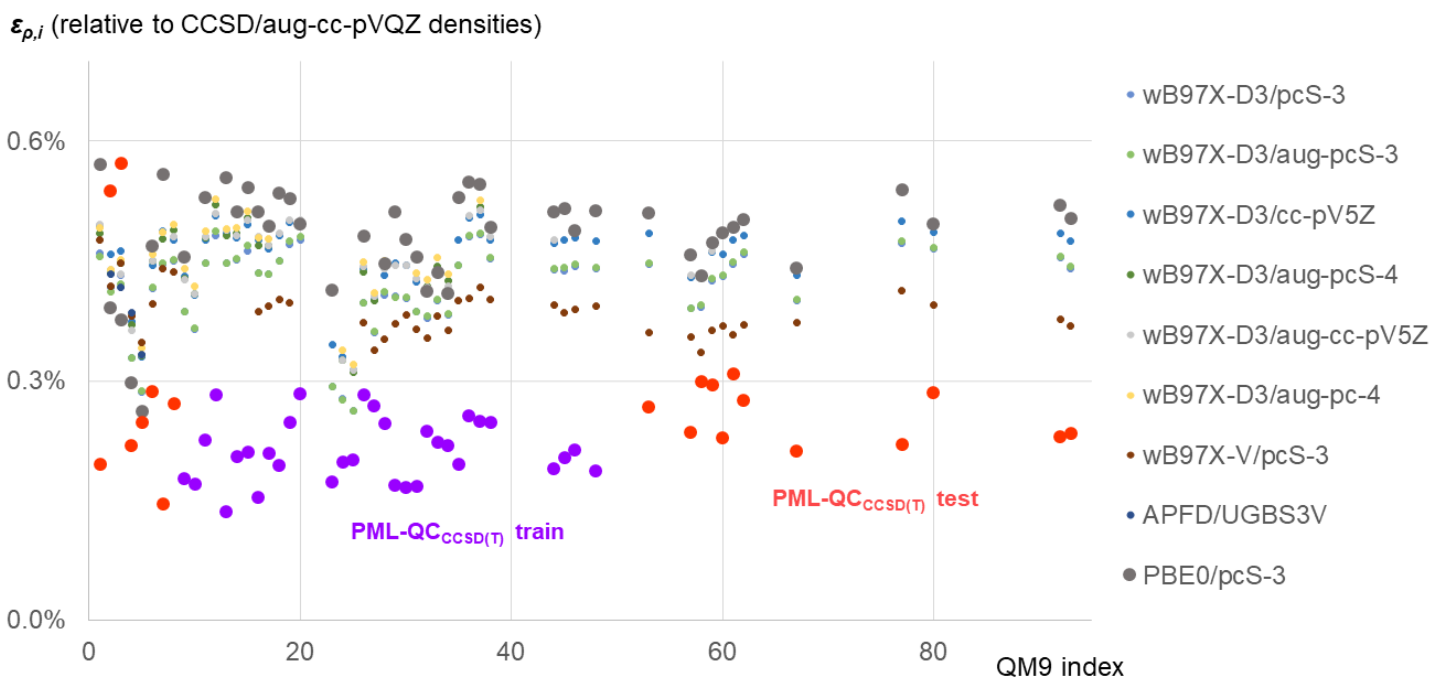

**Fig. S6**. The PML-QC$_{CCSD(T)}$ model outperforms all DFT methods, not only PBE0/pcS-3, in predicting electron densities of 52 molecules for which we had reference RI-CCSD/aug-cc-pVQZ electron densities. This figure supplements Fig. 3a from the main text. Only combinations of DFT methods and basis sets with the lowest values of $\varepsilon_\rho$ are shown.

*S4.5. Scaling*

DNN computations run much faster on the same hardware (1 GPU and 2 CPUs, see Appendix A2 for details) than DFT (PBE0/pcS-3) computations for every molecule in the QM9 dataset (Fig. 2f). For the overwhelming majority of the database entries (with indices ~100 and higher) the DNN is at least two orders of magnitude faster than DFT. HF computations in the small basis set, required to generate the DNN input, are also much faster than DFT computations for the same molecules. The bottleneck in the computational pipeline turned out to be the step of generation of input cube files from HF fchk files. To generate cube files, we used cubegen utility from Gaussian, which is currently implemented only on a single CPU. We expect that this stage can be significantly speeded up in the future with parallelization and the use of GPUs. Even in the current setup, the DNN computations, together with HF computations and generation of the input cube files, are faster than DFT (PBE0/pcS-3) computations for all expect the first four molecules in the QM9 database. Closer to the end of the database, the gain in the speed (measured by wallclock time) reaches a factor of ~30. (Only relative wallclock times are shown in this work; absolute values of wallclock times for computations performed in Gaussian are not disclosed according to the Gaussian licensing agreement.)

Wallclock time to run the reported DNN stays virtually the same over the whole QM9 dataset due to the architecture of the reported DNN. Wallclock time of HF computations increases with the number of elections $N$ as $\sim N^{1.62}$, while PBE0/pcS-3 wallclock time scales less favorably, as $\sim N^{3.26}$. Due to this difference in scaling of the DNN, HF and DFT cost with the size of the molecule, the relative efficiency of the approach proposed here increases with the size of a modeled molecule. The ratio of the wallclock time for HF to that for DFT is ~10% for the first several molecules in the database (entries 1-6), falls to ~1% by entry ~100, and falls even further to ~0.3% closer to the end of the dataset. The wallclock time for the DNN relative to the wallclock time for DFT falls from ~30% for the first several molecules to ~1% by entry ~100, and down to ~0.1-0.3% by the end of the database. As for the relative price of the HF and DNN parts, the DNN stage dominates for smaller systems with up to ~30 electrons. For larger systems, HF becomes consistently more expensive (pronouncedly from ~60 electrons), but the gap between the HF and DFT costs in this limit increases. We do not consider here the scaling of wallclock time for the cube file generation, because, as stated earlier, we expect that this part of computations can be significantly accelerated in comparison to the current version.

## S5. Additional discussion

*S5.1. Comparison to previous work on ML in Chemistry*

Over the last years, ML has been actively applied to various problems in Quantum Chemistry (for reviews, see Refs. [11,13,67-72]). Below, we briefly overview the previous literature focusing on several aspects important for a proper understanding of our work and its relationship to the current state of the field.

**Energy prediction.** In most cases, comparisons in the literature are made only to energies computed with DFT,[8-10,12,15,18,49,54-56,58-65,71,73-95] which are typically several kcal/mol away from the corresponding exact values. Under these circumstances, claims in the literature that certain ML models reach thermochemical accuracy (that is, MAE comparable to or less than 1 kcal/mol,[9,12,54,55,58-65] or even as low as 0.2-0.3 kcal/mol[12,15,59,60,64,71,85,87,93,94]) are misleading. This problem may be mitigated in models trained and tested on post-HF energies (e.g., CCSD(T)/CBS extrapolations for the ANI-1ccx dataset[18,96] or a set of 15 molecules by Nakai et al.,[17] MP2/cc-pVTZ and CCSD(T)/cc-pVDZ energies for QM7b-T molecules and MP2/cc-pVTZ energies for GDB-13-T molecules[14]). Among ML models, we expect the best performance (in comparison to the exact solutions) for those trained on CCSD(T) results in large basis sets, as in Refs. [17,18].

There are two different aspects of the problem of energy predictions: (1) predicting energies of various molecules "at equilibrium" (that is, at minima of their potential energy surfaces), and (2) predicting energies of non-equilibrium (sampled at nonzero temperatures) conformations of molecules. Most of the past work focused on either of these aspects. In particular, some models were trained on datasets that contain only optimized geometries of molecules (e.g., the QM9 dataset),[17,85] or even insensitive to geometries by construction (e.g., use only graph-based descriptors),[57,93,95] and therefore can address only the first problem. Some other papers reported ML models trained to predict non-equilibrium energies of molecules (e.g., undergoing thermal fluctuations at 500 K).[8,9,12,14,15,36,37,81-85,90,97-100] In these cases, the reported accuracy may be comparable to or below 0.1 kcal/mol (even as low as 0.002 kcal/mol for a water molecule at 350 K,[14] or 0.02 kcal/mol for a hydrogen molecule[8]). However, these models are molecule-specific ("dedicated ML force fields": to model a new molecule, one has to train a new model from scratch).[8,9,12,15,36,37,81-85,90,97-100] Only a few models, to the best of our knowledge, can address both problems and predict non-equilibrium geometries of *various* molecules with the *same* model, including the ANI family of models,[18,89,96,101] MOB-ML models trained on the QM7b-T dataset,[14,102] the models using FCHL18 and FCHL19 representations trained on the same dataset,[15,83] and the DTNN models trained on eMol9_CM dataset.[94] In this work, we chose to focus on the first problem, because it is more challenging (residual MAEs are at least one order of magnitude greater than those for the second problem) and practically useful (due to the applicability of the same model to multiple compounds). Extension of our model to non-equilibrium geometries should be possible merely by expanding the training set, without any changes in the architecture or methodology of modeling.

**Electron density prediction.** By now, the only (to the best of our knowledge) ML model capable of predicting electron densities of a wide range of molecules was reported by Ceriotti, Corminboeuf et al. in Ref. [31]. In this model, the full electron density is analytically represented as a linear combination of basis set functions (namely, cc-pVQZ/JKFIT), and ML is used to predict coefficients in that expansion. The training set included 2000 dimers of amino acids formed by sidechain-sidechain interactions extracted from the BioFragment Database (BFDb). Ground truth electron densities were computed with an advanced DFT functional in a relatively large basis set (ωB97X-D/cc-pVQZ), and then fitted with analytical expressions in terms of cc-pVQZ/JKFIT basis sets. The relative error in predicted electron densities $\varepsilon_\rho$, as defined by Eq. (S13), for the test set (229 dimers from the same database) was reported to be ~0.2-0.3% relative to ωB97X-D/cc-pVQZ densities. On our test set of molecules, for which we have been able to compute CCSD/aug-cc-pVQZ densities, the relative error of our method (~0.2% in comparison to CCSD/aug-cc-pVQZ densities, see Section S4.4, Figure S6) is nearly the same as the error of the cited model (in comparison to ωB97X-D/cc-pVQZ densities), with largest errors also concentrated near the nuclei. Note, however, that electron densities computed with CCSD/aug-cc-pVQZ should be closer to exact ones than those computed with ωB97X-D/cc-pVQZ (see section S2.2), making the predictions of our model closer to the real electron densities. For example, over the first six molecules from the QM9 dataset, $\varepsilon_\rho$ for CCSD/aug-cc-pVQZ and ωB97X-D/cc-pVQZ electron densities are 0.20% and 0.48%, respectively (with CCSD/cc-pCV5Z electron densities used as reference 'exact' densities). Also, note that the cited model, unlike ours, does not predict energies.

Some other ML models predicting electron densities have been reported, but they have narrower fields of applicability. An earlier model by Ceriotti, Corminboeuf et al. was trained on several hydrocarbon molecules ($C_2H_6$, $C_2H_4$, $C_4H_6$, $C_4H_{10}$, multiple conformations sampled at 300 K) and tested on other conformations of the same molecules, as well as $C_8H_{10}$ and $C_8H_{18}$.[16] Yet earlier, ML models predicting electron densities in dihydrogen, water, benzene, ethane, and malonaldehyde (one model per molecule),[8] sulfur-crosslinked carbon nanotubes,[56] Ni/Al alloys,[73] Cu/Au and Fe/Pt allows,[92] polyethylene chains and Al slab[90] have been developed. Also, computations of electron densities of benzene, ethanol, malonaldehyde, toluene and water were reported as a part of energy computations.[37] All of these works used DFT electron densities (with $\varepsilon_\rho > 0.35\%$) for training and testing, which limits their accuracy.

In addition to models directly predicting electron densities, some other ML models were reported to predict wave functions. We also consider them here, because electron densities can be computed from wave functions by Eq. (S2). Such models fall into two categories:

1) Models yielding explicitly correlated wave functions, which may exceed in their accuracy even CCSD in large basis sets. In these models, neural networks are used in essence as wave functions ansätze, and variational solution of the Schrodinger equation is technically performed in the form of training an ML model, with the loss function identified with the energy computed with a probe wave function coded by the neural network. An architecture of such a model is molecule-specific, because the dimensionalities of wave functions are different in molecules with different numbers of electrons. "Training" such a model (that is, variationally minimizing the expectation value of the Hamiltonian for a given molecule) coincides with using it for a prediction. For this reason, we doubt that such models can be classified as true ML models, at least from the practical viewpoint, because training, which is computationally very expensive, is not supposed to be repeated every time when an ML model is asked to make a prediction. In any case, we are not aware of any models of this sort that would be capable of predicting *correlated* wave functions for *wide classes* of molecules.

2) Models predicting wave functions in a one-electron (mean field) approximation.[81] In this case, the same model with the same architecture might be applied to molecular systems of arbitrary sizes (note, however, that Ref. [81] reports separate neural networks for water, ethanol, malondialdehyde, and uracil). An HF mean field approximation severely limits the accuracy of predicted wave functions, making them incompetitive with models explicitly predicting electron densities and trained on post-HF electron densities. In the case of Kohn-Sham DFT, the total density computed from the orbitals is in principle exact. However, training an ML model on approximate DFT methods (such as PBE in the cited work) makes their predictions inferior in comparison to the PML-QC$_{CCSD(T)}$ model reported in this work.

We follow the existing approach to benchmark electron densities against those computed *ab initio* with CCSD in large basis sets.[5,6,103] With our computational resources, we were able to compute CCSD/cc-pCV5Z electron densities for the first six molecules from the QM9 dataset, and made comparisons of other methods to them (Section S2.1, Tables S1 and S2). This comparisons demonstrated that DFT (even with such advanced functionals as ωB97X-D) does not approach the accuracy level of CCSD with quadruple-zeta-level basis sets, in agreement with the previous literature.[5] Note that little work has been done in the past on measuring the actual accuracy of DFT in predicting electron densities of molecular systems.[5,6,35,39,104] However, such benchmark studies, even less representative than benchmark studies for energies, would be important for understanding the benefits and drawback of DFT, relative comparisons of various DFT functionals, as well as comparisons of DFT to ML and to other quantum chemical methods. A recent study of the performance of various DFT functionals and *ab initio* methods on several atoms and ions revealed possible problems with overfitting of recent DFT functionals on energies;[5] this is however an area of active discussion.[35,39,103,104] Benchmarking various DFT functionals on electron densities of 52 molecules from the QM9 dataset, for which we have been able to compute CCSD/aug-cc-pVQZ densities, confirms that the relative accuracy of various DFT methods stays stable for this larger set of molecules (Section S2.2, Table S4). For these reasons, we conclude that our PML-QC$_{CCSD(T)}$ model predicts electron densities closer to the real physical electron densities than any other ML model reported in the literature so far. Building datasets of high-theory-level *ab initio* electron densities for training and testing ML models is of vital need, but goes beyond the scope of this work.

**Choice of descriptors**. One of the most influential approaches to featurizing molecular systems goes back to the paper by Behler and Parrinello in 2007.[53] Symmetry functions $G_i^{1,2}$ proposed by these authors encode the composition and geometry of a molecule into a set of parameters characterizing the radial and angular distribution of each atom in space (written below for the *i*-th atom):

$$G_i^1 = \sum_{j \neq i}^{all} \exp\left(-\eta(R_{ij} - R_s)^2\right) f_c(R_{ij}), \tag{S97}$$

$$G_i^2 = 2^{1-\zeta} \sum_{j,k \neq i}^{all} (1 + \lambda \cos\theta_{ijk})^\zeta \exp\left(-\eta(R_{ij}^2 + R_{ik}^2 + R_{jk}^2)\right) f_c(R_{ij}) f_c(R_{ik}) f_c(R_{jk}), \tag{S98}$$

where summation is performed over all neighboring atoms with indices $j$ and $k$; $\eta$, $R_s$ and $\zeta$ are parameters (multiple values of each of these parameters may be used to provide multiple descriptors of a molecule), $\lambda = 1$ or $-1$, $R_{ij}$ is the distance between atoms $i$ and $j$, $\theta_{ijk}$ is the angle between atoms $i$-$j$-$k$, and $f_c$ is a cutoff function defined as

$$f_c(R_{ij}) = \begin{cases} 0.5 \left[ \cos\left(\frac{\pi R_{ij}}{R_c}\right) + 1 \right] & for\ R_{ij} \leq R_c \\ 0 & for\ R_{ij} > R_c \end{cases} \quad (S99)$$

where $R_c$ is the cutoff distance. Various modifications of these symmetry functions have been suggested, such as introducing radial shells and an arbitrary number of shifts in the angular symmetry functions,[10,18,89] prefactors for faster decay with distance and the use of log-normal distributions,[15,60,83,105-107], accounting for external electric field,[83,106] encoding chemical elements by their period and group numbers[60,107] or nuclear charges,[61] etc. An important advantage of such descriptors is the easiness and speed with which they can be computed from atomic coordinates. Also, this approach may intuitively seem reasonable, because it resembles a basis set expansion in Quantum Mechanics (e.g., by radial and angular components), as well as the use of radial distribution functions (pair correlation functions) to characterize condensed phase systems in Statistical Mechanics. However, such a resemblance does not make these descriptors theoretically justified. For example, coefficients $\eta$ are chosen by a researcher's discretion or optimized as hyperparameters during training of an ML model, by contrast to the constants in the radial components of atomic orbitals, which have a clear physical interpretation in terms of a (screened) nuclear charge. Symmetry functions and related descriptors, to the best of our knowledge, have never been derived from the Schrodinger equation, or any other physical equations, even with rough approximations. We only know that atomic coordinates and nuclear charges of atoms in a molecule uniquely determine the external potential, and therefore the solutions to the Schrodinger equation, as discussed above in Section S3.4. Presumably, symmetry functions uniquely determine these atomic coordinates and charges, too, at least locally, but it is not clear whether they provide a physically justified transformation of these degrees of freedom that would simplify a mathematical solution to the problem.

Another popular way of molecular featurization is the use of graph-based descriptors that treat a molecule as a graph (molecular fingerprints, etc.).[57,93,95] Such descriptors may be computed even faster than symmetry functions, and their interpretation is easier from the viewpoint of General Chemistry (though not from the viewpoint of Quantum Mechanics). An evident drawback of these descriptors is that they cannot be used to explicitly model spatial degrees of freedom in molecules (including thermal fluctuations and geometries of transition states). The use of graph-based descriptors pushes ML models in the direction of empirical additive schemes, which may not be well transferable between different classes of chemical compounds.

A few papers suggested the use of results of lower-level QM calculations as descriptors to predict results of high-level methods with ML. In this way, HF matrix elements (but not the electron densities or SCF solutions) were used to predict MP2 and CCSD energies,[14,102,108] MP2 amplitudes to predict coupled cluster energies,[109] approximate HF electron densities and some other 3D variables computed with HF to predict CCSD(T) correlation energy,[17] and coefficients in a basis set expansion of DFT (namely, PBE) electron densities to predict CCSD(T) energies.[36] Such approaches are computationally more expensive at the stage of featurization, in comparison to descriptors easily computed from the molecular geometry, but latest advances in speeding up HF and DFT computations should mitigate this issue. This approach is more physical than the use of symmetry functions or graph-based descriptors: In principle, HF results serve as immediate input to post-HF methods, while for an approximate exchange-only DFT density is was proven that it can serve as a descriptor to compute a correction to energy.[21] However, previous papers on the use of HF (or other low-theory-level) computations to featurize a molecule do not clarify the question of which specific variables computed with low-level QM methods can serve as features, and why they do so. In this paper, we demonstrate that it is an approximate (HF or DFT) electron density that can serve as a universal way to featurize a molecule (Section S3).

3D representations of potentials[8,36,90] or electron densities[17] were also used as descriptors for ML models. Note that the representation of electrostatic potentials in Ref. [8] via Gaussian functions is pretty rough, and their 3D input to an ML model may be equally considered as an approximate electron density, which, in the light of theorems proven in Section S3, might explain the success of this representation. DNNs were also used to automatically build descriptors for atoms, instead of defining such descriptors by hand.[9,12,81,87,90,94] A conceptually close approaches are to perform automatic feature selection with random forest regression (from HF matrix elements, after sorting them in the order of increasing distance between molecular orbitals),[14] or the automatic descriptors.[98]

Regardless of a specific definition, descriptors are often aimed at representing atoms as embedded in chemical environment (either by explicitly defining them, or allowing an ML model to implicitly learn them from simpler geometric representations).[9,12,31,57,81,83,85-87,89,93,94,98,99,101,102,107] Usually, this feature is considered as an advantage of such descriptors; however, we think this may unintentionally lead to negative consequences. Large numbers of descriptors and reference molecules used in ML models (e.g., 1000 reference atomic environments in the best performing model in Ref. [31], or 70 cluster-specific regressions for off-diagonal contributions to energy in Ref. [102]) may push an ML model towards memorizing atom types, rather than learning the physics of molecular systems, which may result in overfitting an ML model, and thereby undermine the transferability of the model to new classes of compounds. Such descriptors of chemical environment also typically use arbitrarily chosen cutoff distances in their definitions, which limits their physicality.

Most descriptors in the literature are not meaningfully transferable to new elements without retraining the corresponding ML models. For example, in SchNet the charge of a nucleus is encoded in a set of coefficients that are initialized randomly and updated during training. The resulting coefficients are not directly defined by physical laws, and adding a new element to the model requires retraining the whole model, including optimization of a new set of coefficients to represent this element.[9,12,15,81] Note, however, that there are exceptions to this limitation, and several descriptors allow for applying ML models to compounds with new elements without retraining; numerical performance of such models is usually noticeably worse than those not admitting straightforward extensions to new elements.[8,14,60]

Other descriptors of molecular systems have also been suggested in the literature.[16,19,31,54-56,82,84,93,97-100,110-112] These descriptors share some or all of the drawbacks with the descriptors discussed above. Specifically, they have not been derived from main physical equations, they provide an opportunity for an ML model to follow an easy path and memorize numerous atom types instead of learning the physics, and they are usually not transferable to new elements without retraining.

In contrast to all these descriptors, an approximate (HF or DFT) electron density as a descriptor has a clear physical interpretation, is not limited to specific elements or equilibrium geometries, and does not imply arbitrary cutoff distances.

**Choice of ML models**. The vast majority of models in the field of ML in molecular modeling are based on the use of kernels.[7,8,15,16,22,31,36,60,82,83,97,98,100,105-107,110-113] In the simplest version (a kernel ridge regression model),[7,8,15,22,60,83,98,105-107,111-113] a predicted property $p$ (e.g., an energy, or a coefficient in a basis set expansion of an electron density / wave function) is represented in the following form as a function of molecular descriptors **d** characterizing the molecular system:

$$p = \sum_{j=1}^{m} \alpha_j k(\mathbf{d}, \mathbf{d}_j), \tag{S100}$$

where $j$ enumerates molecules/conformations in the training set, $m$ is the total number of the molecules in the training set, $\alpha_j$ are coefficients to be learned (fitted), and the kernel $k$ is usually chosen in the form of a Gaussian function of the distance between the descriptors $\|\mathbf{d} - \mathbf{d}_j\|$, with a constant or second-order polynomial prefactor $f(\|\mathbf{d} - \mathbf{d}_j\|)$:

$$k(\mathbf{d}, \mathbf{d}_j) = f(\|\mathbf{d} - \mathbf{d}_j\|) \exp\left(-c\|\mathbf{d} - \mathbf{d}_j\|^2\right), \tag{S101}$$

where $c$ is a constant. Some other approaches, such as ML-HK maps,[8,36] gradient-domain machine learning,[82,84,97,100] and Gaussian process regression,[16,19,31,102,110] use similar expressions based on kernels. For example, in Gaussian process regression, summation is performed over not only kernels, but also their derivatives. Thus, *kernel-based methods explicitly retain information on a finite number of cases* for certain molecules/conformations with known exact answers. By construction, this ML method may provide efficient interpolation within the set of molecules / conformations used for training (and extrapolation close to the training set), but does not learn the physics of the modeled system. It is easy to see this from the following limiting case: Suppose that a kernel-based model is used to make a prediction for a molecule/conformation significantly different from all molecules/conformations in the training set. Then all the distances between the molecular descriptors $\|\mathbf{d} - \mathbf{d}_j\|$ will be large, and the corresponding kernels (as well as their derivatives) will be negligibly small because of their Gaussian decay with the distance. As a result, the predicted value of the property will go to zero (or a constant, if the summation involves a constant term), regardless of what system the model is applied to, and what property $p$ is predicted. It was noted[36] that this property of kernel-based methods is useful for numerical stability if ML is used to predict corrections to an approximate physical method whose behavior is better understood (such as HF or DFT), ensuring that outside the training set it does not create artifacts. However, this is in a striking contrast to the transferability of the Schrodinger equation (or other general physical equations), which are universally applicable, and do not lead to vanishing corrections to approximate models far from the training set.

DNNs are more flexible than kernel-based models. They do not memorize specific molecules or conformations, do not automatically make zero predictions in cases dissimilar from those used for training, and are overall in a better position to learn the physics.[23,27,114] Nakai et al. reported a DNN that used five input channels (electron density, density gradient, kinetic energy density, HF exchange energy density, and electron density calculated using the fractional occupation number of MOs) to predict correlation energies. An attempt to use the electron density as the only input led to significantly worse results (MAE of 29 kcal/mol instead of 1.1 kcal/mol for the ML5 model with all five input channels).[17] The neural network used in the cited work was not a convolutional network, and could not directly compute derivatives of the density and some other variables depending on it, which may explain the need for multiple input channels. This difference is important not only from the theoretical viewpoint, but also for applications: the time limiting step in predicting correlation energies in Ref. [17] was calculating HF exchange energy density, required for the ML5 and ML4 models demonstrating reasonable MAE, which were slower than electron density computations. The DNN that we report in this work did not require multiple input channels and, unlike the cited paper by Nakai et al., predicted not only energies, but also electron densities.

An important idea in the field of ML in molecular modeling is to build multi-task models, that is models simultaneously predicting several properties of a molecular system. Numerous ML models in the literature were trained to predict properties related to electron densities, such as forces, partial charges, dipole moments, polarizabilities, hyperpolarizabilities,[12,15,44,54,55,60,74,81,83,89,91-93,100,110,115-117] as well as one-electron wave functions,[81] simultaneously with predicting energies. Predicting multiple properties regularizes ML models, makes them more physical and more resistant to overfitting to a specific task of predicting energies. Note that in order to ensure regularization, predictions of energies and other properties should be done from a shared pool of highly processed features, and will not happen if such predictions are performed by independent neural networks, as often done in the literature. The model we suggest in this work follows this philosophy, but goes beyond the current state of affairs by using electron densities, which are more information-rich than all the properties listed above. A possible extension of this work in the future may include the use of 3D correlation energy density, as in Ref. [17], instead of a single value of the energy per molecule.

**Learning density functionals.** So far in this Section S5.1, we have been discussing ML models that directly predict energies or electron densities (or corrections to them relative to quantum chemical methods, such as HF theory). However, an important part of research follows a different philosophy. It uses ML to directly learn the exact Hohenberg-Kohn functional $E[n]$ (or one of its nontrivial components) from data on specific molecular systems (e.g., high theory level *ab initio* solutions for datasets of molecules).[8,13,16,22,37,49,69,118-124] It is supposed that $E[n]$ learned in this way could be subsequently used to compute energies and electron densities of molecules. However, practical implementations of this approach may face technical problems. Solutions may become unphysical due to oscillations and sensitivity to basis sets.[22,125] Optimization tasks for DNNs are known to be associated with artifacts, such as "adversarial examples" (slightly perturbed images that are wrongly classified by DNNs, though the original images were classified correctly),[126-128] and optimization procedures with functions or functionals approximated by DNNs in Quantum Chemistry may encounter similar problems, ending up in unphysical solutions ("adversarial electron densities"). The use of $E[n]$ in an iterative minimization procedure should be much more computationally expensive than a single-shot computation of the density and energy as suggested in this work. Training on ground state solutions does not ensure the correct behavior of $E[n]$ away from the minima. (In many cases, DFT energies can be improved if computed with DFT functionals on HF densities, demonstrating that these functionals are good at minima, but provide wrong behavior away from them. If an ML model is trained only on minima of $E[n]$, it will likely suffer from the same issue.) To circumvent all these difficulties, we propose in this work to directly learn the functional $\Delta E[\cdot]$ and the function-to-function mapping $\Delta \rho[\cdot]$, rather than the Hohenberg-Kohn functional $E[\cdot]$ or its unknown parts.

*S5.2. Additional discussion of the proposed PML-QC approach*

In the previous Section S5.1, we compared the key features of our PML-QC$_{DFT}$ and PML-QC$_{CCSD(T)}$ models to ML models in the literature. We now proceed with the discussion, focusing on some aspects that were not covered in that format.

Predicting not only energies, but also 3D electron densities, and the use of chemically diverse test sets enable us to check the performance of the models (in particular, the PML-QC$_{DFT}$ model) not only in terms of aggregate measures of errors, like $L1$ measures, but also with 3D detailization for various molecules and functional groups in them. Such analysis reveals that our models predicted multiple nontrivial features in the density differences, though the input densities did not contain anything similar to such features (Figs. 2c and 3b, Movies S2 and S3). For example, toroidal parts of isosurfaces around C–C and C–H bonds (Fig. 2c and Movie S2, $\Delta\rho = -0.003$) have been placed by the PML-QC$_{DFT}$ model in correct positions, perpendicular to the corresponding chemical bonds, and at a distance where the input density is small by absolute value and does not include any toroidal shapes. Also, our models learned to distinguish all heavy atoms in the training and test sets (C, N, O, F), as seen from correct density difference isosurfaces predicted around them (correspond to 0, 1, 2 and 3 lone pairs, respectively), though no information on the chemical nature of these atoms or their nuclear charges has been explicitly passed to the neural network. As for the PML-QC$_{CCSD(T)}$ model, even though in this case a comparison to ground truth electron densities is possible for a smaller number of molecules, it confirms a chemically meaningful prediction of features of electron densities around various functional groups (Fig. 3b and Movie S3). In all cases of significant discrepancies between the PML-QC predictions and ground truth results, we found that errors were in the scale of predicted $\Delta\rho$, but not its local spatial behavior. This might be corrected in future work by adding additional layers to the DNNs to allow for a more flexible nonlinear transformation of the output (though this would slow down calculations).

Due to the chosen architecture of the DNN, the cost of the PML-QC computations stays nearly constant as a function of the size of a molecule (Fig. 2f). However, this relationship holds only for molecules that fit into the cube grid we used in this work. In general, for molecules much larger than those in the QM9 database, larger grids would be required, and a new DNN with a different architecture would have to be trained. We speculate that the cost of DNN predictions in this regime will scale in the range between $\sim N \log N$ and $\sim N^3 \log N$, depending on the shape of molecules. The first factor ($N$ to $N^3$) comes from an increase of the required 3D grid size: for compact molecules, roughly linearly with the molecule volume, hence roughly linearly with $N$, but for elongated molecules, roughly cubically with the molecule length, hence cubically with $N$. The second factor, $\log N$, estimates possible increase in the depth of the network that may be required to process larger grids. The scaling could be made more favorable (maybe down to $\sim N \log N$) with more flexible designs of DNNs, enabling them to work with non-cubic inputs of variable size, e.g., scanning window architectures. This would allow for $\sim N$, not $\sim N^3$, scaling of the input size for elongated molecules. However, such an optimization of PML-QC models might not be a top priority now, because for large molecular systems, which are of more practical interest, HF computations required to generate an input to the DNN are more expensive than DNN computations (Fig. 2f).

It is important to emphasize that ML does not replace or invalidate the methods of Quantum Chemistry. In fact, ML reinforces these methods and extends them to a wider realm of practical applicability. Unlike *ab initio* methods, predictions of electron densities and energies with PML-QC are expected to have much more favorable scaling, as discussed above. However, this favorable scaling can be achieved only on the basis of running high-level quantum chemical computations for molecules in a sufficiently large training database. The use of PML-QC models therefore separates the problem of increasing the accuracy of quantum chemical methods from the problem of routine applicability of such methods. With ML, it may become not required that an accurate quantum chemical method works fast enough for every new molecule that an end user may be interested in. Instead, the focus shifts to generating highly accurate results only for a finite dataset to be used for training, while the efficiency in practical applications is to be achieved via improvements in PML-QC models to make them faster and more accurate.

In connection to the problem of training datasets, a question that remains open is how the performance of various quantum chemical methods, in combinations with various basis sets, relates to each other. This question is very important from the practical viewpoint, because it determines the strategy of generating datasets for ML. Which theory level and basis set to use, given limited computational resources? What measures of errors to use for comparisons? In this work, we generated the database of DFT (PBE0/pcS-3) solutions for the QM9 dataset, and RI-CCSD/aug-cc-pVQZ densities and RI-CCSD(T)/aug-cc-pVTZ energies for a subset of QM9. However, these electron densities and energies are still noticeably different from the exact ones, as follows from Tables S1–S4. What will be the next level of accuracy for QM9 or other quantum mechanical databases of a similar size? This question is of major practical importance, and answering it may require large-scale benchmark computations for various methods and basis sets. Training on higher level data is a separate problem. Though there is some evidence that ML methods may work well with higher level quantum chemical methods,[14,19,97,100,102] this is yet to be demonstrated. One should not assume as granted that if a certain ML model can be trained on not-so-accurate DFT functionals, it will perform equally well (or even better) on high-quality data.

Other possible directions for further research include the following. Databases of electron densities for a large number of inorganic molecules, organic molecules with more elements than in QM9, ions, noncovalently bound molecular complexes and other molecular systems not included in the QM9 database should be built and used for validation of the proposed approach of physical machine learning. New architectures of DNNs, allowing for piecewise scanning of input and output, could be developed to remove a restriction on the size of a modeled system that exists in the reported DNN. Also, the present approach needs to be extended to modeling geometries of molecules far from equilibrium, processes of bond formation and breaking, and excited electronic states.

We would like to end this discussion by recapping that the key element of success in applications of ML in Quantum Chemistry, in our opinion, lies in involving as much physics as possible into ML models. We demonstrate in this work that the use of HF (or DFT) electron densities looks very promising, because such a representation is more physical than other descriptors. Training ML models not only on energies, but also on electron densities may serve as a possible strategy to make models more physical, and stimulate learning the physics of modeled phenomena, rather than curve fitting.

# APPENDICIES

## A1. Brief characterization of the QM9 dataset

Out of several quantum chemical databases of molecules, we chose to work with QM9.[28] This database includes 133,885 organic molecules made up of C, H, N, O and F elements. All molecules in QM9 have up to nine heavy (non-hydrogen) atoms, zero electric charge, and zero spin. The QM9 database includes the geometries of all molecules optimized with B3LYP DFT functional and 6-31G(2df,p) basis set. Unlike other existing quantum chemical databases,[10,129-131] QM9 is at the same time easy to work with (all data files can be easily downloaded and processed locally); contains more molecules than most other databases, but still a manageable number of molecules to run quantum chemical computations for all of them with our resources; includes F (not only C, H, N, O), and focuses on conformations close to equilibrium (we chose to investigate how to do ML of electron densities in molecules near equilibrium before proceeding to non-equilibrium geometries, because, as discussed in Section S5.1, the problem of building a neural network applicable to multiple molecules is much more difficult than the problem of building a network predicting energies of thermally sampled non-equilibrium geometries of one or several molecules). In this work, we make the results of our quantum chemical computations for the QM9 dataset available to all researchers by publishing it in a public repository, hoping that these data on the wavefunctions and electron densities of more than 133 thousand molecules will foster further progress in the field of ML in Quantum Chemistry.

## A2. Quantum chemical computations

Production DFT (**PBE0/pcS-3**) computations for all molecules in the QM9 database were carried out in Gaussian 16, revision A.03 (Gaussian, Inc.). Each computation used one GPU (NVIDIA Kepler GK210, in NVIDIA Tesla K80) and two CPUs (in Intel Xeon CPU E5-2680 v2) on XStream, a Cray CS-Storm GPU compute cluster at Stanford University (http://xstream.stanford.edu). The geometry of all molecules was taken from the QM9 database. No changes in the geometry, such as energy minimization, were made. Computations in Gaussian were run with DFT functional "PBE1PBE" and keywords "NoSymmetry Output=WFX Density=Current Population=Full". The pcS-3 basis set was downloaded from EMSL Basis Set Library (https://bse.pnl.gov).[132,133] Both wfx and chk files were recorded; fchk files were subsequently generated from chk files with formchk utility from Gaussian. Cube files for the total electron density were generated from fchk files with cubegen utility from Gaussian with keyword "FDensity=SCF". The grid had 256 points per each side (cubic $256 \times 256 \times 256$ grid), and the step size of 0.1 bohr in each direction. Each molecule was positioned in the middle of a cube (i.e., shifted in space relative to the position in the original QM9 database); no rotations of the molecules were performed. The resulting $256 \times 256 \times 256$ cube files were coarse-grained to $64 \times 64 \times 64$ cube files using a simple C++ code written by us.

Production **HF/cc-pVDZ** computations for all molecules in the QM9 database were carried out in the same way as PBE0/pcS-3 computations (see above), except that the method keyword was "HF", and the cc-pVDZ basis set internally implemented in Gaussian was used.

Reference **CCSD/cc-pCV5Z** computations for the first six molecules from the QM9 database were carried out in Gaussian (as above). Each computation used one CPU (in Intel Xeon E5-4640, E5-4650v4 or E5-2697Av4) and up to 400 GB of memory on Sherlock, a high-performance computing cluster at Stanford University (https://www.sherlock.stanford.edu). To make the full use of the symmetry of these small molecules, Z-matrices accounting for their high symmetry were manually generated. The values of the bond lengths, angles and dihedral angles were computed from the corresponding Cartesian coordinates in the QM9 database; in the cases of small differences between such values computed from different subsets of atoms related by symmetry operations, arithmetic averages of the values were used. No other changes in the geometry, such as energy minimization, were performed. The basis set was taken from EMSL Basis Set Library (as above). Keywords "Output=WFX Density=Current Population=Full" were used. Both wfx and chk files were recorded; fchk files were subsequently generated from chk files with formchk utility from Gaussian. Cube files for the total electron density were generated from fchk files with cubegen utility from Gaussian with keyword "FDensity=CC". The grid had 161 points per each side (cubic 161 × 161 × 161 grid), and the step size of 0.1 bohr in each direction. No shifts or rotations of the molecules in the cube files were performed (due to the use of Z-matrices rather than coordinates from the QM9 database for these computations).

Various **HF, MP2, CCSD and DFT computations with various basis sets** for the purpose of generating data shown in Tables S1 and S2 were carried out in Gaussian (as above) or Q-Chem, version 5.1.0 (Q-Chem, Inc.). Computations were run on XStream (as above) with the use of one CPU and one GPU (as above), or on Sherlock (as above) with the use of one CPU (as above). Z-matrices, the same as in CCSD/cc-pCV5Z computations, were used. Basis sets internally implemented in Gaussian or Q-Chem were used. For computations in Gaussian, chk files were saved, converted to fchk files, and then cube files (161 × 161 × 161 grid, step size of 0.1 bohr) were computed (as above). For computations in Q-Chem, cube files of the same size were directly generated, with the use of "make_cube_files true" keyword in the Q-Chem input files. No frozen cores were used in any of these computations.

The following DFT functionals were screened (keywords for the methods in the corresponding software are given): APFD, B3LYP, B3PW91, BLYP, HSEH1PBE, M062X, mPW3PBE, OHSE1PBE, OHSE2PBE, PBE1PBE, PBEh1PBE, TPSSh, ωB97X, ωB97XD (in Gaussian), B3LYP, B3PW91, B97-D3, B97M-rV, BLYP, M06-2X, M06-L, PBE, revPBE, revPBE0, TPSS, TPSSh, ωB97M-V, ωB97X, ωB97X-D, ωB97X-D3, ωB97X-V, wM05-D (in Q-Chem). These functionals were selected because they demonstrated high performance in various benchmark studies.[5,24,44]

The list of screened basis sets included the following: Apr-cc-pV5Z, Apr-cc-pV6Z, Apr-cc-pVDZ, Apr-cc-pVQZ, Apr-cc-pVTZ, AUG-cc-pV5Z, AUG-cc-pV6Z, AUG-cc-pVDZ, AUG-cc-pVQZ, AUG-cc-pVTZ, aug-pc-3, aug-pc-4, aug-pcJ-3, aug-pcJ-4, aug-pcS-3, aug-pcS-4, aug-pcseg-3, aug-pcseg-4, CBSB7, cc-pV5Z, cc-pV6Z, cc-pVDZ, cc-pVQZ, cc-pVTZ, CEP-121G, CEP-31G, CEP-4G, D95, D95V, dAug-cc-pV5Z, dAug-cc-pV6Z, dAug-cc-pVDZ, dAug-cc-pVQZ, dAug-cc-pVTZ, Def2QZV, Def2QZVP, Def2QZVPP, Def2SV, Def2SVP, Def2SVPP, Def2TZV, Def2TZVP, Def2TZVPP, DGDZVP, DGDZVP2, DGTZVP, EPR-II, EPR-III, Jul-cc-pV5Z, Jul-cc-pV6Z, Jul-cc-pVDZ, Jul-cc-pVQZ, Jul-cc-pVTZ, Jun-cc-pV5Z, Jun-cc-pV6Z, Jun-cc-pVDZ, Jun-cc-pVQZ, Jun-cc-pVTZ, LanL2DZ, LanL2MB, May-cc-pV5Z, May-cc-pV6Z, May-cc-pVDZ, May-cc-pVQZ, May-cc-pVTZ, MidiX, MTSmall, pc-3, pc-4, pcJ-3, pcJ-4, pcS-3, pcS-4, pcseg-3, pcseg-4, QZVP, SDD, SDDAll, SHC, spAug-cc-pV5Z, spAug-cc-pV6Z, spAug-cc-pVDZ, spAug-cc-pVQZ, spAug-cc-pVTZ, STO-3G, SV, SVP, TApr-cc-pV5Z, TApr-cc-pV6Z, TApr-cc-pVDZ, TApr-cc-pVQZ, TApr-cc-pVTZ, TJul-cc-pV5Z, TJul-cc-pV6Z, TJul-cc-pVDZ, TJul-cc-pVQZ, TJul-cc-pVTZ, TJun-cc-pV5Z, TJun-cc-pV6Z, TJun-cc-pVDZ, TJun-cc-pVQZ, TJun-cc-pVTZ, TMay-cc-pV5Z, TMay-cc-pV6Z, TMay-cc-pVDZ, TMay-cc-pVQZ, TMay-cc-pVTZ, TZV, TZVP, UGBS, UGBS1O, UGBS1P, UGBS1V, UGBS2O, UGBS2P, UGBS2V, UGBS3O, UGBS3P, UGBS3V, 3-21G, 4-31G, 6-21G, 6-311+G, 6-311G, 6-31G (in Gaussian), aug-cc-pCV5Z, aug-cc-pCVDZ, aug-cc-pCVQZ, aug-cc-pCVTZ, aug-cc-pV5Z, aug-cc-pVDZ, aug-cc-pVQZ, aug-cc-pVTZ, aug-pc-1, aug-pc-2, aug-pc-3, aug-pc-4, aug-pcS-0, aug-pcS-1, aug-pcS-2, aug-pcS-3, aug-pcS-4, aug-pcseg-0, aug-pcseg-1, aug-pcseg-2, aug-pcseg-3, aug-pcseg-4, cc-pCV5Z, cc-pCVDZ, cc-pCVQZ, cc-pCVTZ, cc-pV5Z, cc-pVDZ, cc-pVQZ, cc-pVTZ, crenbl, def2-QZVP, def2-QZVPD, def2-QZVPP, def2-QZVPPD, def2-SVP, def2-SVPD, def2-TZVP, def2-TZVPD, def2-TZVPP, def2-TZVPPD, DZ, DZ+, DZ++, G3LARGE, G3MP2LARGE, hwmb, lacvp, lanl2dz, lanl2dz-sv, pc-0, pc-1, pc-2, pc-3, pc-4, pcJ-0, pcJ-1, pcJ-2, pcJ-3, pcJ-4, pcS-0, pcS-1, pcS-2, pcS-3, pcS-4, pcseg-0, pcseg-1, pcseg-2, pcseg-3, pcseg-4, r64G, racc-pVDZ, racc-pVQZ, racc-pVTZ, rcc-pVQZ, rcc-pVTZ, sbkjc, srlc, srsc, STO-2G, STO-3G, STO-6G, SV, TZ, TZ+, TZ++, TZV, UGBS, VDZ, VTZ, 3-21+G, 3-21G, 4-31G, 6-31+G, 6-311+G, 6-311G, 6-31G (in Q-Chem).

In total, for DFT, we obtained nontrivial results for 1259 combinations of a functional and a basis set in Gaussian, and 1744 combinations in Q-Chem (two lists of combinations partially overlapped). For MP2, we successfully ran computations for at least one molecule out of six with 97 basis sets, and for CCSD, with 86 basis sets. These numbers exclude screened combinations of functionals and basis sets (for DFT) or basis sets (for MP2 and CCSD) for which quantum chemical computations have not converged for various reasons [e.g., insufficient wallclock time (up to 2 days allocated), insufficient memory, diverged SCF iterations, etc.].

$\varepsilon_\rho$ measures reported in Tables S1 and S2 were computed as follows. Cube files for two compared combinations of method and basis set were calculated, either from fchk files (as described) or directly (CCSD and MP2 computations in Q-Chem), with the same size, position and orientation of the grids as for the reference electron densities. After that, differences of two cube files were computed with cubman utility from Gaussian, and sums of absolute values of all elements in each difference cube files were computed with a simple C++ code written by us. The values of $L1_i$ were computed as the products of sums of all values multiplied by the grid spacing cubed (essentially, with a 3D generalization of the rectangle rule). Simultaneously, an integral of each electron density over the whole cube was computed to check the accuracy of such integration and the sufficiency of the cube size. We tried different grid spacings and concluded that 0.1 bohr (but not 0.2 bohr) is sufficient to get at least two correct significant figures in the values of $L1_i$ measures (data not shown). With this grid spacing, it is sufficient to use a 161 × 161 × 161 grid to fit any of six molecules shown in Table S1 and S2. $L1_i$ values were converted to $\varepsilon_\rho$ values by Eq. (S13).

To compute the effective time shown in Fig. S2, we ran DFT computations for each combination of a functional and basis set for QM9 entries 8 000, 16 000 and 32 000, as described above, and recorded total wallclock time for each computation to complete. Whenever possible, ratios of total times for molecules 16 000 and 8 000 were computed, and a median value of these ratios across all combinations of a functional and basis set was found. Similarly, a median value of the ratio of wallclock times for molecules 32 000 and 8 000 was computed. Finally, for every functional/basis set combination, the wallclock time for molecule 8 000, the wallclock time for 16 000 divided by the median for the 16000/8000 ratio, and the wallclock time for 32 000 divided by the median for the 32000/8000 ratio were computed whenever possible. The effective time for every functional/basis set combination was computed as a geometric average of those of three variables that were available.

## A3. Illustration for Taylor series expansion

This Appendix provides an illustration for the analysis presented in Section S3.3 for a much simpler example of a function of a single scalar variable. Consider an infinitely differentiable real-valued function $y(x)$ of a real-valued variable $x$. Suppose we know the values of all derivatives $y(x)$ of at a certain value $x_0$. Then, the values of $y$ at any $x$ within the radius of convergence can be found from the Taylor series:

$$y(x) = y(x_0) + y'(x_0)\Delta x + \frac{y''(x_0)}{2!}\Delta x^2 + \frac{y'''(x_0)}{3!}\Delta x^3 + \dots, \tag{S102}$$

where $\Delta x = x - x_0$. Further, Eq. (S102) can be used to find all minima of $y(x)$ within the radius of convergence with an arbitrary accuracy as a function of the values of derivatives of $y(x)$ at $x_0$ only.

Cutting this series at the second order terms, we arrive at

$$y(x) \simeq y(x_0) + y'(x_0)\Delta x + \frac{y''(x_0)}{2!}\Delta x^2. \tag{S103}$$

This equation is analogous to Eq. (S71) in Section S3.3. The corresponding minimum [assuming $y''(x_0) > 0$] is achieved at

$$y'(x_0) + y''(x_0)\Delta x = 0, \tag{S104}$$

$$\Delta x = -\frac{y'(x_0)}{y''(x_0)}. \tag{S105}$$

This equation corresponds to Eq. (S73) in Section S3.3. Cutting Eq. (S102) at the third-order term, we get a more accurate approximation:

$$y(x) \simeq y(x_0) + y'(x_0)\Delta x + \frac{y''(x_0)}{2!}\Delta x^2 + \frac{y'''(x_0)}{3!}\Delta x^3. \tag{S106}$$

At the local minimum, the following condition is satisfied:

$$y'(x_0) + y''(x_0)\Delta x + \frac{y'''(x_0)}{2}\Delta x^2 = 0. \tag{S107}$$

This equation has two solutions:

$$\Delta x = -\frac{y''(x_0)}{y'''(x_0)}\left[1 \pm \sqrt{1 - \frac{2y'(x_0)y'''(x_0)}{\left(y''(x_0)\right)^2}}\right]. \tag{S108}$$

In the limit of $y'''(x_0) \to 0$, the solution with the plus sign diverges, while the solution with the minus sign approaches the right hand side of Eq. (S105), following the asymptotic expansion given below:

$$\Delta x = -\frac{y'(x_0)}{y''(x_0)} - \frac{(y'(x_0))^2}{2(y''(x_0))^3} y'''(x_0) + O\left((y'''(x_0))^2\right). \tag{S109}$$

This result corresponds to Eq. (S75) in Section S3.3. Note that Eq. (S109) can also be obtained by rearranging Eq. (S107) in the following way:

$$y''(x_0)\Delta x = -y'(x_0) - \frac{y'''(x_0)}{2}\Delta x^2, \tag{S110}$$

$$\Delta x = -\frac{y'(x_0)}{y''(x_0)} - \frac{y'''(x_0)}{2y''(x_0)}\Delta x^2, \tag{S111}$$

rewriting it in an iterative form:

$$\Delta x_{n+1} = -\frac{y'(x_0)}{y''(x_0)} - \frac{y'''(x_0)}{2y''(x_0)}\Delta x_n^2, \tag{S112}$$

using the approximate solution, Eq. (S105), as the initial value for iterations $\Delta x_0$, and carrying out one iteration with Eq. (S112). This approach [unlike the asymptotic expansion of Eq. (S108) leading to Eq. (S109)] can be directly generalized from the analysis of a real-valued function $y(x)$ to the density functionals as given in Section S3.3.

To sum up, for an infinitely differentiable real-valued function of a real-valued variable, to find a minimum of the function, it is sufficient to know the values of the derivatives only at a certain point $x_0$ that is close enough to the minimum (more strictly, the minimum must be within the radius of convergence of the Taylor series about $x_0$). Our analysis in Section S3.3 provides an analogue of this simple result for the density functional $E[\cdot]$ in Quantum Chemistry.

### A4. Data availability

The fchk files for all QM9 molecules computed with PBE0/pcS-3 and HF/pcS-3, as well as a file with the corresponding energies, are available from Stanford Digital Repository by the following link: https://purl.stanford.edu/kf921gd3855 The code for the DNN (in Tensorflow) and the cube files used for training will be made publicly available after the paper is accepted.

### References

1   Hohenberg, P. & Kohn, W. Inhomogeneous Electron Gas. *Phys. Rev.* **136**, B864 (1964).
2   Kohn, W. & Sham, L. J. Self-Consistent Equations Including Exchange and Correlation Effects. *Phys. Rev.* **140**, A1133, doi:10.1103/PhysRev.140.A1133 (1965).
3   Parr, R. G. & Yang, W. *Density-functional theory of atoms and molecules*. (Oxford University Press; Clarendon Press, 1989).
4   Becke, A. D. Perspective: Fifty years of density-functional theory in chemical physics. *J. Chem. Phys.* **140**, 18A301, doi:10.1063/1.4869598 (2014).
5   Medvedev, M. G., Bushmarinov, I. S., Sun, J., Perdew, J. P. & Lyssenko, K. A. Density functional theory is straying from the path toward the exact functional. *Science* **355**, 49-52, doi:10.1126/science.aah5975 (2017).
6   Brorsen, K. R., Yang, Y., Pak, M. V. & Hammes-Schiffer, S. Is the Accuracy of Density Functional Theory for Atomization Energies and Densities in Bonding Regions Correlated? *J Phys Chem Lett* **8**, 2076-2081, doi:10.1021/acs.jpclett.7b00774 (2017).

7   Rupp, M., Tkatchenko, A., Müller, K. R. & von Lilienfeld, O. A. Fast and accurate modeling of molecular atomization energies with machine learning. *Phys Rev Lett* **108**, 058301, doi:10.1103/PhysRevLett.108.058301 (2012).
8   Brockherde, F. *et al.* Bypassing the Kohn-Sham equations with machine learning. *Nat. Commun.* **8**, 872, doi:10.1038/s41467-017-00839-3 (2017).
9   Schutt, K. T., Arbabzadah, F., Chmiela, S., Muller, K. R. & Tkatchenko, A. Quantum-chemical insights from deep tensor neural networks. *Nat. Commun.* **8**, 13890, doi:10.1038/ncomms13890 (2017).
10  Smith, J. S., Isayev, O. & Roitberg, A. E. ANI-1: an extensible neural network potential with DFT accuracy at force field computational cost. *Chem. Sci.* **8**, 3192-3203, doi:10.1039/c6sc05720a (2017).
11  Butler, K. T., Davies, D. W., Cartwright, H., Isayev, O. & Walsh, A. Machine learning for molecular and materials science. *Nature* **559**, 547-555, doi:10.1038/s41586-018-0337-2 (2018).
12  Schutt, K. T., Sauceda, H. E., Kindermans, P. J., Tkatchenko, A. & Muller, K. R. SchNet - A deep learning architecture for molecules and materials. *J. Chem. Phys.* **148**, 241722, doi:10.1063/1.5019779 (2018).
13  von Lilienfeld, O. A. Quantum Machine Learning in Chemical Compound Space. *Angew. Chem. Int. Ed. Engl.* **57**, 4164-4169, doi:10.1002/anie.201709686 (2018).
14  Cheng, L., Welborn, M., Christensen, A. S. & Miller, T. F. A universal density matrix functional from molecular orbital-based machine learning: Transferability across organic molecules. *J Chem Phys* **150**, 131103, doi:10.1063/1.5088393 (2019).
15  Christensen, A. S., Bratholm, L. A., Faber, F. A., Glowacki, D. R. & von Lilienfeld, O. A. FCHL revisited: faster and more accurate quantum machine learning. *arXiv*, 1909.01946v01941 (2019).
16  Grisafi, A. *et al.* Transferable Machine-Learning Model of the Electron Density. *ACS Cent Sci* **5**, 57-64, doi:10.1021/acscentsci.8b00551 (2019).
17  Nudejima, T., Ikabata, Y., Seino, J., Yoshikawa, T. & Nakai, H. Machine-learned electron correlation model based on correlation energy density at complete basis set limit. *J Chem Phys* **151**, 024104, doi:10.1063/1.5100165 (2019).
18  Smith, J. S. *et al.* Approaching coupled cluster accuracy with a general-purpose neural network potential through transfer learning. *Nature Commun.* **10**, 2903, doi:10.1038/s41467-019-10827-4 (2019).
19  Wilkins, D. M. *et al.* Accurate molecular polarizabilities with coupled cluster theory and machine learning. *Proc Natl Acad Sci U S A* **116**, 3401-3406, doi:10.1073/pnas.1816132116 (2019).
20  Lemonick, S. Is machine learning overhyped? *Chemical & Engineering News* **96**, 16-20 (2018).
21  Levy, M. & Görling, A. Correlation-energy density-functional formulas from correlating first-order density matrices. *Phys Rev A* **52**, R1808-R1810, doi:10.1103/physreva.52.r1808 (1995).
22  Snyder, J. C., Rupp, M., Hansen, K., Müller, K. R. & Burke, K. Finding density functionals with machine learning. *Phys Rev Lett* **108**, 253002, doi:10.1103/PhysRevLett.108.253002 (2012).
23  Goodfellow, I., Bengio, Y. & Courville, A. *Deep learning*.  (2015).
24  Mardirossian, N. & Head-Gordon, M. Thirty years of density functional theory in computational chemistry: an overview and extensive assessment of 200 density functionals. *Mol. Phys.* **115**, 2315-2372 (2017).
25  Sinitskiy, A. V. & Pande, V. S. Deep Neural Network Computes Electron Densities and Energies of a Large Set of Organic Molecules Faster than Density Functional Theory (DFT). *arXiv*, 1809.02723, doi:arXiv:1809.02723 (2018).
26  Ronneberger, O., Fischer, P. & Brox, T. in *Medical Image Computing and Computer-Assisted Intervention – MICCAI 2015. MICCAI 2015. Lecture Notes in Computer Science* Vol. 9351  (eds Navab N, Hornegger J, Wells W, & Frangi A)  (Springer, Cham, 2015).
27  Farimani, A. B., Gomes, J. & Pande, V. S. Deep Learning the Physics of Transport Phenomena. *arxiv*, 1709.02432v02431 (2017).
28  Ramakrishnan, R., Dral, P. O., Rupp, M. & von Lilienfeld, O. A. Quantum chemistry structures and properties of 134 kilo molecules. *Sci. Data* **1**, 140022, doi:10.1038/sdata.2014.22 (2014).
29  Adamo, C. & Barone, V. Toward reliable density functional methods without adjustable parameters: The PBE0 model. *J. Chem. Phys.* **110**, 6158-6169, doi:10.1063/1.478522 (1999).
30  Jensen, F. Basis Set Convergence of Nuclear Magnetic Shielding Constants Calculated by Density Functional Methods. *J. Chem. Theory Comput.* **4**, 719-727, doi:10.1021/ct800013z (2008).
31  Fabrizio, A., Grisafi, A., Meyer, B., Ceriotti, M. & Corminboeuf, C. Electron density learning of non-covalent systems. *Chem. Sci.* **10**, 9424-9432, doi:10.1039/c9sc02696g (2019).
32  Koch, W. & Holthausen, M. C. *A chemist's guide to density functional theory*. 2nd edn,  (Wiley-VCH, 2001).
33  Jones, R. O. Density functional theory: Its origins, rise to prominence, and future. *Rev. Mod. Phys.* **87**, 897 (2015).
34  Wasserman, A. *et al.* The Importance of Being Inconsistent. *Annu Rev Phys Chem* **68**, 555-581, doi:10.1146/annurev-physchem-052516-044957 (2017).
35  Vuckovic, S., Song, S., Kozlowski, J., Sim, E. & Burke, K. Density Functional Analysis: The Theory of Density-Corrected DFT. *J Chem Theory Comput* **15**, 6636-6646, doi:10.1021/acs.jctc.9b00826 (2019).
36  Bogojeski, M., Vogt-Maranto, L., Tuckerman, M. E., Mueller, K.-R. & Burke, K. Density Functionals with Quantum Chemical Accuracy: From Machine Learning to Molecular Dynamics. *ChemRxiv*, 8079917.v8079911, doi:10.26434/chemrxiv.8079917.v1 (2019).
37  Dick, S. & Fernandez-Serra, M. Machine Learning Accurate Exchange and Correlation Functionals of the Electronic Density. *ChemRxiv*, 9947312.v9947312, doi:10.26434/chemrxiv.9947312.v2 (2019).
38  Medvedev, M. G., Bushmarinov, I. S., Sun, J., Perdew, J. P. & Lyssenko, K. A. Response to Comment on "Density functional theory is straying from the path toward the exact functional". *Science* **356**, 496, doi:10.1126/science.aam9550 (2017).

39 Sim, E., Song, S. & Burke, K. Quantifying Density Errors in DFT. *J. Phys. Chem. Lett.* **9**, 6385−6392, doi:10.1021/acs.jpclett.8b02855 (2018).
40 Kim, M. C., Sim, E. & Burke, K. Understanding and reducing errors in density functional calculations. *Phys Rev Lett* **111**, 073003, doi:10.1103/PhysRevLett.111.073003 (2013).
41 Chai, J. D. & Head-Gordon, M. Long-range corrected hybrid density functionals with damped atom-atom dispersion corrections. *Phys. Chem. Chem. Phys.* **10**, 6615-6620, doi:10.1039/b810189b (2008).
42 Lin, Y. S., Li, G. D., Mao, S. P. & Chai, J. D. Long-Range Corrected Hybrid Density Functionals with Improved Dispersion Corrections. *J. Chem. Theory Comput.* **9**, 263-272, doi:10.1021/ct300715s (2013).
43 Perdew, J. P., Burke, K. & Ernzerhof, M. Generalized Gradient Approximation Made Simple. *Phys. Rev. Lett.* **77**, 3865-3868, doi:10.1103/PhysRevLett.77.3865 (1996).
44 Bleiziffer, P., Schaller, K. & Riniker, S. Machine Learning of Partial Charges Derived from High-Quality Quantum-Mechanical Calculations. *J. Chem. Inf. Model.* **58**, 579-590, doi:10.1021/acs.jcim.7b00663 (2018).
45 Kupka, T. *et al.* From CCSD(T)/aug-cc-pVTZ-J to CCSD(T) complete basis set limit isotropic nuclear magnetic shieldings via affordable DFT/CBS calculations. *Magn. Reson. Chem.* **49**, 231-236, doi:10.1002/mrc.2738 (2011).
46 Hedegard, E. D., Jensen, F. & Kongsted, J. Basis Set Recommendations for DFT Calculations of Gas-Phase Optical Rotation at Different Wavelengths. *J. Chem. Theory Comput.* **8**, 4425-4433, doi:10.1021/ct300359s (2012).
47 Reid, D. M., Kobayashi, R. & Collins, M. A. Systematic Study of Locally Dense Basis Sets for NMR Shielding Constants. *J. Chem. Theory Comput.* **10**, 146-152, doi:10.1021/ct4007579 (2014).
48 Neese, F. & Valeev, E. F. Revisiting the Atomic Natural Orbital Approach for Basis Sets: Robust Systematic Basis Sets for Explicitly Correlated and Conventional Correlated ab initio Methods? *J Chem Theory Comput* **7**, 33-43, doi:10.1021/ct100396y (2011).
49 Ji, H. & Jung, Y. A local environment descriptor for machine-learned density functional theory at the generalized gradient approximation level. *J Chem Phys* **148**, 241742, doi:10.1063/1.5022839 (2018).
50 Payne, P. W. Density functionals in unresticted Hartree–Fock theory. *J. Chem. Phys.* **71**, 490, doi:10.1063/1.438124 (1979).
51 Theophilou, A. K. The energy density functional formalism for excited states. *J. Phys. C: Solid State Physics* **12**, 5419, doi:10.1088/0022-3719/12/24/013 (1979).
52 Ernzerhof, M. Taylor-series expansion of density functionals. *Phys Rev A* **50**, 4593-4607, doi:10.1103/physreva.50.4593 (1994).
53 Behler, J. & Parrinello, M. Generalized neural-network representation of high-dimensional potential-energy surfaces. *Phys Rev Lett* **98**, 146401, doi:10.1103/PhysRevLett.98.146401 (2007).
54 Pronobis, W., Tkatchenko, A. & Müller, K. R. Many-Body Descriptors for Predicting Molecular Properties with Machine Learning: Analysis of Pairwise and Three-Body Interactions in Molecules. *J Chem Theory Comput* **14**, 2991-3003, doi:10.1021/acs.jctc.8b00110 (2018).
55 Collins, C. R., Gordon, G. J., von Lilienfeld, O. A. & Yaron, D. J. Constant size descriptors for accurate machine learning models of molecular properties. *J. Chem. Phys.* **148**, 241718, doi:10.1063/1.5020441 (2018).
56 Alred, J. M., Bets, K. V., Xie, Y. & Yakobson, B. I. Machine learning electron density in sulfur crosslinked carbon nanotubes. *Composites Science and Technology* **166**, 3-9, doi:10.1016/j.compscitech.2018.03.035 (2018).
57 Grambow, C. A., Li, Y. P. & Green, W. H. Accurate Thermochemistry with Small Data Sets: A Bond Additivity Correction and Transfer Learning Approach. *J Phys Chem A* **123**, 5826-5835, doi:10.1021/acs.jpca.9b04195 (2019).
58 Faber, F. A. *et al.* Prediction Errors of Molecular Machine Learning Models Lower than Hybrid DFT Error. *J. Chem. Theory Comput.* **13**, 5255-5264, doi:10.1021/acs.jctc.7b00577 (2017).
59 Schutt, K. T. *et al.* SchNet: A continuous-filter convolutional neural network for modeling quantum interactions. *Adv. Neural Inf. Process. Syst.* **30**, 1-11 (2017).
60 Faber, F. A., Christensen, A. S., Huang, B. & von Lilienfeld, O. A. Alchemical and structural distribution based representation for universal quantum machine learning. *J. Chem. Phys.* **148**, 241717, doi:10.1063/1.5020710 (2018).
61 Gastegger, M., Schwiedrzik, L., Bittermann, M., Berzsenyi, F. & Marquetand, P. wACSF-Weighted atom-centered symmetry functions as descriptors in machine learning potentials. *J. Chem. Phys.* **148**, 241709, doi:10.1063/1.5019667 (2018).
62 Gubaev, K., Podryabinkin, E. V. & Shapeev, A. V. Machine learning of molecular properties: Locality and active learning. *J Chem Phys* **148**, 241727, doi:10.1063/1.5005095 (2018).
63 Kranz, J. J., Kubillus, M., Ramakrishnan, R., von Lilienfeld, O. A. & Elstner, M. Generalized Density-Functional Tight-Binding Repulsive Potentials from Unsupervised Machine Learning. *J Chem Theory Comput* **14**, 2341-2352, doi:10.1021/acs.jctc.7b00933 (2018).
64 Lubbers, N., Smith, J. S. & Barros, K. Hierarchical modeling of molecular energies using a deep neural network. *J Chem Phys* **148**, 241715, doi:10.1063/1.5011181 (2018).
65 Unke, O. T. & Meuwly, M. A reactive, scalable, and transferable model for molecular energies from a neural network approach based on local information. *J. Chem. Phys.* **148**, 241708, doi:10.1063/1.5017898 (2018).
66 Sinitskiy, A. V. & Pande, V. S. Physical machine learning outperforms "human learning" in Quantum Chemistry. *arXiv*, 1908.00971v00971, doi:arXiv:1908.00971v1 (2019).
67 Goh, G. B., Hodas, N. O. & Vishnu, A. Deep learning for computational chemistry. *J. Comput. Chem.* **38**, 1291-1307, doi:10.1002/jcc.24764 (2017).
68 Huang, B., Symonds, N. O. & von Lilienfeld, O. A. The fundamentals of quantum machine learning. *arxiv*, 1807.04259v04252 (2018).

69  Rupp, M., von Lilienfeld, O. A. & Burke, K. Guest Editorial: Special Topic on Data-Enabled Theoretical Chemistry. *J Chem Phys* **148**, 241401, doi:10.1063/1.5043213 (2018).
70  Noé, F., Tkatchenko, A., Müller, K.-R. & Clementi, C. Machine learning for molecular simulation. *arXiv*, 1911.02792v02791 (2019).
71  von Lilienfeld, O. A., Müller, K.-R. & Tkatchenko, A. Exploring Chemical Compound Space with Quantum-Based Machine Learning. *arXiv*, 1911.10084v10082 (2019).
72  Mater, A. C. & Coote, M. L. Deep Learning in Chemistry. *J Chem Inf Model* **59**, 2545-2559, doi:10.1021/acs.jcim.9b00266 (2019).
73  Schmidt, E., Fowler, A. T., Elliott, J. A. & Bristowe, P. D. Learning models for electron densities with Bayesian regression. *Computational Materials Science* **149**, 250-258 (2018).
74  Bereau, T., DiStasio, R. A., Jr., Tkatchenko, A. & von Lilienfeld, O. A. Non-covalent interactions across organic and biological subsets of chemical space: Physics-based potentials parametrized from machine learning. *J. Chem. Phys.* **148**, 241706, doi:10.1063/1.5009502 (2018).
75  Nguyen, T. T. *et al.* Comparison of permutationally invariant polynomials, neural networks, and Gaussian approximation potentials in representing water interactions through many-body expansions. *J Chem Phys* **148**, 241725, doi:10.1063/1.5024577 (2018).
76  Imbalzano, G. *et al.* Automatic selection of atomic fingerprints and reference configurations for machine-learning potentials. *J. Chem. Phys.* **148**, 241730, doi:10.1063/1.5024611 (2018).
77  Wu, Z. *et al.* MoleculeNet: a benchmark for molecular machine learning. *Chem. Sci.* **9**, 513-530, doi:10.1039/c7sc02664a (2018).
78  Yao, K., Herr, J. E., Toth, D. W., McKintyre, R. & Parkhill, J. The TensorMol-0.1 model chemistry: a neural network augmented with long-range physics. *Chem. Sci.* **9**, 2261-2269, doi:10.1039/c7sc04934j (2018).
79  Smith, J. S., Nebgen, B., Lubbers, N., Isayev, O. & Roitberg, A. E. Less is more: Sampling chemical space with active learning. *J. Chem. Phys.* **148**, 241733, doi:10.1063/1.5023802 (2018).
80  Zaspel, P., Huang, B., Harbrecht, H. & von Lilienfeld, O. A. Boosting quantum machine learning models with multi-level combination technique: Pople diagrams revisited. *arxiv*, 1808.02799v02792 (2018).
81  Schütt, K. T., Gastegger, M., Tkatchenko, A., Müller, K. R. & Maurer, R. J. Unifying machine learning and quantum chemistry with a deep neural network for molecular wavefunctions. *Nat Commun* **10**, 5024, doi:10.1038/s41467-019-12875-2 (2019).
82  Chmiela, S. *et al.* Machine learning of accurate energy-conserving molecular force fields. *Sci Adv* **3**, e1603015, doi:10.1126/sciadv.1603015 (2017).
83  Christensen, A. S., Faber, F. A. & von Lilienfeld, O. A. Operators in Quantum Machine Learning: Response Properties in Chemical Space. *J. Chem. Phys.* **150**, 064105 (2019).
84  Chmiela, S., Sauceda, H. E., Poltavsky, I., Müller, K.-R. & Tkatchenko, A. sGDML: Constructing accurate and data efficient molecular force fieldsusing machine learning. *Computer Physics Communications* **240**, 38–45, doi:10.1016/j.cpc.2019.02.007 (2019).
85  Schütt, K. T., Tkatchenko, A. & Müller, K.-R. Learning representations of molecules and materials with atomistic neural networks. *arXiv*, 1812.04690v04691 (2018).
86  Schütt, K. T., Gastegger, M., Tkatchenko, A. & Müller, K.-R. in *Explainable AI: Interpreting, Explaining and Visualizing Deep Learning* Vol. 11700  (eds W. Samek *et al.*)  311-330  (Springer, Cham, 2019).
87  Schütt, K. T. *et al.* SchNetPack: A Deep Learning Toolbox For Atomistic Systems. *J Chem Theory Comput* **15**, 448-455, doi:10.1021/acs.jctc.8b00908 (2019).
88  Dick, S. & Fernandez-Serra, M. Learning from the density to correct total energy and forces in first principle simulations. *J Chem Phys* **151**, 144102, doi:10.1063/1.5114618 (2019).
89  Zubatyuk, R., Smith, J. S., Leszczynski, J. & Isayev, O. Accurate and transferable multitask prediction of chemical properties with an atoms-in-molecules neural network. *Sci Adv* **5**, eaav6490, doi:10.1126/sciadv.aav6490 (2019).
90  Chandrasekaran, A. *et al.* Solving the electronic structure problem with machine learning. *npj Computational Materials* **5**, 22 (2019).
91  Lot, R., Pellegrini, F., Shaidu, Y. & Kucukbenli, E. PANNA: Properties from Artificial Neural Network Architectures. *arXiv*, 1907.03055v03051 (2019).
92  Pasini, M. L. *et al.* Fast, accurate predictions of total energy of solid solution alloys. *arXiv*, 1912.11152v11151 (2019).
93  Chen, C., Ye, W., Zuo, Y., Zheng, C. & Ong, S. P. Graph Networks as a Universal Machine Learning Framework for  Molecules and Crystals   *Chem. Mater.* **31**, 3564−3572, doi:10.1021/acs.chemmater.9b01294  (2019).
94  Lu, J., Wang, C. & Zhang, Y. Predicting Molecular Energy Using Force-Field Optimized Geometries and Atomic Vector Representations Learned from an Improved Deep Tensor Neural Network. *J Chem Theory Comput* **15**, 4113-4121, doi:10.1021/acs.jctc.9b00001 (2019).
95  Chen, G. *et al.* Alchemy: A Quantum Chemistry Dataset for Benchmarking AI Models. *arXiv*, 1906.09427v09421 (2019).
96  Smith, J. S. *et al.* The ANI-1ccx and ANI-1x Data Sets, Coupled-Cluster and Density Functional Theory Properties for Molecules. *ChemRxiv*, 10050737.v10050731, doi:10.26434/chemrxiv.10050737.v1 (2019).
97  Sauceda, H. E., Poltavsky, I., Chmiela, S., Müller, K.-R. & Tkatchenko, A. Construction of Machine Learned Force Fields with Quantum Chemical Accuracy: Applications and Chemical Insights. *arXiv*, 1909.08565v08561 (2019).
98  Unke, O. T. & Meuwly, M. Machine Learning Potential Energy Surfaces. *arXiv*, 1909.08027v08021 (2019).
99  Zhang, Y., Hu, C. & Jiang, B. Embedded Atom Neural Network Potentials: Efficient and Accurate Machine Learning with a Physically Inspired Representation. *J Phys Chem Lett*, 4962-4967, doi:10.1021/acs.jpclett.9b02037 (2019).
100    Sauceda, H. E., Chmiela, S., Poltavsky, I., Müller, K. R. & Tkatchenko, A. Molecular force fields with gradient-domain machine learning: Construction and application to dynamics of small molecules with coupled cluster forces. *J Chem Phys* **150**, 114102, doi:10.1063/1.5078687 (2019).

101	Zubatyuk, T. *et al.* Machine Learned Hückel Theory: Interfacing Physics and Deep Neural Networks. *arXiv*, 1909.12963v12961 (2019).
102	Cheng, L., Kovachki, N. B., Welborn, M. & Miller, T. F. Regression Clustering for Improved Accuracy and Training Costs with Molecular-Orbital-Based Machine Learning. *J Chem Theory Comput* **15**, 6668-6677, doi:10.1021/acs.jctc.9b00884 (2019).
103	Wang, Y., Wang, X., Truhlar, D. G. & He, X. How Well Can the M06 Suite of Functionals Describe the Electron Densities of Ne, Ne. *J Chem Theory Comput* **13**, 6068-6077, doi:10.1021/acs.jctc.7b00865 (2017).
104	Gould, T. What Makes a Density Functional Approximation Good? Insights from the Left Fukui Function. *J Chem Theory Comput* **13**, 2373-2377, doi:10.1021/acs.jctc.7b00231 (2017).
105	Mezei, P. D. & von Lilienfeld, O. A. Non-covalent quantum machine learning corrections to density functionals. *arXiv*, 1903.09010v09011 (2019).
106	Christensen, A. S. & von Lilienfeld, O. A. Operator quantum machine learning: Navigating the chemical space of response properties. *arXiv*, 1910.14418v14411 (2019).
107	Westermayr, J., Faber, F. A., Christensen, A. S., von Lilienfeld, O. A. & Marquetand, P. Neural networks and kernel ridge regression for excited states dynamics of $CH_2NH_2^+$: From single-state to multi-state representations and multi-property machine learning models. *arXiv*, 1912.08484v08481 (2019).
108	Welborn, M., Cheng, L. & Miller III, T. F. Transferability in Machine Learning for Electronic Structure via the Molecular Orbital Basis. *arxiv*, 1806.00133v00132 (2018).
109	Margraf, J. T. & Reuter, K. Making the Coupled Cluster Correlation Energy Machine-Learnable. *J Phys Chem A* **122**, 6343-6348, doi:10.1021/acs.jpca.8b04455 (2018).
110	Grisafi, A., Wilkins, D. M., Csányi, G. & Ceriotti, M. Symmetry-Adapted Machine Learning for Tensorial Properties of Atomistic Systems. *Phys Rev Lett* **120**, 036002, doi:10.1103/PhysRevLett.120.036002 (2018).
111	Janet, J. P. *et al.* Designing in the Face of Uncertainty: Exploiting Electronic Structure and Machine Learning Models for Discovery in Inorganic Chemistry. *Inorg Chem* **58**, 10592-10606, doi:10.1021/acs.inorgchem.9b00109 (2019).
112	Veit, M. *et al.* Equation of State of Fluid Methane from First Principles with Machine Learning Potentials. *J Chem Theory Comput* **15**, 2574-2586, doi:10.1021/acs.jctc.8b01242 (2019).
113	Faber, F. A. & von Lilienfeld, O. A. in *Materials Informatics: Methods, Tools and Applications* (eds Olexandr Isayev, Alexander Tropsha, & Stefano Curtarolo) 171-179 (Wiley-VCH Verlag GmbH & Co. KGaA, 2019).
114	Pakravan, S., Mistani, P. A., Aragon-Calvo, M. A. & Gibou, F. Solving inverse-PDE problems with physics-aware neural networks. *arXiv*, 2001.03608v03601 (2020).
115	Montavon, G. *et al.* Machine learning of molecular electronic properties in chemical compound space. *New Journal of Physics* **15**, 095003 (2013).
116	Nebgen, B. *et al.* Transferable Dynamic Molecular Charge Assignment Using Deep Neural Networks. *J. Chem. Theory Comput.* **14**, 4687–4698, doi:10.1021/acs.jctc.8b00524 (2018).
117	Sifain, A. E. *et al.* Discovering a Transferable Charge Assignment Model Using Machine Learning. *J. Phys. Chem. Lett.* **9**, 4495–4501, doi:10.1021/acs.jpclett.8b01939 (2018).
118	Yao, K. & Parkhill, J. Kinetic Energy of Hydrocarbons as a Function of Electron Density and Convolutional Neural Networks. *J. Chem. Theory Comput.* **12**, 1139-1147, doi:10.1021/acs.jctc.5b01011 (2016).
119	Seino, J., Kageyama, R., Fujinami, M., Ikabata, Y. & Nakai, H. Semi-local machine-learned kinetic energy density functional with third-order gradients of electron density. *J. Chem. Phys.* **148**, 241705, doi:10.1063/1.5007230 (2018).
120	Hollingsworth, J., Li, L., Baker, T. E. & Burke, K. Can exact conditions improve machine-learned density functionals? *J Chem Phys* **148**, 241743, doi:10.1063/1.5025668 (2018).
121	Golub, P. & Manzhos, S. Kinetic energy densities based on the fourth order gradient expansion: performance in different classes of materials and improvement via machine learning. *Phys Chem Chem Phys* **21**, 378-395, doi:10.1039/c8cp06433d (2018).
122	Nagai, R., Akashi, R., Sasaki, S. & Tsuneyuki, S. Neural-network Kohn-Sham exchange-correlation potential and its out-of-training transferability. *J Chem Phys* **148**, 241737, doi:10.1063/1.5029279 (2018).
123	Nagai, R., Akashi, R. & Sugino, O. Completing density functional theory by machine-learning hidden messages from molecules. *arXiv*, 1903.00238v00235 (2019).
124	Ryczko, K., Strubbe, D. A. & Tamblyn, I. Deep learning and density-functional theory. *Phys. Rev. A* **100**, 022512, doi:10.1103/PhysRevA.100.022512 (2019).
125	Kanungo, B., Zimmerman, P. M. & Gavini, V. Exact exchange-correlation potentials from ground-state electron densities. *Nat Commun* **10**, 4497, doi:10.1038/s41467-019-12467-0 (2019).
126	Szegedy, C. *et al.* Intriguing properties of neural networks. *arxiv*, 1312.6199v1314 (2013).
127	Kurakin, A., Goodfellow, I. & Bengio, S. Adversarial examples in the physical world. *arxiv*, 1607.02533v02534 (2016).
128	McCloskey, K., Taly, A., Monti, F., Brenner, M. P. & Colwell, L. J. Using attribution to decode binding mechanism in neural network models for chemistry. *Proc Natl Acad Sci U S A* **116**, 11624-11629, doi:10.1073/pnas.1820657116 (2019).
129	Nakata, M. & Shimazaki, T. PubChemQC Project: A Large-Scale First-Principles Electronic Structure Database for Data-Driven Chemistry. *J Chem Inf Model* **57**, 1300-1308, doi:10.1021/acs.jcim.7b00083 (2017).
130	Smith, J. S., Isayev, O. & Roitberg, A. E. ANI-1, A data set of 20 million calculated off-equilibrium conformations for organic molecules. *Sci. Data* **4**, 170193, doi:10.1038/sdata.2017.193 (2017).
131	Ghahremanpour, M. M., van Maaren, P. J. & van der Spoel, D. The Alexandria library, a quantum-chemical database of molecular properties for force field development. *Sci. Data* **5**, 180062, doi:10.1038/sdata.2018.62 (2018).

132	Feller, D. The Role of Databases in Support of Computational Chemistry Calculations. *J. Comp. Chem.* **17**, 1571-1586 (1996).
133	Schuchardt, K. L. *et al.* Basis set exchange: a community database for computational sciences. *J. Chem. Inf. Model.* **47**, 1045-1052, doi:10.1021/ci600510j (2007).